\documentstyle[twoside]{article}

\catcode`\@=11
\long\def\@makefntext#1{
\protect\noindent \hbox to 3.2pt {\hskip-.9pt  
$^{{\eightrm\@thefnmark}}$\hfil}#1\hfill}		

\def\thefootnote{\fnsymbol{footnote}}
\def\@makefnmark{\hbox to 0pt{$^{\@thefnmark}$\hss}}	
	
\def\ps@myheadings{\let\@mkboth\@gobbletwo
\def\@oddhead{\hbox{}
\rightmark\hfil\eightrm\thepage}   
\def\@oddfoot{}\def\@evenhead{\eightrm\thepage\hfil
\leftmark\hbox{}}\def\@evenfoot{}
\def\sectionmark##1{}\def\subsectionmark##1{}}

\oddsidemargin=\evensidemargin
\renewcommand{\thefootnote}{\fnsymbol{footnote}}

\newcounter{sectionc}\newcounter{subsectionc}\newcounter{subsubsectionc}
\renewcommand{\section}[1] {\vspace{12pt}\addtocounter{sectionc}{1} 
\setcounter{subsectionc}{0}\setcounter{subsubsectionc}{0}\noindent 
	{\tenbf\thesectionc. #1}\par\vspace{5pt}}
\renewcommand{\subsection}[1] {\vspace{12pt}\addtocounter{subsectionc}{1} 
	\setcounter{subsubsectionc}{0}\noindent 
	{\bf\thesectionc.\thesubsectionc. {\kern1pt \bfit #1}}\par\vspace{5pt}}
\renewcommand{\subsubsection}[1] {\vspace{12pt}\addtocounter{subsubsectionc}{1}
	\noindent{\tenrm\thesectionc.\thesubsectionc.\thesubsubsectionc.
	{\kern1pt \tenit #1}}\par\vspace{5pt}}

\newcounter{appendixc}
\newcounter{subappendixc}[appendixc]
\newcounter{subsubappendixc}[subappendixc]
\renewcommand{\thesubappendixc}{\Alph{appendixc}.\arabic{subappendixc}}
\renewcommand{\thesubsubappendixc}
	{\Alph{appendixc}.\arabic{subappendixc}.\arabic{subsubappendixc}}

\renewcommand{\appendix}[1] {\vspace{12pt}
        \refstepcounter{appendixc}
        \setcounter{figure}{0}
        \setcounter{table}{0}
        \setcounter{lemma}{0}
        \setcounter{theorem}{0}
        \setcounter{corollary}{0}
        \setcounter{definition}{0}
        \setcounter{equation}{0}
        \renewcommand{\thefigure}{\Alph{appendixc}.\arabic{figure}}
        \renewcommand{\thetable}{\Alph{appendixc}.\arabic{table}}
        \renewcommand{\theappendixc}{\Alph{appendixc}}
        \renewcommand{\thelemma}{\Alph{appendixc}.\arabic{lemma}}
        \renewcommand{\thetheorem}{\Alph{appendixc}.\arabic{theorem}}
        \renewcommand{\thedefinition}{\Alph{appendixc}.\arabic{definition}}
        \renewcommand{\thecorollary}{\Alph{appendixc}.\arabic{corollary}}
        \renewcommand{\theequation}{\Alph{appendixc}.\arabic{equation}}
        \noindent{\tenbf Appendix \theappendixc #1}\par\vspace{5pt}}
\newcommand{\subappendix}[1] {\vspace{12pt}
        \refstepcounter{subappendixc}
        \noindent{\bf Appendix \thesubappendixc. {\kern1pt \bfit #1}}
	\par\vspace{5pt}}
\newcommand{\subsubappendix}[1] {\vspace{12pt}
        \refstepcounter{subsubappendixc}
        \noindent{\rm Appendix \thesubsubappendixc. {\kern1pt \tenit #1}}
	\par\vspace{5pt}}

\newcommand{\textlineskip}{\baselineskip=13pt}
\newcommand{\smalllineskip}{\baselineskip=10pt}

\def\eightcirc{
\begin{picture}(0,0)
\put(4.4,1.8){\circle{6.5}}
\end{picture}}
\def\eightcopyright{\eightcirc\kern2.7pt\hbox{\eightrm c}} 

\newcommand{\copyrightheading}[1]
	{\vspace*{-2.5cm}\smalllineskip{\flushleft
	{\footnotesize International Journal of Modern Physics E, #1}\\
	{\footnotesize $\eightcopyright$\, World Scientific Publishing
	 Company}\\
	 }}

\newcommand{\publisher}[2]{{\begin{center}\footnotesize\smalllineskip 
	Received #1\\
	Revised #2
	\end{center}
	}}

\renewenvironment{thebibliography}[1]
	{\frenchspacing
	 \ninerm\baselineskip=11pt
	 \begin{list}{\arabic{enumi}.}
        {\usecounter{enumi}\setlength{\parsep}{0pt}     
	 \setlength{\leftmargin 12.7pt}{\rightmargin 0pt} 
         \setlength{\itemsep}{0pt} \settowidth
	{\labelwidth}{#1.}\sloppy}}{\end{list}}

\newcounter{itemlistc}
\newcounter{romanlistc}
\newcounter{alphlistc}
\newcounter{arabiclistc}

\newcommand{\fcaption}[1]{
        \refstepcounter{figure}
        \setbox\@tempboxa = \hbox{\footnotesize Fig.~\thefigure. #1}
        \ifdim \wd\@tempboxa > 5in
           {\begin{center}
        \parbox{5in}{\footnotesize\smalllineskip Fig.~\thefigure. #1}
            \end{center}}
        \else
             {\begin{center}
             {\footnotesize Fig.~\thefigure. #1}
              \end{center}}
        \fi}

\newcommand{\tcaption}[1]{
        \refstepcounter{table}
        \setbox\@tempboxa = \hbox{\footnotesize Table~\thetable. #1}
        \ifdim \wd\@tempboxa > 5in
           {\begin{center}
        \parbox{5in}{\footnotesize\smalllineskip Table~\thetable. #1}
            \end{center}}
        \else
             {\begin{center}
             {\footnotesize Table~\thetable. #1}
              \end{center}}
        \fi}

\def\@citex[#1]#2{\if@filesw\immediate\write\@auxout
	{\string\citation{#2}}\fi
\def\@citea{}\@cite{\@for\@citeb:=#2\do
	{\@citea\def\@citea{,}\@ifundefined
	{b@\@citeb}{{\bf ?}\@warning
	{Citation `\@citeb' on page \thepage \space undefined}}
	{\csname b@\@citeb\endcsname}}}{#1}}

\newif\if@cghi
\def\cite{\@cghitrue\@ifnextchar [{\@tempswatrue
	\@citex}{\@tempswafalse\@citex[]}}
\def\citelow{\@cghifalse\@ifnextchar [{\@tempswatrue
	\@citex}{\@tempswafalse\@citex[]}}
\def\@cite#1#2{{$\null^{#1}$\if@tempswa\typeout
	{IJCGA warning: optional citation argument 
	ignored: `#2'} \fi}}

\def\pmb#1{\setbox0=\hbox{#1}
	\kern-.025em\copy0\kern-\wd0
	\kern.05em\copy0\kern-\wd0
	\kern-.025em\raise.0433em\box0}

\def\fnt#1#2{\footnotetext{\kern-.3em
	{$^{\mbox{\scriptsize #1}}$}{#2}}}

\def\fpage#1{\begingroup
\voffset=.3in
\thispagestyle{empty}\begin{table}[b]\centerline{\footnotesize #1}
	\end{table}\endgroup}

\font\tenrm=cmr10
\font\tenit=cmti10 
\font\tenbf=cmbx10
\font\bfit=cmbxti10 at 10pt
\font\ninerm=cmr9

\font\eightrm=cmr8

\def\qed{\hbox{${\vcenter{\vbox{			
   \hrule height 0.4pt\hbox{\vrule width 0.4pt height 6pt
   \kern5pt\vrule width 0.4pt}\hrule height 0.4pt}}}$}}

\renewcommand{\thefootnote}{\fnsymbol{footnote}}	

\def\bsc{{\sc a\kern-6.4pt\sc a\kern-6.4pt\sc a}}	
\def\bflatex{\bf L\kern-.30em\raise.3ex\hbox{\bsc}\kern-.14em 
T\kern-.1667em\lower.7ex\hbox{E}\kern-.125em X} 

\begin{document}


\normalsize\textlineskip
\thispagestyle{empty}
\setcounter{page}{1}

\copyrightheading{}			

\vspace*{0.88truein}

\fpage{1}
\centerline{\bf ELECTROMAGNETIC MESON PRODUCTION}
\centerline{\bf IN THE NUCLEON RESONANCE REGION}
\vspace*{0.035truein}
\vspace*{0.37truein}
\centerline{\footnotesize V. D. Burkert\footnote{
email:burkert@jlab.org}}
\vspace*{0.015truein}
\centerline{\footnotesize\it Thomas Jefferson National Accelerator Facility}
\baselineskip=10pt
\centerline{\footnotesize\it 12000 Jefferson Avenue, Virginia 23606, USA }
\vspace*{10pt}
\centerline{\footnotesize and}
\vspace*{10pt}
\centerline{\footnotesize T.-S. H. Lee\footnote{email:lee@theory.phy.anl.gov}}
\vspace*{0.015truein}
\centerline{\footnotesize\it Physics Division, Argonne National Laboratory}
\baselineskip=10pt
\centerline{\footnotesize\it Argonne, Illinois 60439, USA}
\vspace*{0.225truein}
\publisher{(received date)}{(revised date)}
\today
\vspace*{0.21truein}
\begin{abstract}

Recent experimental and theoretical advances in investigating 
 electromagnetic meson production reactions in the nucleon resonance
region are reviewed. The article gives a description 
of current experimental facilities with electron and photon 
beams and presents an unified derivation of most of the phenomenological 
approaches being used to extract the resonance parameters from the data.
The  analyses of $\pi$ and $\eta$ production data and the
 resulting transition form factors for the $\Delta(1232)P_{33}$, N(1535)$S_{11}$,
 N(1440)$P_{11}$, and N(1520)$D_{13}$
resonances are discussed in detail. 
 The status of our understanding of
the reactions with production of two pions, kaons, and vector mesons
is also reviewed.
\end{abstract}

\vspace*{1pt}\textlineskip	

\textheight=7.8truein
\setcounter{footnote}{0}
\renewcommand{\thefootnote}{\alph{footnote}}

\section{Introduction}
\label{sec:introduction}

The quest for understanding the structure and interaction of hadrons has been 
the motivation of strong interaction physics for decades. 
The advent of Quantum Chromodynamics (QCD) \cite{qcd} has led to a general 
theoretical description of the strong interaction in terms of the fundamental 
constituents, quarks and gluons. 
At very high energies, perturbative methods have proven very effective in the
description of many processes. 
However, because of the complexity of the 
theory, we are still a long way from being able to describe the strong force as 
it is manifest in the structure  of  baryons and mesons. 
The most fundamental approach  to resolve this difficulty is 
to develop accurate numerical simulations of QCD on the Lattice
 (Lattice QCD)\cite{lqcd}.
Alternatively, hadron models with effective degrees of freedom have been constructed for interpreting data.  For example, near threshold
pion-pion scattering, pion-nucleon scattering, and pion 
photoproduction 
can be successfully described by chiral perturbation
theory\cite{chpt} which is formulated in terms of hadron degrees 
of freedom and constrained only by the symmetry properties of QCD.
The constituent quark model\cite{isgkar,capstick86} is another successful, though not fully understood, example.
In some cases, the results from these two different theoretical efforts 
are complementary in understanding the data and making predictions for
future experiments.

For heavy quark systems, Lattice QCD (LQCD) can now predict accurate  
quantities for interpreting the data from, for example, $B$ meson facilities.  
For light-quark systems, the small quark
masses are difficult to implement, and approximations have been 
necessary in Lattice QCD calculations. 
Nevertheless, significant progress has been made in calculating some
basic properties of baryons, such
as masses of ground states, as well as of low lying excited
states \cite{lqcd1,lqcd2,lqcd3,lqcd4}. Even the first  LQCD calculation of
the electromagnetic transition form factors from the ground state proton to
the first excited state, the $\Delta(1232)$,  has been attempted 
recently\cite{alexandrou03}.
However, reliable Lattice QCD calculations
 for electromagnetic meson production reactions, 
the subject of this article, seem to be in the distant future. 
In the foreseeable future, models of
hadron structure and reactions will likely continue to play an important role and provide
theoretical guidance for experimenters.  

The development of hadron models for the nucleon and nucleon 
resonances ($N^*$) has a long history.
In the past three decades, the constituent quark model has
been greatly refined to account for 
residual quark-quark interactions 
due to one-gluon-exchange\cite{isgkar,capstick86} and/or
Goldstone boson exchange\cite{risglz,graz}. 
Efforts are underway to re-formulate the model within the relativistic
quantum mechanics\cite{desp,coeris,graz1}.
Conceptually completely different models have also 
been developed, such as bag models\cite{mitbag},
 chiral bag models\cite{brown,thomas1}, 
algebraic models\cite{bijker}, soliton models\cite{tdlee},
color dielectric models\cite{neil} 
Skyrme models\cite{schwes}, and 
covariant models based on Dyson-Schwinger equations\cite{robe}. 
With suitable phenomenlogical procedures,
most of these models
are comparable in reproducing the {\it low-lying} $N^*$
spectra as determined by
the amplitude analyses of elastic $\pi N$ scattering.
However they have rather different predictions on the number and ordering of the
highly excited $N^*$ states. They also differ significantly in predicting
some dynamical quantities such as the
electromagnetic and mesonic $N$-$N^*$ transition form factors.
Clearly, accurate experimental information for these $N^*$ observables is needed to
distinguish these models.  This information can be extracted from
the data of electromagnetic meson production reactions.
In the past few years, such data with high precision have been extensively
accumulated at Thomas Jefferson National Accelerator Facility(JLab), MIT-Bates, and 
LEGS of Brookhaven National Laboratory in the United States,
MAMI of Mainz and ELSA of Bonn in Germany, GRAAL of Grenoble in France, and LEPS of
Spring-8  in Japan.
In this paper we will review these xperimental developments and the status of
our understanding of the data accumulated in recent years.
Our focus will be on the study of $N^*$ excitations. The 
use of these data for other investigations
will not be covered.

It is useful to briefly describe here the recent advances in using the
new data to address
some of the long-standing problems in the study of $N^*$ physics.
The first one is the so-called {\it missing resonance problem}.
This problem originated from the observation
that some of the $N^*$ states predicted by the constituent quark model are not seen
in the baryon spectra determined mainly from the amplitude analyses of
$\pi N$ elastic scattering. There are two possible solutions for
this problem.  First, it is possible that the constituent quark model
has wrong effective degrees of freedom of QCD in describing the
highly excited baryon states. 
Other models with fewer degrees of freedom, such as quark-diquark models
or  models based on alternative symmetry schemes\cite{kirchbach}, 
could be more accurate in reproducing the 
baryon spectra.
The second  possibility is
that these missing resonances do not couple strongly with the $\pi N$ channel
and can only be observed in other processes,  as suggested
by Isgur and Koniuk\cite{koniuk80} in 1980.
Data from the experiments measuring  as many meson-baryon
channels as possible
are needed to resolve the missing resonance problem. 

The second outstanding problem in the study of $N^*$ physics is that the
partial decay widths of baryon resonances compiled and published periodically by the
Particle Data Group (PDG) have very large uncertainties in most 
 cases~\cite{pdg04}. For some decay channels, such as $\eta N$, $K\Sigma$ and $\omega N$, the
large uncertainties are mainly due to insufficient data. But the discrepancies between
the results from using different amplitude analysis methods is also a source of the uncertainties.
This problem can be resolved only with a sufficiently large data base that 
allows much stronger constraints on amplitude analyses, and a strong reduction of the model dependence 
of the extracted partial decay widths as well as other $N^*$ parameters.
This requires that the data must be precise and must
cover very large kinematic regions in scattering angles,
energies, and momentum transfers. The data of polarization observables must also be 
as extensive as possible.

The above two experimental challenges have been met with the operations of the
electron and photon facilities mentioned above.
These facilities are also equipped with sophisticated detectors
for measuring not only the dominant single pion channel but also 
kaon, vector meson, and two-pion channels.
The CEBAF Large Acceptance Spectrometer (CLAS) 
at JLab is the most complete and advanced detector in the field.

The third long-standing problem is in the theoretical interpretations of the $N^*$ parameters
listed by the PDG. Most of the model 
predictions on $N^* \rightarrow \gamma N$ helicity
are only in a very qualitative agreement with the PDG values.
In some cases, they disagree even in signs.
One could attribute this to the large experimental 
uncertainties, as discussed above.
However, the well 
determined empirical values of the simplest and most unambiguous
$\Delta\rightarrow \gamma N$ helicity amplitudes are 
about $40 \%$ larger than the predictions from practically all of the
hadron models mentioned above.
This raises the question about how the hadron models as well as the Lattice QCD calculations
are related to the  $N^*$ parameters extracted from empirical amplitude
analyses.
We need to evaluate critically their relationships from the
point of view of fundamental reaction theory.
 The discrepancies in the $\Delta(1232)$ region must be understood before
meaningful comparisons between theoretical predictions and empirical values can
be made. Much progress has been made in this area.
The results, as will be detailed in section 5.1, strongly indicate that it is
necessary to apply an appropriate reaction theory in making
 meaningful comparisons
of the empirical values from amplitude analyses and
the predictions from hadron models and LQCD.

\begin{figure}[tbh]
\vspace{8cm}
\includegraphics{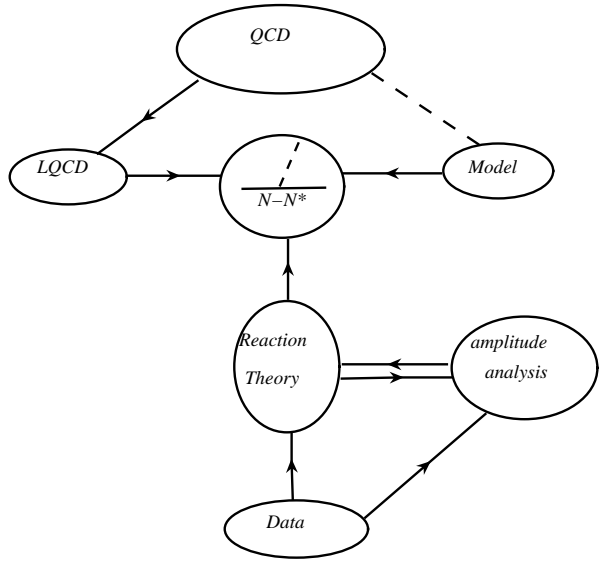}
\caption{Scheme for $N^*$ study }
\label{fig:flow}
\end{figure}

Summing up the above discussions, it is clear that 
in the absence of a fundamental solution of QCD in the resonance regions,
the study of $N^*$ excitations needs close collaborations between theoretical and experimental
efforts. This is illustrated in Fig.~\ref{fig:flow}.
On the theoretical side, we need to use Lattice QCD calculations and/or 
hadron structure models
to predict properties of nucleon resonances, such as the $N$-$N^*$ transition
form factors indicated in Fig.~\ref{fig:flow}. On the experimental side, we need to
accumulate sufficiently extensive and precise data of 
meson production reactions.
We then must develop reaction models for interpreting the data in terms of
hadron structure calculations. The development of
empirical amplitude analyses of the data
is an important part of this task. 

In section 2, the current experimental facilities will be reviewed. 
The general formulation for calculating cross sections of
electromagnetic meson production is presented in section 3.
Section 4 is devoted to provide a unified derivation of models 
used in the interpretation of the data. Results are presented in
section 5. Concluding remarks and outlook are given in section 6.

\newpage

\section{Experimental Facilities}

\subsection{Thomas Jefferson National Accelerator Facility }
\label{sec:jlab}
\noindent
The Thomas Jefferson National Accelerator Facility (JLab)
in Newport News, Virginia, 
operates a CW electron accelerator with energies in the range up to 6 GeV~\cite{grunder}. 
Three experimental Halls receive highly polarized electron beams with the same, or with different but correlated energies, simultaneously. Beam currents in the range from 0.1 nA to 
150 $\mu$A can be delivered to the experiments, simultaneously. 

\subsubsection{Experimental Hall A - HRS$^2$}
\label{sec:halla}
Hall A houses a pair of identical focussing magnetic spectrometers~\cite{halla} with high 
resolution (HRS$^2$), each
with a momentum resolution of $\Delta p /p \sim 2\times 10^{-4}$, one of them 
is instrumented with a gas Cerenkov counter and a shower counter for 
the identification of electrons. The hadron arm is instrumented with 
a proton recoil polarimeter. 
The detector package allows 
identification of charged pions, kaons, and protons.  
A polarized $^3He$ target is used for experiments that require polarized 
neutron targets.
The HRS$^2$ spectrometers have been used to measure the 
reaction $\vec{e} p \rightarrow e\vec{p}\pi^{\circ}$ in the $\Delta(1232)$ 
region and to extract various single and double polarization response functions.

\subsubsection{Experimental Hall B - CLAS}

\begin{figure}[thb]
\vspace{8cm}
\includegraphics{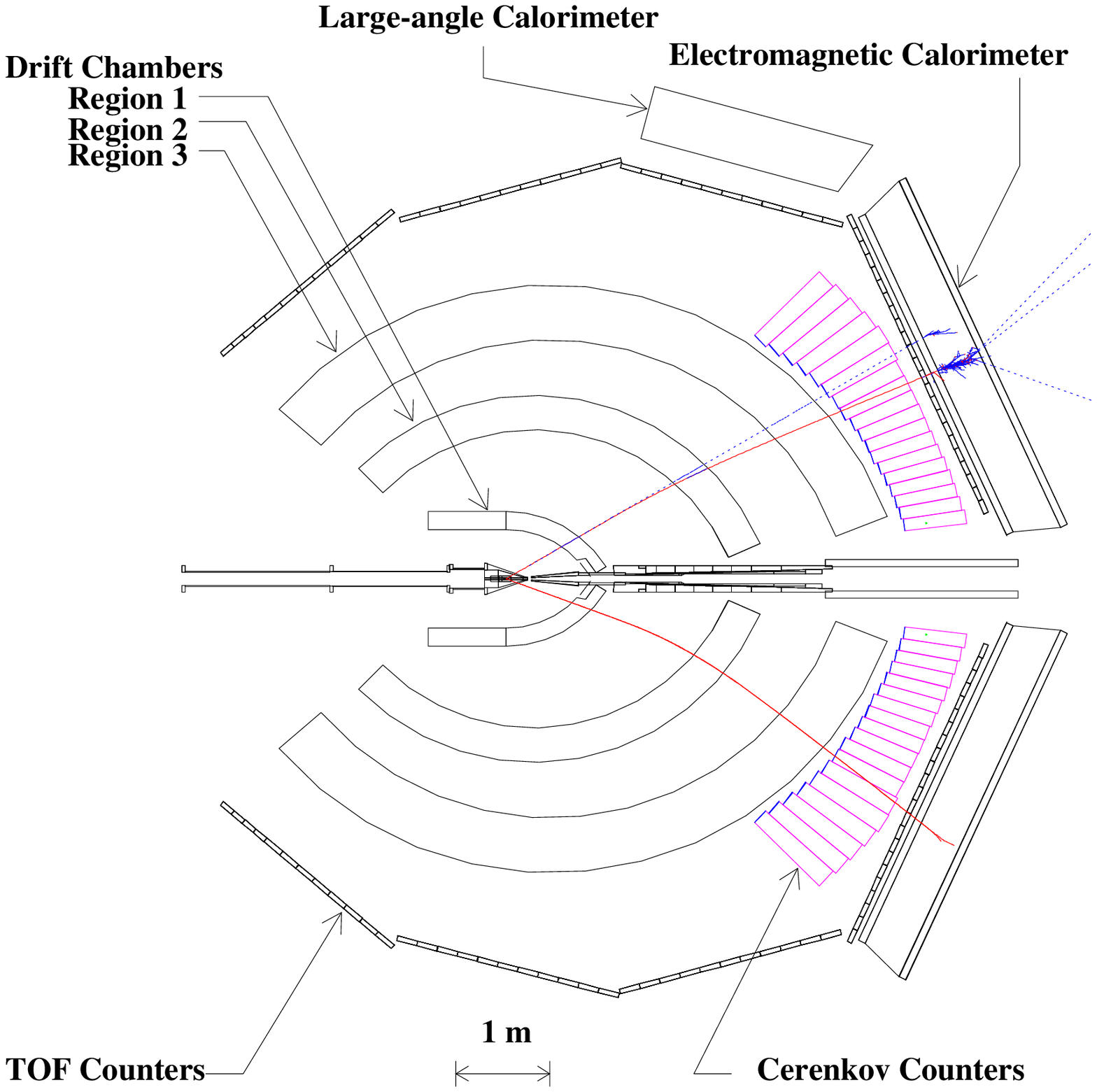}   
\includegraphics{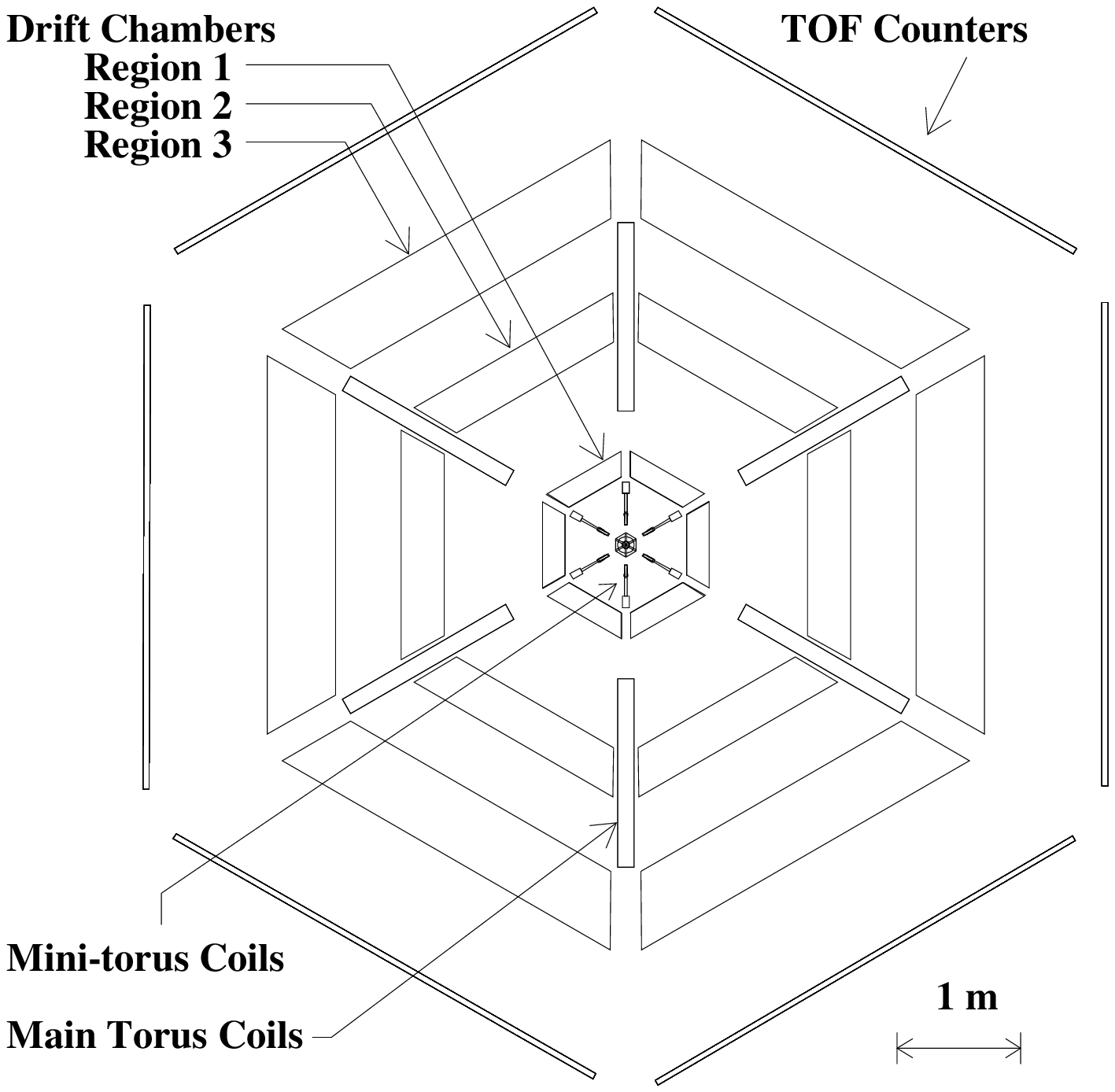}   
\caption{\small{The CLAS detector at JLab.  Left panel: The longitudinal cut along the
beam line shows the 3 drift chamber regions, the Cherenkov counter at forward angles, the time-of-flight (TOF) system, 
and the electromagnetic calorimeters.  A simulated event shows an 
electron (upper) and a 
positively charged hadron. Right panel: Transverse cut through CLAS. The six 
superconducting coils provide a six sector structure with independent detectors. 
A polarized target can be inserted into the large bore near the center.}}
\label{fig:clas_detector}
\end{figure}

\begin{figure}[thb]
\vspace{6cm}
\includegraphics{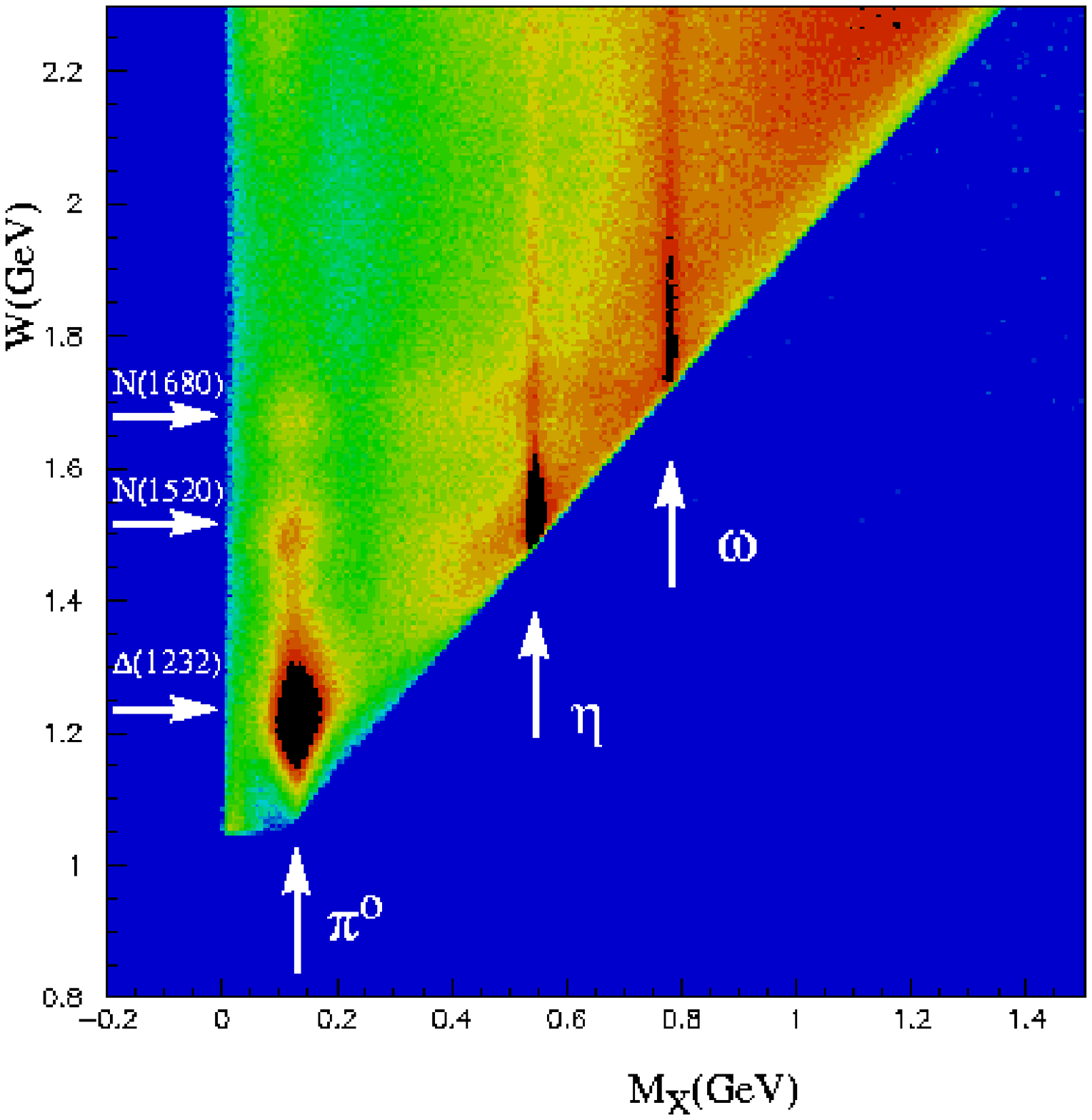}   
\includegraphics{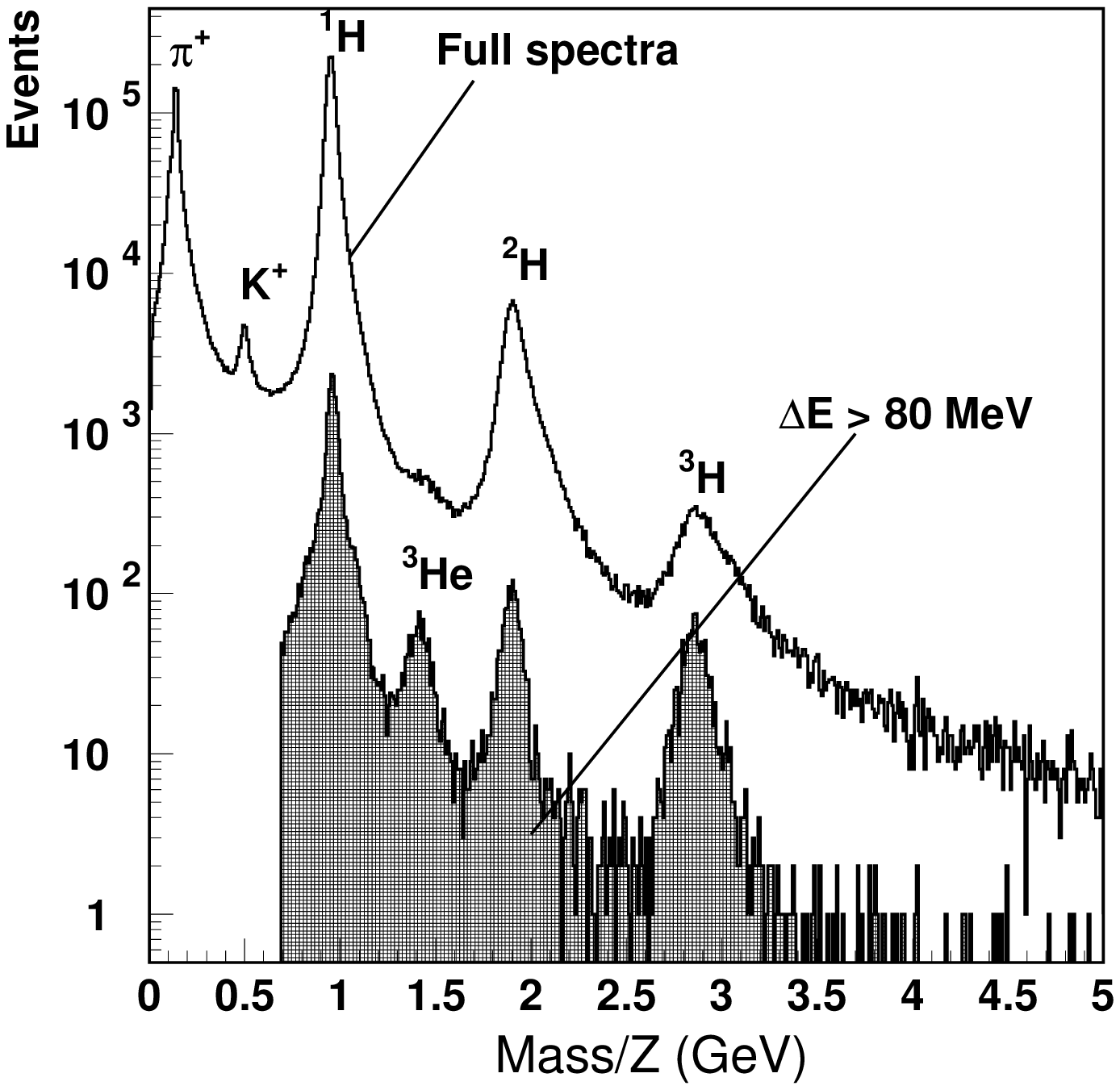} 
\caption{\small{Left panel: 
Charged particle identification in CLAS. 
The reconstructed mass/Z(charge number) 
for positive tracks from a carbon target is shown. Additional sensitivity to 
high-mass particles is obtained by requiring large energy loss in the 
scintillators (shaded histogram). Right panel: Invariant mass versus missing mass for 
$ep\rightarrow epX$ at an electron beam energy of 4 GeV. }}
\label{fig:clas_pid}
\end{figure}

Hall B houses the CEBAF Large Acceptance Spectrometer (CLAS)
 detector, and a photon energy tagging facility~\cite{clas}. 
CLAS can be operated with electron beams and with energy tagged photon beams.
The photon beam can be either unpolarized or can be linearly or circularly 
 polarized.
The detector system was designed specifically with the detection of 
multiple particle final states in mind. The driving motivation for the construction of CLAS 
was the nucleon resonances ($N^*$) program, with the emphasis on the
 study of the $\gamma NN^*$ and 
$\gamma N\Delta^*$ transition form factors, and the 
search for {\it missing resonances}.
Figure \ref{fig:clas_detector} shows 
the CLAS detector. 
At the core of the detector is a toroidal magnet consisting of six superconducting coils 
symmetrically arranged around the beam line. Each of the six sectors is instrumented as an independent 
spectrometer with 34 layers of tracking chambers allowing
for the full reconstruction of the charged particle 3-momentum vectors. 
Charged hadron identification is accomplished by combining momentum and 
time-of-flight, and the measured path length from the target to the 
plastic scintillation counters which surround the entire tracking chambers.
Timing resolutions of 
$\Delta{T} = 120 - 200$~psec (rms) are achieved, depending on the length of the
 scintillator bar which ranges from 30~cm to 350~cm. Mass and charge number (Z)
 reconstruction 
is shown in the left panel of Fig.~\ref{fig:clas_pid}. 
Protons and pions can be separated for 
momenta up to 4 GeV/c, and pions and kaons up to about 2 GeV/c. 
The wide range of particle identification allows to study the complete 
range of reactions relevant to the $N^*$ program. In the polar angle range 
of up to 70$^{\circ}$ photons and neutrons can be 
detected using the electromagnetic calorimeters.
The forward angular range from about 10$^{\circ}$ to 50$^{\circ}$ is 
instrumented with gas Cerenkov counters for the identification of electrons.  

In the $N^*$ program, CLAS is often used as a ``missing mass'' spectrometer, 
where all final state particles except one particle are detected. The undetected particle 
is inferred through the overdetermined kinematics, making use of the good
momentum and angle resolution. The right panel in figure~\ref{fig:clas_pid} shows an 
example of the kinematics covered in the reaction $ep \rightarrow epX$.
It shows the invariant hadronic mass $W$  versus the missing mass $M_X$. 
The undetected
particles $\pi^{\circ}$, $\eta$, and $~\omega$ are clearly visible as 
bands of constant $M_X$. The
correlation of certain final states with specific resonance excitations
is also clearly seen.

\subsubsection{Experimental Hall C - HMS and SOS}
\label{section:hallc}

Hall C houses the high momentum spectrometer (HMS) and the short orbit 
spectrometer (SOS). The HMS reaches a maximum momentum of 7 GeV/c, while 
the SOS is limited to about 1.8 GeV/c. 
The spectrometer pair has been used to measure the $\gamma^* N\Delta(1232)$ and 
$\gamma^*NN^*(1535)$ transition
at high $Q^2$ values. For these kinematics the SOS was 
used as electron spectrometer and the HMS to detect the proton. To achieve 
a large kinematics coverage, the spectrometers have to be moved in angles, and 
the spectrometer optics has to be adjusted to accomodate different particle momenta. 
This makes such a two spectrometer setup most useful for studying meson production 
at high momentum transfer, or close to threshold. 
In either case, the Lorentz boost guarantees that particles are produced in a relatively 
narrow cone around the virtual photon, and can be detected in magnetic spectrometers 
with relatively small solid angles.

\subsection{MAMI-B}

The MAMI-B microtron electron accelerator~\cite{mami} at Mainz in Germany reaches a maximum beam 
energy of 850 MeV. 
There are experimental areas for electron scattering experiments  with
three focussing magnetic spectrometers with high resolution~\cite{mami_a1}. 
A two-spectrometer configuration has been 
used in cross section and polarization asymmetry measurements of $\pi^{\circ}$ 
electroproduction from protons in the $\Delta(1232)$ 
region.  

Another experimental area is equipped for physics with an energy-tagged photon bremsstrahlung 
beam~\cite{mami_a2}. Experimental setups with $BaF_2$ crystals (TAPS) have been employed for 
measurements of differential cross sections for $\pi^{\circ}$ and $\eta$ 
production and for beam asymmetry measurements using a 
linearly polarized coherent bremsstrahlung beam.

\subsection{MIT-Bates}

The Bates 850 MeV linear electron accelerator has been used to study 
$\pi^{\circ}$ production in the 
$\Delta(1232)$ region using an out-of-plane spectrometer setup~\cite{bates_oops}. A set of 
four independent focussing spectrometers was used to measure various response 
functions, including the beam helicity-dependent out-of-plane response 
function. 
Because of the small solid angles covered by this seteup, a limited 
range of the polar angles in the center of mass frame of the
$p\pi^{\circ}$ subsystem  could be covered. These spectrometers are no longer in use, but data are still being analyzed.

\subsection{Laser backscattering photon facilities}
\noindent
Electron storage rings built as light sources for material science studies 
are often used parasitically to produce high energy photons for nuclear physics
applications. An intense laser beam is directed tangentially at the electron beam 
circulating in the storage ring producing high energy Compton backscattered 
photons in an energy range dependent upon the wavelength of the laser light.
While the photon intensities are quite modest, the energy spectrum is peaked
at the high energy end providing an efficient source of high energy photons for 
nuclear physics experiments. The laser light is easily polarized linearly 
or circularly. In the Compton backscattering process the polarization of the laser light 
is transferred to the high energy photon beam providing a convenient source of 
polarized photons.

\subsubsection{The Graal Tagged Photon Facility}

The Grenoble Synchrotron Light Source facility is used 
to generate a laser backscattered polarized photon beam of up to 
1470 MeV energy for nuclear physics applications. 
A BGO crystal detector is used for the detection of photons~\cite{graal} 
covering a large portion 
of 4$\pi$. Multi-wire proportional chambers allow charged particle tracking. 
Particle identification
is achieved by time-of-flight measurements at forward angles, and by energy loss measurements at large angles.   
The large solid angle coverage allows the study of reactions with multiple photons 
in the final state which is important for nucleon resonance studies in 
$\pi^{\circ}$ and $\eta$ production~\cite{graal_nstar}. 

\subsubsection{The LEGS at Brookhaven National Laboratory}

Brookhaven National Laboratory operates an electron synchrotron as a 
light source with an energy of 2.8 GeV/c. A laser backscattered real 
photon beam  with an energy up to 470 MeV is used 
for nuclear physics experiments~\cite{legs}. 
A tagging system measures the energy of the Compton-scattered 
electron from which the photon energy is inferred. 
The photon beam is used with an unpolarized hydrogen or nuclear 
target, and with a polarized HD target~\cite{legs_hd}. Several arrays of NaI(Tl) 
crystal detectors have been used to measure Compton scattering and $\pi^{\circ}$ 
production off protons in the $\Delta(1232)$ region.

\subsubsection{LEPS at Spring-8}

SPring-8 operates an 8 GeV electron synchrotron near
Osaka in Japan. A laser-backscattered, energy-tagged polarized photon 
beam with an energy up to 2.4 GeV is produced for 
nuclear and particle physics applications~\cite{nakano}. 
The LEPS detector consists of a plastic scintillator to detect charged 
particles produced in the target, an aerogel Cerenkov counter for particle identification, 
charged-particle tracking counters, a large dipole magnet, and a time-of-flight wall for particle
identification. The LEPS detector has been used for $\omega$, and near-threshold $\phi$ production, and
for strange particle production.

\subsection{Electron Stretcher and  Accelerator (ELSA).}

The University of Bonn operates a 2.5 GeV electron synchrotron and
a stretcher ring and post accelerator to obtain a high duty factor beam and 
an energy of  3 GeV. Three experimental setups have been used 
for meson production experiments during the past decade. 
  
\subsubsection{The SAPHIR Detector}

SAPHIR is a large acceptance detector with  2$\pi$ azimuthal coverage.  
An external electron beam was used to generate an energy-tagged real photon
beam for experiments with the SAPHIR detector~\cite{saphir}. 
At the core of the detector is a large-gap dipole magnet. Tracking is provided by a 
central drift chamber located 
inside the dipole magnet, and additional chambers outside the magnetic 
field region. Scintillation counters are used for triggering and to provide  
time-of-flight information for particle identification.

\subsubsection{The Crystal Barrel Detector at ELSA}
The Crystal Barrel (CB-ELSA) detector was originally used at the LEAR $p\bar{p}$ 
ring at CERN. The detector was recently brought to 
ELSA for operation in an energy-tagged
bremsstrahlung photon beam~\cite{cb-elsa}. 
The detector consists of CsI crystals providing nearly full 
solid angle coverage for neutral particle detection. 
The main focus is the detection of multiple neutral particle final states.

\subsubsection{The Elan Apparatus}

The Elan apparatus has been used for studies of single pion electroproduction 
in the $\Delta(1232)$ region. A focussing magnetic spectrometer detects the 
scattered electrons, and electromagnetic shower detectors measure photons from 
$\pi^{\circ}$ decays. Protons and charged pions are detected as well.  
Charged hadrons are not magnetically analyzed. This setup has been used for 
measurements of $\pi^{\circ}$ and $\pi^+$ in the $\Delta(1232)$ region.

\section{General Formalism}

The bulk of data from the facilities described in the previous section
are from experiments with a single meson and baryon
in the final state. We therefore only present the formulation for such 
reactions. The
generalization of the formulation to the cases that the final states are
three-body states is straightforward.

\begin{figure}[tb]
\vspace{3.5cm}
\centering{\includegraphics{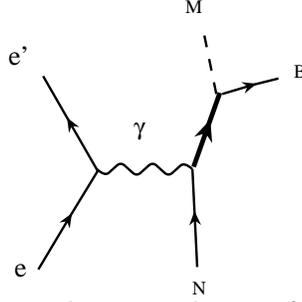}}
\caption{One-photon exchange mechanism for meson electroproduction from a nucleon.}
\label{fig:meson_electroproduction}
\end{figure}

We consider the process $N(e,e' M)B$ illustrated in Fig.~\ref{fig:meson_electroproduction}. 
The final meson-baryon states are two-body states, such as
$\pi N,\eta N, K\Lambda$, $\omega N$ and $\phi N$.
Within the Relativistic Quantum Field Theory, the
Hamiltonian density for describing this process can be concisely written as
\begin{eqnarray}
H_{em}(x) = e A_\mu(x) [ j^\mu(x) + J^\mu(x) ] \, ,
\end{eqnarray}
where $A_\mu$ is the photon field,
\begin{eqnarray}
j^\mu(x) = \bar{\psi}_e(x)\gamma^\mu \psi_e(x)
\end{eqnarray}
is the lepton current, and the
electromagnetic interactions involving hadrons are induced by
the hadron current $J^\mu$.

With the convention of Bjorken and Drell\cite{bjor}, the
Hamiltonian density Eq.(1) leads to
\begin{eqnarray}
< k p' \mid \int d x A^\mu(x)J_\mu(x) \mid q p > =
(2\pi)^4\delta^{4}(p+q-k-p')
<k p' \mid \epsilon_\mu(q)J^\mu(0) \mid q p > \, , 
\end{eqnarray}
where $q$, $p$, $k$, and $p^\prime$ are the momenta for the intial
photon, intial nucleon, final meson, and final nucleon, respectively,
$\epsilon_\mu (q)$ is the photon polarization vector. 
Throughout this paper, we will suppress the spin and isospin indices unless
they are needed for detailed explanations.

\begin{figure}[htbn{figure}[tbh]
\vspace{5cm}
\centering{\includegraphics{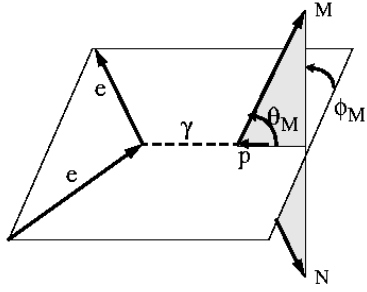}}
 \caption[]{Kinematics of meson electroproduction reaction.}
\label{fig:diag_2}
\end{figure}

It is convenient
to write
\begin{eqnarray}
<k p' \mid \epsilon_\mu(q)J^\mu(0) \mid q p >
&=&\frac{1}{(2\pi)^6}\sqrt{\frac{m_N}{E_N(p')}}\frac{1}{\sqrt{2E_\pi (k)}}
\epsilon_\mu(q)J^\mu(k,p';q,p) \nonumber \\
&\times& \sqrt{\frac{m_N}{E_N(p)}}\frac{1}{\sqrt{2\omega}} \, .
\end{eqnarray}
The expression for calculating electromagnetic meson production
cross sections can be expressed
in terms of $J^\mu(k,p';q,p)$.
For evaluating electroproduction cross sections,
it is common and convenient to choose a
coordinate system that the virtual photon is
in the quantization $z-$direction, and the angle between the
$e-e'$ plane and $M-B$ plane is $\phi_M$, as illustrated in 
Fig.~\ref{fig:diag_2}.
 With some straightforward but lengthy
derivations, it is possible to write
the differential cross section of $N(\vec{e},e' M)B$ reaction  in 
the following form \begin{eqnarray}
\frac{d^5\sigma_h}{dE_{e^\prime} d\Omega_{e^\prime} d\Omega_M^*}
&=&\Gamma\frac{d\sigma_h}{d\Omega^*_M} \, \nonumber 
\end{eqnarray}
with
\begin{eqnarray}
\frac{d\sigma_h}{d\Omega^*_M}&=&\Gamma [\frac{d\sigma_{unpol}}
{d\Omega_M^*}+h\sqrt{2\epsilon(1-\epsilon)}
\frac{d\sigma_{LT'}}{d\Omega^*_M} sin\phi_M] \, ,
\end{eqnarray}
where $h$ is the helicity of the incoming electron.
The kinematic factors associated with the incoming and outgoing
electrons are only contained in the following two variables
\begin{eqnarray}
\epsilon&=&
\{1+\frac{2\mid {\bf q} \mid^2}{Q^2} tan^2\frac{\theta_e}{2} \}^{-1} \,, \\
\Gamma&=&\frac{\alpha K_H}{2\pi^2Q^2}\frac{E_e^\prime}{E_e}
\frac{1}{1-\epsilon} \,,
\end{eqnarray}
where $K_H=\omega - Q^2/2m_N$ is the virtual photon flux, $\alpha=1/137$ is
the electromagnetic coupling constant, $q^\mu=(\omega, {\bf q})$ is the
momentum of the photon, and  $Q^2 = -q^2 = \mid {\bf q}\mid ^2 - \omega^2 $.
The incident and outgoing electron energies are related to 
$q^\mu$ by
\begin{eqnarray}
\omega&=& E_e - E^\prime_e \,, \\
Q^2 &=& 4 E_e E^\prime_e sin^2 \frac{\theta_e}{2} \,,
\end{eqnarray}
where $\theta_e$ is the angle between the incident and outgoing electrons.

For investigating $N^*$ excitations, the kinematics is often characterized by the initial $\gamma^* N$ invariant
mass $W$ and $Q^2$. For such a choice, the energy transfer is then defined by
\begin{eqnarray}
\omega = \frac{W^2+Q^2-m_N^2}{2m_N} .
\end{eqnarray}
The corresponding electron kinematics for conducting experiments with a
given $(Q^2, W)$ can then be evaluated by using Eqs.(8)-(9).

Note that the differential cross sections in the right-hand side of Eq.(5)
is defined in the center of mass (c.m.) 
frame  of the initial $\gamma^* N$ and the  
final $MB$ systems. These quantities must be evaluated in terms of the momenta
in that c.m. frame. For the coordinate system chosen as
 ${\bf q} \parallel \hat{z}$, all momenta needed in calculating
$d\sigma/d\Omega^*_{M}$ must be transformed by a Lorentz boost
with ${\bf \beta } = \hat{z} \mid {\bf q} \mid /(\omega + m_N)$. In terms of
variables $W$ and $Q^2$,  
a momentum $p^\mu_c$  in the considered  c.m. frame is related to a 
momentum $p^\mu$ in the laboratory frame by
\begin{eqnarray}
p^0_c &=& \frac{m_N+\omega}{W} p^0 - \frac{ \mid {\bf q} \mid} {W}p^z \,, \\
p^x_c &=& p^x \,, \\
p^y_c &=& p^y \, \\
p^z_c &=&-\frac{ \mid {\bf q} \mid} {W} p^0 +  \frac{m_N+\omega}{W}p^z .
\end{eqnarray}
Specifically, we have for the virtual photon 
\begin{eqnarray}
\mid{\bf q}_c\mid &=&\frac{m_N}{W}\mid {\bf q} \mid \,, \\
\omega_c &=&\frac{\omega m_N - Q^2}{W} \,.
\end{eqnarray}
It is easy to see that $Q^2 = {\bf q}^2 - \omega^2 = {\bf q}^2_c - \omega^2_c$

We next present formulae for calculating the c.m. 
differential cross sections in the right-hand-side of Eq.(5).
The unpolarized cross section is given by
\begin{eqnarray}
\frac{d\sigma_{unpol}}{d\Omega^*_M} =
\frac{d\sigma_T}{d\Omega^*_M}+\epsilon \frac{d\sigma_L}{d\Omega^*_M}
+\epsilon\frac{d\sigma_{TT}}{d\Omega^*_M}cos2\phi_M
+\sqrt{2\epsilon(1+\epsilon)}\frac{d\sigma_{LT}}{d\Omega^*_M}cos\phi_M \,,
\end{eqnarray}
where $\sigma_T$, $\sigma_L$, $\sigma_{TT}$, and $\sigma_{LT}$ are called the
transverse, longitudinal, polarization, and interference cross sections.
These four cross sections and the $d\sigma_{LT'}/d\Omega^*$ in Eq.(5)
can be written as
\begin{eqnarray}
\frac{d\sigma_\beta}{d\Omega^*_M} = \frac{\mid {\bf k_c} \mid}
{ q^\gamma_c}
M_\beta (k_c,p_c';q_c,p_c) \,,
\end{eqnarray}
where $q^\gamma_c = (W^2-m_N^2)/(2m_N) = K_H$ is the effective
photon c.m. momentum,
$\beta= T, L, TT, LT $ and $LT^\prime$, and the c.m. momenta 
$k_c$, $p'_c$, $q_c$, and $p_c$
can be calculated from the corresponding  momenta
in the laboratory frame by using Eqs.(11)-(14). 
Obviously $q^\gamma_c = \mid {\bf q_c} \mid$ at the photon point $Q^2=0$.

The meson production dynamics is contained in $M_\beta$ of Eq.(18).
They are  calculated from  various combinations of
current matrix elements evaluated on the $\phi_M=0$ plane
(see Fig.\ref{fig:diag_2}):
\begin{eqnarray}
M_T(k_c,p_c';q_c,p_c) &=& \frac{F}{4}\sum_{spins}
[\mid J^x(k_c,p'_c;q_c,p_c) \mid^2
 +\mid J^y(k_c,p'_c;q_c,p_c)\mid^2]_{\phi_M=0} \nonumber \\
M_L(k_c,p_c';q_c,p_c) &=& \frac{F}{2}\sum_{spns}
\frac{Q^2}{\omega^2}[\mid J^z(k_c,p'_c;q_c,p_c) \mid^2]_{\phi_M=0} \nonumber \\
M_{TT}(k_c,p_c';q_c,p_c) &=& \frac{F}{4}\sum_{spins}
[\mid J^x(k_c,p'_c;q_c,p_c) \mid^2
-\mid J^y(k_c,p'_c;q_c,p_c)\mid^2]_{\phi_M=0} \nonumber \\
M_{LT}(k_c,p_c';q_c,p_c) &=&- \frac{F}{2}\sum_{spins}
\sqrt{\frac{Q^2}{\omega^2}} Re\{J^z(k_c,p'_c;q_c,p_c)
J^{x*}(k_c,p'_c;q_c,p_c)\}_{\phi_M=0} \nonumber \\
M_{LT'}(k_c,p_c';q_c,p_c) &=& \frac{F}{2}\sum_{spins}
\sqrt{\frac{Q^2}{\omega^2}} Im\{J^z(k_c,p'_c;q_c,p_c)
J^{x*}(k_c,p'_c;q_c,p_c)\}_{\phi_M=0} \nonumber \\
\end{eqnarray}
with

\begin{eqnarray}
F= \frac{e^2}{(2\pi)^2 }
\frac{1}{2E_M(k_c)}\frac{m_N}{E_B(p'_c)}
\frac{m_N}{E_N(p_c)}\frac{E_M(k_c)E_B(p'_c)}{2W}
\end{eqnarray}
where $E_a(p)=\sqrt{{\bf p}^2+m_a^2}$ with $m_a$ denoting
the mass of particle $a$.

The differential cross sections of $N(\vec{e},e'M)B$ are often expressed
in terms of response functions\cite{donnelly} $R_\alpha$
which are related to the
differential cross sections of Eq.(18) by
\begin{eqnarray}
\frac{d\sigma_T}{d\Omega^*_M} &=& \frac{\mid {\bf k_c} \mid}{q^\gamma_c} R_T \,
\nonumber \\
\frac{d\sigma_{TT}}{d\Omega^*_M} &=& \frac{\mid {\bf k_c} \mid}{q^\gamma_c} R_{TT}
\, \nonumber \\
\frac{d\sigma_L}{d\Omega^*_M} &=& \frac{\mid {\bf k_c} \mid}{q^\gamma_c} 
\frac{Q^2}{\omega_c^2}R_L \, \nonumber \\
\frac{d\sigma_{LT}}{d\Omega^*_M} &=& \frac{\mid {\bf k_c} \mid}{q^\gamma_c}
\sqrt{\frac{Q^2}{\omega_c^2}} R_{LT} \,  \nonumber \\
\frac{d\sigma_{LT^\prime}}{d\Omega^*_M} 
&=& \frac{\mid {\bf k_c} \mid}{q^\gamma_c} 
\sqrt{\frac{Q^2}{\omega_c^2}} R_{LT^\prime} . \nonumber \\
\end{eqnarray}

The above formulation can be readily used to calculate various polarization
observables with a polarized initial nucleon.
For observables with a polarized recoiled final baryon, the situation is more
complicated. They have been
explicitly derived for pseudo-scalar meson production~\cite{chiang,knoc}.
Formulations for analyzing spin observables of vector meson production 
were developed in Ref.\cite{tabakin}.

We also note that the unpolarized photoproduction cross section is given by
$d\sigma_T/d\Omega^*_M$
evaluated at $Q^2=0$ and $q^\gamma_c \rightarrow  \mid {\bf q_c} \mid$.
For polarized photons, one needs to choose an appropriate combination of
$J^x$ and $J^y$. For instance,
$d\sigma_\perp/d\Omega^*_M$($d\sigma_\parallel/d\Omega^*_M$)for the
 photon polarization normal (parallel) to the hadron plane
is calculated
from keeping only $J^y$ ($J^x$) contribution and multipling the
resulting cross section by a factor of 2.
 The photon asymmetry is defined as
\begin{eqnarray}
\Sigma_\gamma=\frac{\sigma_\perp-\sigma_\parallel}
{\sigma_\perp+\sigma_\parallel} .
\end{eqnarray}
Calculations of other photoproduction
polarization observables are given, for example, in the
appendix C of Ref.\cite{nbl}.

We next present formulae which are often used in analyzing the
production of pseudo-scalar mesons, such as $MB=\pi N, \eta N, KY$.
The Lorentz invariance and gauge invariance allow us to write
the hadron current matrix elements as
\begin{eqnarray}
\epsilon_\mu(q) J^\mu(k, p^\prime; q, p) = \sum_{i=1,6}
\bar{u}[({\bf p}^\prime)A_i(s,t,u)  M_i ]u({\bf p}) \,,
\end{eqnarray}
where $u({\bf p})$ is the Dirac spinor, $A_i(s,t,u)$ are Lorentz invariant
functions, and $M_i$ are independent invariances formed from 
$\gamma^\mu$, $\gamma_5$, and momenta variables. The expressions for
$M_i$ are irrelevant to this paper and hence are omitted here. But they can be
found, for example, on page 5 of Ref.\cite{donn}.
For $\pi$ production,
the amplitudes defined above can be further classified by isospin
quantum numbers. There are $A^{(0)}$ for the isoscalar photon, and for the
isovector the two amplitudes $A^{(1/2)}$ and $A^{(3/2)}$ for the
final $\pi N$ system with total isospin $I=1/2$ and $I=3/2$ respectively.
Each
invariant amplitude in Eq.(23) can be expanded as
\begin{eqnarray}
A_i = \frac{1}{2}A_i^{(-)} [\tau_\alpha, \tau_3] +A^{(+)}_i\delta_{\alpha,3}
+A_i^{(0)} \tau_\alpha \,,
\end{eqnarray}
where $\tau$ is the isospin Pauli operator,
and $\alpha$ is the isospin quantum number associated with the produced pion.
Eq.(24) then leads to  $A_i^{(1/2)}=A_i^{(+)} + 2 A_i^{(-)}$ and
$A_i^{(3/2)}=A_i^{(+)} - A_i^{(-)}$. 
It is useful to further define proton $_{p}A^{(1/2)}$ and neutron
$_{n}A^{(1/2)}$ amplitudes with total isospin $I=1/2$
\begin{eqnarray}
_pA_i^{(1/2)} &=& A_i^{(0)} +\frac{1}{3}A_i^{(1/2)} \,,  \nonumber \\
_nA_i^{(1/2)} &=& A_i^{(0)} -\frac{1}{3}A_i^{(1/2)} \,.
\end{eqnarray}
Then the amplitudes for four physical processes can be written as
\begin{eqnarray}
A_i(\gamma^* p \rightarrow n\pi^+) &=&\sqrt{2}[_pA_i^{(1/2)} -\frac{1}{3}A_i^{(3/2)}]
 \,, \nonumber \\
A_i(\gamma^* p \rightarrow n\pi^0) &=&_pA_i^{(1/2)} +\frac{2}{3}A_i^{(3/2)}
\,, \nonumber \\
A_i(\gamma^* n \rightarrow n\pi^-) &=&\sqrt{2}[_nA_i^{(1/2)} +\frac{1}{3}A_i^{(3/2)}]
\,, \nonumber \\
A_i(\gamma^* n \rightarrow n\pi^0) &=&-_nA_i^{(1/2)} +\frac{2}{3}A_i^{(3/2)}\, .
\end{eqnarray}
The above invariant functions $A_i^{(\pm,0)}$ 
are the starting point for developing
dispersion relation approach which will be given in section 4.7.
The  isospin relations Eqs.(24)-(26)
are valid for all of the amplitudes we are going
to discuss.  However, the isospin quantum numbers as well as spin
quantum numbers will be suppressed in the remainder of this article.

For investigating nucleon resonances, it is useful to
have a formulation expressing
the meson production cross sections in terms of multipole amplitudes.
If the final hadron state consists of 
only a pseudo-scalar and a spin 1/2 baryon,
such as $\pi N$, $K Y$ and $\eta N$ states, such a formulation
has been well developed. 
This is accomplished by casting Eq.(23) into the 
Chew, Goldberger, Low, and Nambu (CGLN)\cite{cgln}
 form defined in the c.m. frame of the final meson-baryon system
\begin{eqnarray}
\epsilon^\mu J^\mu(k_c,p'_c;q_c,p_c) = \sum_{i=1,6}
F_i(s,t,Q^2) \bar{u}({\bf p_c}^\prime) O_i {u}({\bf p_c}) \,,
\end{eqnarray}
where $F_i(s,t,u)$ are the Lorentz invariant CGLN amplitudes 
and $O_i$ are operators
defined in the baryon spin space
\begin{eqnarray}
O_1&=& i{\bf \sigma}\cdot {\bf b} \,, \\
O_2&=& {\bf \sigma}\cdot {\bf \hat{k_c}}
{\bf \sigma}\cdot ({\bf \hat{q}}\times {\bf b}) \,, \\
O_3&=& i{\bf \sigma}\cdot {\bf \hat{q_c}}
{\bf \hat{k}_c}\cdot {\bf b} \,, \\
O_4&=& i{\bf \sigma}\cdot {\bf \hat{k_c}}
{\bf \hat{k}_c}\cdot {\bf b} \,,  \\
O_5&=& -i{\bf \sigma}\cdot {\bf \hat{k_c}}b_0 \,, \\
O_6&=&-i{\bf \sigma}\cdot {\bf \hat{q_c}}b_0 \,,
\end{eqnarray}
with 
\begin{eqnarray}
b^\mu = \epsilon^\mu(q_c) -\frac{\vec{\epsilon}\cdot\vec{q}_c}
{\mid {\bf q}_c \mid } q_c^\mu
\end{eqnarray}
Obviously we have ${\bf b}\cdot {\bf q_c}=0$.
The CGLN amplitudes $F_i(s,t,Q^2)$
can be expanded in terms of multipole amplitudes characterized by the
angular momentum quantum numbers of the initial
$\gamma ^* N$ and the final $MB$
systems. The relations
are found to be
\begin{eqnarray}
F_1 &=& \sum_{\ell}[P^\prime_{\ell+1}(x) E_{\ell+}+P^\prime_{\ell-1}(x) 
E_{\ell-} +P^\prime_{\ell+1}(x) M_{\ell+}+(\ell+1)
P^\prime_{\ell-1}(x) M_{\ell-}] \\
F_2 &=& \sum_{\ell}[(\ell+1) P^\prime_{\ell}(x) M_{\ell+}
+\ell P^\prime_{\ell}(x) M_{\ell}] \\
F_3 &=& \sum_{\ell}[P^{\prime\prime}_{\ell+1}(x) E_{\ell+}
+P^{\prime\prime}_{\ell-1}(x) E_{\ell-}
-P^{\prime\prime}_{\ell+1}(x) M_{\ell+}+P^{\prime\prime}_{\ell-1}(x) 
M_{\ell-}] \\
F_4 &=& \sum_{\ell}[-P^{\prime\prime}_{\ell}(x) E_{\ell+}
-P^{\prime\prime}_{\ell}(x) E_{\ell-}
+P^{\prime\prime}_{\ell}(x) M_{\ell+}-P^{\prime\prime}_{\ell}(x) 
M_{\ell-}] \\
F_5 &=& \sum_{\ell}[-(\ell+1)P^\prime_{\ell}(x) 
S_{\ell+}+\ell P^\prime_{\ell}(x) S_{\ell-}] \\
F_6 &=& \sum_{\ell}[(\ell+1)P^\prime_{\ell+1}(x) S_{\ell+}
-\ell P^\prime_{\ell-1}(x) S_{\ell-}]
\end{eqnarray}
In the above equations, the multipole amplitudes $E_{\ell\pm}$, $M_{\ell\pm}$
and $S_{\ell\pm}$ are functions of $W$ and $Q^2$ only. They
describe the transitions which can be classified according to
the character of the photon, transverse or scalar(or longitudinal), 
and the total angular momentum $J=\ell\pm 1/2$ of the final
state. In addition, the transverse photon states can either be electric with
parity $(-1)^{L_\gamma}$, or magnetic, with parity $(-1)^{L_\gamma+1}$, 
where $L_\gamma$ is the  orbital angular momentum of the $\gamma^* N$ system. 
In Table \ref{tab:multipoles} 1, we list
how each multipole amplitude with $ J \leq 3/2$  is related to the initial $L_\gamma$ and
final $(\ell,J)$ angular momentum quantum numbers.
The longitudinal multipoles are related to the scalar multipoles by
$L_{\ell\pm}=(\omega/\mid {\bf q} \mid )S_{\ell\pm}$.

\begin{table}[htb]
\caption{Angular momentum quantum numbers associated with
$\gamma ^* N \rightarrow \pi N$ multipole amplitudes. See text
for the explanations.}
\centerline{\smalllineskip
\begin{tabular}{lccccl}
   $\ell$           & J   & $L_\gamma$   & Notation \\
\hline
0 &     1/2  &     1   &        $E_{0+}$ \\
1 &     3/2  &     2   &        $E_{1+}$ \\
1 &     1/2  &     1   &        $M_{1-}$ \\
1 &     3/2  &     1   &        $M_{1+}$ \\
0  &    1/2  &     1   &        $S_{0+}$ \\
1   &   1/2  &     0   &        $S_{1-}$ \\
1    &  3/2  &     2   &        $S_{1+}$ \\
                \hline
\end{tabular}}
\label{tab:multipoles}
\end{table}

We now note that the matrix elements $J^i$ with $i=x,y,z$ 
for evaluating  Eq.(19) can be
obtained from Eq.(27) by setting
 $\epsilon^\mu=(0,\hat{x})$,$\epsilon^\mu=(0,\hat{y})$,
$\epsilon^\mu=(0,\hat{z})$, respectively.
By further using  the relations Eqs.(35)-(40), 
the differential cross sections Eq.(5) or Eq.(18) can
 then be expressed
in terms of multipole amplitudes.
For example, Eq.(5) can lead to the total inclusive cross section
\begin{eqnarray}
\frac{d\sigma}{dE_e^\prime d\Omega_e^\prime} =\Gamma [\sigma_T 
+\frac{Q^2}{\mid {\bf q} \mid ^2} \epsilon\sigma_L] \,,
\end{eqnarray}
where 
\begin{eqnarray}
\sigma_T&=&\frac{2\pi k_c}{\mid {\bf q_c} \mid }\sum_{\ell}\ell(\ell+1)[\mid M_{\ell+}\mid^2
+\mid E_{(\ell+1)-}\mid^2 +\ell^2(\ell+1)[\mid M_{\ell-}\mid^2
+\mid E_{(\ell-1)+}\mid^2] \nonumber \\
\sigma_L&=&\frac{4\pi k_c}{\mid {\bf q_c} \mid}\sum_{\ell}
 [(\ell+1)^3 \mid L_{(\ell+1)-} \mid^2 +\ell^3 \mid L_{(\ell-1)+} \mid ^2 ]
\, .
\end{eqnarray}

With the above formulation, we then turn to describe various theoretical models for
analyzing electromagnetic meson production reactions.

\section{Theoretical Models}

The development of theoretical models for investigating electromagnetic pion
 production reactions began in 1950's with the pioneering
work by Chew, Goldberger,
Low, and Nambu(CGLN)\cite{cgln}. In the subsequent years, their
dispersion-relation approach was the basis of many analyses\cite{donn1} of
pion production data in the $\Delta$ excitation region.
This approach has been revived\cite{hans,azn} recently and
extended\cite{azn1,azn2} to also analyze $\eta$ production. 
For investigating the data at higher energies where the
production of two pions and  other mesons($\eta$ and $K$,
$\omega$, and $\phi$) could arise 
, the isobar models\cite{walk}
were developed to
extract the parameters of higher mass nucleon resonances.
During the years around 1980, the K-matrix effective Lagrangian
models\cite{olson,rpi}
 were developed to study the $\Delta$ excitation.
The K-matrix method and isobar parameterization have been used subsequently
to develop tools for performing amplitude analyses of the data and 
determining the resonance parameters. Examples are the
very useful dial-in codes SAID\cite{said} and MAID\cite{maid}. 
Progress has also been made in extracting resonance parameters using
the multi-channel K-matrix method\cite{kent,kent1,giessen} 
and the unitary coupled-channel isobar model\cite{cmb,zagreb,pittanl}.

In recent years, a rather different theortical point of view has been
taken to develop dynamical 
models\cite{tanabe,yang,nbl,gros,sl1,yosh,sl2,kamyan,chen,dmt,chitab,fuda,lms,kais1,oset1}
 of meson production reactions.
These models account for the off-shell scattering effects and
can therefore provide a much more direct way to interpret
the resonance parameters in terms of the existing hadron structure models.
So far, 
the dynamical reaction model has been able to
interprete the resonance parameters, in particular the $\Delta$
resonance,  in terms of constituent quark models.
Its connection with the results from
quenched and unquenched Lattice QCD calculations remains to be established.
                                                                                
In the first part of this section, we will give a general derivation of
most of the exisiting models in order to clarify their differences.
We then give some detailed formula for the dynamical model
which are needed for discussing the results in section 5.
The analyses based on the dispersion relation approach will be described at
 the end of this section.

\subsection{Hamiltonian Formulation}

\begin{figure}[htbn{figure}[tbh]
\vspace{50mm}
\centering{\includegraphics{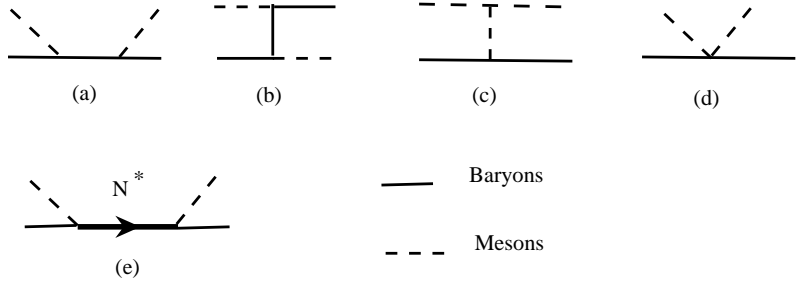}}
 \caption[]{ Tree diagrams for meson-baryon interactions. 
$N^*$ is a nucleon
resonance.  }
\label{fig:mechanism_1}
\end{figure}

Most of the existing models for analyzing 
the data of electromagnetic meson production reactions
can be schematically derived from a Hamiltonian formulation of the problem.
The starting point of our derivation is
to assume that the meson-baryon ($MB)$ reactions can be described by a
Hamiltonian of the following form
\begin{eqnarray}
H= H_0 + V \,,
\end{eqnarray}
where $H_0$ is the free Hamiltonian and
\begin{eqnarray}
V = v^{bg} + v^R \,.
\end{eqnarray}
Here $v^{bg}$ is the non-resonant(background) term due to the mechanisms such
as the tree-diagram mechanisms illustrated in
 Fig.~\ref{fig:mechanism_1}(a)-(d), and $v^R$ describes
 the $N^*$ excitation in Fig.~\ref{fig:mechanism_1}(e).
Schematically, the resonant term can be written as
\begin{eqnarray}
v^R(E) = \sum_{N^*_i}\frac{\Gamma^\dagger_i \Gamma_i}{E-M^{0}_i}\,,
\end{eqnarray}
where $\Gamma_i$ defines the decay of the $i$-th $N^*$ state into meson-baryon
states, and $M^0_i$ is a mass parameter related to the resonance position.

The next step is to define a channel space spanned by
the considered  meson-baryon ($MB$)
 channels: $\gamma N$, $\pi N$, $\eta N$, $\pi\Delta$, $\rho N$
$\sigma N$, $\cdot\cdot$. The S-matrix of the
meson-baryon  reaction is defined by
\begin{eqnarray}
S(E)_{a,b} = \delta_{a,b} - 2\pi i \delta(E-H_0) T_{a,b}(E) \,,
\end{eqnarray}
where ($a,b$) denote $MB$ channels, and
the scattering T-matrix is
defined by the following coupled-channel equation
\begin{eqnarray}
T_{a,b}(E) = V_{a,b} + \sum_{c}V_{a,c} g_c(E) T_{c,b}(E) \,.
\end{eqnarray}
Here the meson-baryon propagator of channel $c$ is
\begin{eqnarray}
g_c(E) = < c \mid g(E) \mid c > \nonumber
\end{eqnarray}
with
\begin{eqnarray}
g(E)&=&\frac{1}{E-H_0 +i\epsilon} \, \nonumber \\
&=& g^P(E) -i\pi\delta(E-H_0)  \,,
\end{eqnarray}
where
\begin{eqnarray}
g^P(E) &=& \frac{P}{E-H_0} \,.
\end{eqnarray}
Here $P$ denotes taking the principal-value part of any
integration over the propagator. We can also define K-matrix as
\begin{eqnarray}
K_{a,b}(E) = V_{a,b} + \sum_{c}V_{a,c} g^P_c(E) K_{c,b}(E) \,.
\end{eqnarray}
Eqs.(47)-(50) then define the following
relation between the  K-matrix  and T-matrix 
\begin{eqnarray}
T_{a,b}(E)
= K_{a,b}(E) - \sum_{c}T_{a,c}(E)[i\pi\delta(E-H_0)]_c K_{c,b}(E) \,.
\end{eqnarray}
By using the two potential formulation\cite{gold},
one can cast Eq.(47) into the following form
\begin{eqnarray}
T_{a,b}(E)  &=&  t^{bg}_{a,b}(E) + t^{R}_{a,b}(E)
\end{eqnarray}
with
\begin{eqnarray}
t^{R}_{a,b}(E) &=& \sum_{N^*_i, N^*_j}
\bar{\Gamma}^\dagger_{N^*_i, a}(E) [G(E)]_{i,j}
\bar{\Gamma}_{N^*_j, b}(E)  \,.
\end{eqnarray}
The first term of Eq.(52)
 is determined only by the non-resonant interaction
\begin{eqnarray}
t^{bg}_{a,b}(E)= v^{bg}_{a,b} +\sum_{c} v^{bg}_{a,c}g_c(E) t^{bg}_{c,b}(E)\,.
\end{eqnarray}
The resonant amplitude Eq.(53) is determined by the
dressed vertex
\begin{eqnarray}
\bar{\Gamma}_{N^*,a}(E)  &=&
  { \Gamma_{N^*,a}} + \sum_{b} \Gamma_{N^*,b}
g_{b}(E)t^{bg}_{b,a}(E)\,,
\end{eqnarray}
and the dressed propagator
\begin{eqnarray}
[G(E)^{-1}]_{i,j}(E) = (E - M^0_{N^*_i})\delta_{i,j} - \Sigma_{i,j}(E) \,.
\end{eqnarray}
Here $M_{N^*}^0$ is the bare mass of the resonance state $N^*$, and
the self-energy is
\begin{eqnarray}
\Sigma_{i,j}(E)= \sum_{a}\Gamma^\dagger_{N^*,a} g_{a}(E)
\bar{\Gamma}_{N^*_j,a}(E)\,.
\end{eqnarray}
                                                                                
Note that the meson-baryon propagator $g_a(E)$
for channels including an unstable particle, such as
$\pi \Delta$, $\rho N$ and $\sigma N$,
must be modified
to include a width due to their decay into $\pi\pi N$ channel.
 In the Hamiltonian formulation, this
amounts to the following replacement
\begin{eqnarray}
g_{a}(E) \rightarrow <a \mid\frac{1}{ E - H_0 - \Sigma_V(E)} \mid a> \,,
\end{eqnarray}
where the energy shift is
\begin{eqnarray}
\Sigma_V(E) = \sum_{i} \Gamma^+_V (i)\frac{P_{\pi\pi N}}{E-H_0 + i\epsilon}
\Gamma_V(i) \,.
\end{eqnarray}
Here $\Gamma_V$ describes the decay of $\rho$, $\sigma$ or $\Delta$
 in the quasi-particle channels.
                                                                                
Eq.(47), Eqs.(52)-(59), and Eq.(51)
 are the starting points of our derivations.
>From now on, we consider the formulation in the
partial-wave representation.
The channel labels, ($a,b,c$), will also include the usual
angular momentum and isospin quantum numbers.
                                                                                
\newpage
\subsection{Tree-diagram models}

The tree-diagram models are based on the simplification that $T \sim V = v^{bg} + v^R$.
The resonant effect is included by modifing the mass parameter of
$v^R$, defined in Eq.(45), to include a width, such as $M^0_i = M_{_i} 
- \frac{i}{2}\Gamma^{tot}_i(E)$.
Eq.(47) is then simplifed into
\begin{eqnarray}
T_{a,b}(tree) = v^{bg}_{a,b} 
+\sum_{N^*_i}\frac{\Gamma^\dagger_{i,a} 
\Gamma_{i,b}}{E-M^{0}_i+\frac{i}{2}\Gamma^{tot}_i(E)} \,,
\end{eqnarray}
where $ v^{bg}$ is calculated from the 
tree-diagrams(Fig.~\ref{fig:mechanism_1}(a)-(d)) of a chosen 
Lagrangian, and $\Gamma^{tot}_i$ is the total decay width of the
$i-$th $N^*$. 

In recent years, the tree-diagram models have been applied mainly to 
investigate the photoproduction and electroproduction of 
$K$ mesons\cite{adel,saghai1,will,saghai,saghai2,bennhold00}, vector 
mesons\cite{zhao,otl,ohlee}($\omega$, $\phi$) and
two pions\cite{oset2}.
 At high energies, 
the t-channel amplitudes(Fig.~\ref{fig:mechanism_1}(b)-(c))
 are replaced by the Regge parameterization in 
some tree-diagram models\cite{laget,vend}.
The validity of using the tree-diagram models to
investigate nucleon resonances is obviously very questionable, as
discussed in a study of $\omega$ photoproduction\cite{ohlee} and
kaon photoproduction\cite{chitab}.

\subsection{Unitary Isobar Models ($UIM$)}
\subsubsection{MAID:}

The Unitary Isobar Model developed\cite{maid} by the Mainz group
is based on the on-shell relation Eq.(51).
By including only one hadron channel, $\pi N$  ( or $\eta N$ ), 
Eq.(51) leads to
\begin{eqnarray}
T_{\pi N,\gamma N} &=& \frac{1}{1+i K_{\pi N,\pi N}}K_{\pi N, \gamma N}
\nonumber \\
&=& e^{i\delta_{\pi N}}cos\delta_{\pi N} K_{\pi N, \gamma N}\,.
\end{eqnarray}
Here we have used the relation  $K_{\pi N,\pi N} =- tan \delta_{\pi N}$ with
$\delta_{\pi N}$ being the pion-nucleon scattering phase shift.
By further assuming that
$K=V=v^{bg} + v^R$, one can cast the above equation into the following form
\begin{eqnarray}
T_{\pi N,\gamma N}({\it UIM}) =
e^{\delta_{\pi N}}cos\delta_{\pi N} [v^{bg}_{\pi N, \gamma N}]
+\sum_{N^*_i}T^{N^*_i}_{\pi N, \gamma N}(E) \,.
\end{eqnarray}
                                                                                
Clearly, the  non-resonant multi-channel effects, such as
$\gamma N \rightarrow (\rho N,\pi \Delta) \rightarrow \pi N$, which could be important
in the second and third resonance regions are neglected in MAID.
In addition, they calculate the non-resonant amplitude
$v^{bg}_{\pi N, \gamma N}$ using an energy-dependent
 mixture of PV and PS (pseudo-scalar) $\pi NN$ coupling
\begin{eqnarray}
L_{\pi NN} = \frac{\Lambda_m^2}{\Lambda^2_m+q^2_0} L^{PV}_{\pi NN}
+\frac{q_0^2}{\Lambda^2_m+q^2_0} L^{PS}_{\pi NN} \,,
\end{eqnarray}
where $q_0$ is the on-shell photon momentum.
With cutoff $\Lambda_m= 450$ MeV, one then gets PV coupling at low energies and
PS coupling at high energies.

For resonant terms in Eq.(62), MAID
uses the following Walker's parameterization\cite{walk}
\begin{eqnarray}
T^{N^*_i}_{\pi N, \gamma N}(E) =f^i_{\pi N}(E)
\frac{\Gamma_{tot}M_i e^{i\Phi}}{M^2_i-E^2 - i M_i\Gamma^{tot}}
f^i_{\gamma N}(E)\bar{A}^i \,,
\end{eqnarray}
where $f^i_{\pi N}(E)$ and $f^i_{\gamma N}(E)$ are the form factors describing
the decays of $N^*$, $\Gamma_{tot}$ is the total decay width,
$\bar{A}^i$ is the $\gamma N \rightarrow N^*$ excitation strength.
The phase $\Phi$ is determined by the unitary condition and the assumption
 that the phase $\psi$ of the total amplitude is related to 
$\pi N$ phase shift $\delta_{\pi N}$ and inelasicity $\eta_{\pi N}$ by
\begin{eqnarray}
\psi(E)=tan^{-1}[\frac{1-\eta_{\pi N}(E) cos 2\delta_{\pi N}(E)}
{\eta_{\pi N}(E) sin2\delta_{\pi N}(E)}] \,.
\end{eqnarray}

\subsubsection{JLab/Yeveran UIM:}

The Jlab/Yerevan UIM\cite{azn2} is similar to MAID.
But it implements the
Regge parameterization in calculating the
amplitudes at high energies. It also uses a different procedure to
unitarize the amplitudes.

\vspace{1cm}
                                                                                
Both MAID and JLab/Yeveran UIM have been applied extensively to analyze the
data of $\pi$ and $\eta$ production reactions, as will be discussed in section
5. Very useful new information on $N^*$ have been extracted.

\subsection{Multi-channel K-matrix models}
\subsubsection{SAID:}
                                                                                
The model employed in SAID\cite{said} is based on the on-shell
relation Eq.(51) with three channels:
$\gamma N$, $\pi N$, and $\pi\Delta$ which represents all other open channels.
The solution of the resulting $3\times 3$ matrix equation can be written
as
\begin{eqnarray}
T_{\gamma N,\pi N}(SAID) = A_I(1 + iT_{\pi N,\pi N})
+ A_RT_{\pi N,\pi N}\,,
\end{eqnarray}
where
\begin{eqnarray}
A_I &=& K_{\gamma N, \pi N}
- \frac{K_{\gamma N ,\pi\Delta}K_{\pi N,\pi N}}
{K_{\pi N,\pi\Delta}} \,, \\
A_R&=&\frac{K_{\gamma N, \pi\Delta}}{K_{\pi N, \pi\Delta}}\,.
\end{eqnarray}
In actual analyses, they simply parameterize $A_I$ and $A_R$ as
\begin{eqnarray}
A_I&=& [v_{\gamma N, \pi N}^{bg}]
+ \sum_{n=0}^{M} \bar{p}_n z Q_{l_\alpha +n}(z) \,, \\
A_R&=& \frac{m_\pi}{k_0}(\frac{q_0}{k_0})^{l_\alpha}
\sum_{n=0}^{N} p_n (\frac{E_\pi}{m_\pi})^n \,,
\end{eqnarray}
where $k_0$ and $q_0$ are the on-shell momenta for pion and photon respectively,$z=\sqrt{k^2_0+4m_\pi^2}/k_0$,
$Q_L(z)$ is the legendre polynomial of second kind,
$E_\pi = E_\gamma -m_\pi(1+m_\pi/(2m_N))$, and $p_n$ and
$\bar{p}_n$ are free parameters.
SAID  calculates
$v^{bg}_{\gamma N,\pi N}$ of Eq.(69)
from the standard PS Born term and $\rho$ and $\omega$ exchanges.
The empirical $\pi N$ amplitude $T_{\pi N,\pi N}$ needed to
evaluate Eq.(66) is also available in SAID.
                                                                                
Once the parameters $\bar{p}_n$ and $p_n$ in Eqs.(69)-(70)
are determined,
the $N^*$ parameters are then extracted by
fitting the resulting amplitude $T_{\gamma N,\pi N}$ at energies near the
resonance position to
a Breit-Wigner parameterization(similar to Eq.(64)).
Very extensive data of pion photoproduction have been analyzed by SAID.
The extension of SAID to also analyze pion electroproduction data
is being pursued.
                                                                                
\subsubsection{Giessen Model}
The
coupled-channel model developed by the Giessen group~\cite{giessen}
can be obtained from Eq.(51) by taking the appoximation
$K = V$; namely, neglecting all multiple-scattering effects
included in Eq.(50) for K-matrix.
This leads to a matrix equation involving only the
on-shell matrix elements of $V$
\begin{eqnarray}
T_{a,b}({\it Giessen}) =
\sum_{c}[(1+i V(E))^{-1}]_{a,c} V_{c,b}(E) \,.
\end{eqnarray}
The interaction $V=v^{bg} +v^R$ is evaluated from tree-diagrams
of various effective Lagrangians.
The form factors, coupling constants, and resonance parameters are adjusted to
fit both the $\pi N$ and $\gamma N$ reaction data. They  include up to
5 channels in some fits, and have identified
several new $N^*$ states. But further confirmations are needed to establish
their findings conclusively, as will be discussed later in section 5.6.
                                                                                
\subsubsection{KSU Model}
The Kent State University (KSU) model\cite{kent}
can be derived by noting that the non-resonant amplitude $t^{bg}$,
defined by a $hermitian$ $v^{bg}$ in Eq.(54),  define
a S-matrix with the following properties
\begin{eqnarray}
S^{bg}_{a,b}(E) &=& \delta_{a,b} - 2 \pi i \delta(E-H_0) t^{bg}_{a,b}(E) \\
&=&\sum_{c}\omega^{(+)T}_{a,c}(E)\omega^{(+)}_{c,b}(E) \,,
\end{eqnarray}
where the non-resonant scattering operator is
\begin{eqnarray}
\omega^{(+)}_{a,c}(E) &=& \delta_{a,c} + g_a(E) t^{bg}_{a,c}(E) \,.
\end{eqnarray}
With some derivations, the S-matrix
Eq.(46) and the scattering T-matrix defined by Eqs.(52)-(57) can then
be cast into
following form
\begin{eqnarray}
S_{a,b}(E) =\sum_{c,d} \omega^{(+)T}_{a,c}(E)  R_{c,d}(E)
\omega^{(+)}_{c,b}(E) \,,
\end{eqnarray}
with
\begin{eqnarray}
R_{c,d}(E)=\delta_{c,d} + 2 i T^R_{c,d}(E)\,. \\
\end{eqnarray}
Here we have defined
\begin{eqnarray}
T^R_{c,d}(E) = \sum_{i,j} \Gamma^\dagger_{N^*_i,c}(E)
[G(E)_{i,j}{ \Gamma }_{N^*_j,d}(E) \,.
\end{eqnarray}
The above set of equations is identical to that used
in the  KSU model of Ref.\cite{kent}.
In practice, the KSU model fits the data by parameterizing
 $T^R$ as a Breit-Wigner
resonant form $ T^R \sim x \Gamma/2/(E- M - i\Gamma/2)$
and setting $\omega^{(+)} = B = B_1B_2\cdot\cdot\cdot B_n$, where
$B_i =exp (iX\Delta_i)$ is a unitary matrix.
                                                                                
The KSU model has been applied to $\pi N$ reactions, including pion
photoproduction. It is now being extended to investigate $\bar{K}N$ reactions.

\subsection{The CMB Model}

A unitary multi-channel isobar model with analyticity
 was developed\cite{cmb} in 1970's by the
Carnegie-Mellon Berkeley(CMB) collaboration
for analyzing the $\pi N$ data.
The CMB model can be derived by assuming that the non-resonant
potential $v^{bg}$ is also of the separable form of $v^R$ of Eq.(45)
\begin{eqnarray}
v^{bg}_{a,b} = \frac{\Gamma^\dagger _{L,a} \Gamma_{L,b}}{E-M_L}
+\frac{\Gamma^\dagger _{H,a} \Gamma_{H,b}}{E-M_H} \,.
\end{eqnarray}
The resulting coupled-channel equations are identical to Eqs.(52)-(59), 
except that $t^{bg}_{a,b}=0$ and
the sum over $N^*_i$ is now extended to include these two
distance poles $L$ and $H$.
                                                                                
By changing the integration variables and adding a substraction term,
 Eq.(57) for the self-energy can leads to CMB's dispersion
relations
\begin{eqnarray}
\Sigma_{i,j}(s) &=& \sum_{c} \gamma_{i,c} \Phi_c(s) \gamma_{j,c} \,,
\\
Re [\Phi_c(s)] &=& Re [\Phi_c(s_0)]
+\frac{s-s_{0}}{\pi} \int_{s_{th}}^{\infty}
\frac{Im[\Phi_c(s^\prime)]}{(s^\prime - s)(s^\prime-s_0)} d s^\prime \,,
\end{eqnarray}
where $\gamma_{i,c}$ is a coupling constant defining the decay of 
 $N^*_i$ into channel $c$.
Thus CMB model is analytic in structure which marks its difference
with all K-matrix models described above.
                                                                                
 The CMB model has been revived
in recent years by the Zagreb group\cite{zagreb} and a
Pittsburgh-ANL collaboration\cite{pittanl} to
 extract the $N^*$ parameters
from fitting the recent empirical $\pi N$ and  $\gamma N$ reaction
amplitudes. The resulting $N^*$ parameters have very significant differences
with what are listed by PDG in some partial waves.
 In particular, several important issues concerning the extraction of
 the $N^*$ parameters in $S_{11}$ channel have been analyzed in  detail.
                                                                                
\newpage
                                                                                
\subsection{Dynamical Models}
                                                                                
\begin{center}
{\bf A. In the $\Delta$ region}
\end{center}
                                                                                
Keeping only one resonance $N^*=\Delta$ and two channels 
$a,b= \pi N, \gamma N$,Eqs.(52)-(57) are reduced to what were developed in
the Sato-Lee (SL) model\cite{sl1,sl2}.  

\begin{figure}[thbh]
\vspace{5cm}
\centering{\includegraphics{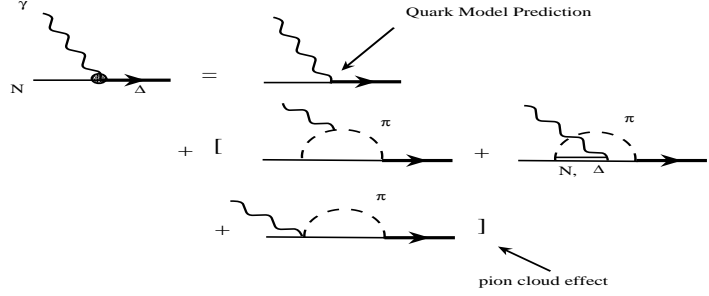}}
\caption{Graphic representation of the dressed $\gamma N \rightarrow \Delta$
vertex.  }
\label{fig:dressgm}
\end{figure}

Explicitly, we have
\begin{eqnarray}
T_{\pi N, \pi N}(E)  &=&  t^{bg}_{\pi N, \pi N}(E)
+
\frac{\bar{\Gamma}^\dagger_{\Delta, \pi N}(E)
\bar{\Gamma}_{\Delta, \pi N}(E)}{E - M^0_{\Delta} - \Sigma_\Delta(E)}\,, \\
T_{\gamma N, \pi N}(E)  &=&  t^{bg}_{\gamma N, \pi N}(E) 
+
\frac{\bar{\Gamma}^\dagger_{\Delta, \gamma N}(E)
\bar{\Gamma}_{\Delta, \pi N}(E)}{E - M^0_{\Delta} - \Sigma_\Delta(E)}  \,,
\end{eqnarray}
with
\begin{eqnarray}
\bar{\Gamma}_{\Delta,\gamma N}(E)  &=&
  { \Gamma_{\Delta,\gamma N}} 
+ \Gamma_{\Delta,\pi N}G_{\pi N}(E)t^{bg}_{\pi N,\gamma N}(E)\,, \\
\bar{\Gamma}_{\Delta,\pi N}(E)  &=&
  { \Gamma_{\Delta,\pi N}}
+ \Gamma_{\Delta,\pi N}G_{\pi N}(E)t^{bg}_{\pi N,\pi N}(E)\,, \\
t^{bg}_{\pi N,\gamma N}(E)&=& v^{bg}_{\pi N, \gamma N}
+t^{bg}_{\pi N, \pi N}(E)G_{\pi N}(E) v^{bg}_{\pi N, \gamma N}\,, \\
t^{bg}_{\pi N,\pi N}(E)&=& v^{bg}_{\pi N, \pi N}
+v^{bg}_{\pi N, \pi N}G_{\pi N}(E) t^{bg}_{\pi N, \pi N}(E)\,, 
\end{eqnarray}
and
\begin{eqnarray}
\Sigma_\Delta(E) &=& \Gamma^\dagger_{\Delta,\pi N}(E)G_{\pi N}(E) 
\bar{\Gamma}_{\Delta,\pi N}(E)\,.
\end{eqnarray}
The above equations clearly indicate how the non-resonant interaction
modify the resonant amplitude. Specifically, Eq.(84) for
the dressed $\Delta\rightarrow \gamma N$ within the SL model
is illustrated in Fig.7.

Alternatively, we can also cast Eq.(47) in the $\Delta$ region as
\begin{eqnarray}
T_{\gamma N, \pi N}(E) = t^B_{\gamma N, \pi N}(E) + t^R_{\gamma N, \pi N}(E)
\end{eqnarray}
with
\begin{eqnarray}
t^B_{\gamma N, \pi N}(E)&=&v^{bg}_{\gamma N, \pi N} + 
v^{bg}_{\gamma N, \pi N}G_{\pi N}(E)T_{\pi N, \pi N}(E) \\
t^R_{\gamma N, \pi N}(E)&=&v^R_{\gamma N, \pi N} 
+ v^R_{\gamma N, \pi N}G_{\pi N}(E)T_{\pi N, \pi N}(E)
\end{eqnarray}
The above equations are used by the Dubna-Mainz-Taiwan (DMT) model\cite{kamyan,dmt}
 except that
they depart from a consistent Hamiltonian formulation and replace
the term $t^R_{\gamma N,\pi N}$
 by the Walker's parameterization\cite{walk} 
\begin{eqnarray}
t^R_{\gamma N,\pi N}(E) =f_{\pi N}(E)
\frac{\Gamma_{tot}M_\Delta e^{i\Phi}}{M^2_\Delta-E^2 - i M_\Delta\Gamma^{tot}}
f_{\gamma N}(E)\bar{A}_{\gamma N} \,.
\end{eqnarray}
Other differences between the SL Model and the DMT model
are in the employed $\pi N$ potential and how the non-resonant 
$\gamma N \rightarrow \pi N$ amplitudes are regularized. In the DMT model,
the non-resonant $\gamma N \rightarrow \pi N$ amplitudes are calculated by
using MAID's mixture Eq.(63) of PS and PV couplings, 
while their $\pi N$ potential is from a model\cite{hung} using $PV$ coupling. 
In the SL model,
the standard $PV$ coupling is used in a consistent derivation of
both the $\pi N$ potential 
and $\gamma N \rightarrow \pi N$ transition interaction
using a unitary transformation method.

We now turn to giving relevant formula which are needed for our discussions in 
section 5.1 on the $\Delta$ resonance.
The $\Delta$ excitation is parameterized in terms of Rarita-Schwinger field.
In the $\Delta$ rest frame where $m_\Delta= q_0+E_N(\vec{q})$,
the resulting $\gamma N \rightarrow \Delta$ vertex function
can be written in the following more transparent form
\begin{eqnarray}
<\Delta\mid \Gamma_{\gamma N\rightarrow \Delta} \mid q>
  &=& -\frac{e}{(2\pi)^{3/2}}\sqrt{\frac{E_N(\vec{q})+m_N
}{2E_N(\vec{q})}}
       \frac{1}{\sqrt{2\omega}}
       \frac{3 (m_\Delta+m_N) }{ 4m_N( E_N(\vec{q})+m_N)}T_3
\nonumber \\
&\times&
{}  [i G_M(q^2)\vec{S}\times\vec{q}\cdot \vec{\epsilon}
      +G_E(q^2)
       (\vec{S}\cdot\vec{\epsilon} \vec{\sigma}\cdot\vec{q}
   +\vec{S}\cdot\vec{q}\vec{\sigma}\cdot\vec{\epsilon}) \nonumber \\
 & & +
        \frac{G_C(q^2)}{m_\Delta}
      \vec{S}\cdot\vec{q} \vec{\sigma}\cdot\vec{q} \epsilon_0], \label{vertgd}
\end{eqnarray}
where $e=\sqrt{4\pi/137}$, $q=(\omega,\vec{q})$ is the photon four-momentum,
and $\epsilon=(\epsilon_0,\vec{\epsilon})$ is the photon polarization
vector.
The transition operators $\vec{S}$ and $\vec{T}$ are defined
by the reduced matrix element $<\Delta ||\vec{S}|| N> =
<\Delta || \vec{T} || N > = 2$ in Edmonds' convention\cite{edmond}.
By using Eq.(93) and the standard definitions\cite{deforest,amaldi}
of the multipole amplitudes, it is straightforward to evaluate the magnetic
M1, electric E2 and Coulomb C2 amplitudes
of the $\gamma N \rightarrow \Delta$ transition.
We find\cite{sl2} that
\begin{eqnarray}
  G_M(q^2)&=& \frac{1}{N}[\Gamma_{\gamma N \rightarrow \Delta}]_{M1} 
, \label{ambare} \\
G_E(q^2)&=&\frac{-1}{N} [\Gamma_{\gamma N \rightarrow \Delta}]_{E2} 
, \label{aebare} \\
G_C(q^2)&=&\frac{2m_\Delta}{\mid \vec{q} \mid N}
[\Gamma_{\gamma N \rightarrow \Delta}]_{C2} , \label{acbare}
\end{eqnarray}
with
\begin{eqnarray}
N= \frac{e}{2m_N}\sqrt{\frac{m_\Delta \mid\vec{q}\mid}{m_N}}
\frac{1}{[1-q^2/(m_N+m_\Delta)^2]^{1/2}}. \nonumber
\end{eqnarray}
At $q^2=0$, the above relations agree with that given in
Appendix A of Ref.\cite{nbl}.
Equations (93)-(96) can  also be used to relate
 the dressed vertex $\bar{\Gamma}_{\gamma N \rightarrow \Delta}$, 
defined by Eq.(84), to
 the corresponding dressed form factors:
\begin{eqnarray}
  G^*_M(q^2)&=& \frac{1}{N}[\bar{\Gamma}_{\gamma N \rightarrow \Delta}]_{M1}
, \nonumber \\
G^*_E(q^2)&=&\frac{-1}{N} [\bar{\Gamma}_{\gamma N \rightarrow \Delta}]_{E2}
, \nonumber \\
G^*_C(q^2)&=&\frac{2m_\Delta}{\mid \vec{q} \mid N}
[\bar{\Gamma}_{\gamma N \rightarrow \Delta}]_{C2}. \nonumber
\end{eqnarray}

At the  $q^2=0$ photon point, we will also compare our
results with the helicity amplitudes defined by PDG\cite{pdg04}.
They are related to the  multipole amplitudes defined above  by
\begin{eqnarray}
A_{3/2}&=&\frac{\sqrt{3}}{2}[[\bar{\Gamma}_{\gamma N \rightarrow \Delta}]_{E2} -
[\bar{\Gamma}_{\gamma N \rightarrow \Delta}]_{M1} ], \\
A_{1/2}&=&-\frac{1}{2}[3[\bar{\Gamma}_{\gamma N \rightarrow \Delta}]_{E2}+
[\bar{\Gamma}_{\gamma N \rightarrow \Delta}]_{M1}].
\end{eqnarray}

At the $\Delta$ resonance position  $E=M_R = 1236$ MeV, the
$\pi N$ phase shift in the $P_{33}$ channel  goes through 90 degrees.
This leads to a relation, as derived in detail in Ref.\cite{sl1}, that
the multipole components of the dressed
vertex $\bar{\Gamma}_{\gamma N \rightarrow \Delta}$ are related to
the imaginary($Im$) parts of the $\gamma N \rightarrow \pi N$
 multipole amplitudes
in the $\pi N$ $P_{33}$ channel
\begin{eqnarray}
{G}_M^*(q^2) &=&
\frac{1}{N}[\bar{\Gamma}^K_{\gamma N \rightarrow \Delta}]_{M1}
=\frac{1}{N}
\sqrt{\frac{8\pi m_\Delta k \Gamma_\Delta}{3 m_N q}}
 \times \mbox{Im}(M_{1+}^{3/2}), \label{am}\\
{G}_E^*(q^2) 
 &=&\frac{1}{N}[\bar{\Gamma}^K_{\gamma N \rightarrow \Delta}]_{E2}=
\frac{1}{N}\sqrt{\frac{8\pi m_\Delta k \Gamma_\Delta}{3 m_N q}}
 \times \mbox{Im}(E_{1+}^{3/2}),  \label{ae}\\
\frac{\mid \vec{q} \mid}{2m_\Delta}{G}_C^*(q^2)  
&=&\frac{1}{N}[\bar{\Gamma}^K_{\gamma N \rightarrow \Delta}]_{C2}=
\frac{1}{N}\sqrt{\frac{8\pi m_\Delta k \Gamma_\Delta}{3 m_N q}} \times
 \mbox{Im}(S_{1+}^{3/2}),   \label{ac}
\end{eqnarray}
where $\Gamma_\Delta$ is the
$\Delta$ width, $k$ and $q$ are respectively
the momenta of the pion and photon in the rest frame of the $\Delta$.
Note that the upper index $K$ in $\Gamma^K_{\Delta,\gamma N}$
in Eqs.(99)-(101) means taking only the principal-value
 integration in evaluating
the second term of Eq.(84). Details are discussed in Ref.\cite{sl1}.

>From the above relations, we obtain a very useful relation that the
$E2/M1$ ratio $R_{EM}$ and $C2/M1$ ratio $R_{SM}$ 
of the dressed
$\gamma N \rightarrow \Delta$ transition at
$W=1232$ MeV can be evaluated directly by
using the $\gamma N \rightarrow \pi N$ multipole amplitudes
\begin{eqnarray}
R_{EM} &=&
\frac{[\bar{\Gamma}^K_{\gamma N \rightarrow \Delta}]_{E2}}
{[\bar{\Gamma}^K_{\gamma N \rightarrow \Delta}]_{M1}}
= \frac{\mbox{Im} (E_{1+}^{3/2})}{\mbox{Im} (M_{1+}^{3/2})},
         \label{rema} \\
R_{SM} &=&
\frac{[\bar{\Gamma}^K_{\gamma N \rightarrow \Delta}]_{C2}}
{[\bar{\Gamma}^K_{\gamma N \rightarrow \Delta}]_{M1}}
 =\frac{\mbox{Im} (S_{1+}^{3/2})}{\mbox{Im} (M_{1+}^{3/2})}.
         \label{rsma}
\end{eqnarray}

Eqs.(99)-(103) can be used in the empirical amplitude analyses 
to extract the form factors and
 the $E2/M1$ and $C2/M1$ ratios of the $\gamma N \rightarrow 
\Delta$ transition.
The extractions of the bare vertices, which can be compared with the
predictions from most of the constituent quark model
calculations, can only be achieved by using the dynamical
 model through Eq.(84).
This indicates why an appropriate reaction theory is needed in the
$N^*$ study, as illustrated in Fig.1.
 
\vspace{1cm}
\begin{center}
{\bf B. In the Second and third resonance regions}
\end{center}
\vspace{0.5cm}
                                                                                
In these regions, we need to include more than $\pi N$ channel to
solve Eq.(47) or Eqs.(52)-(59). In addition, these formula must be extended to include
explicitly the $\pi\pi N$ channel, instead of using the quasi two-particle
 channels
$\pi\Delta$, $\rho N$, and $\sigma N$ to simulate the $\pi\pi N$ continuum.
This however is still being developed\cite{lms}. Here we continue to explain the
current investigations in the second and third resonance
 regions within the formulation defined by
Eqs.(52)-(59).

Eqs.(52)-(59) are used in a 2-$N^*$ and 3-channel ($\pi N$, $\eta N$,
and $\pi \Delta$)  study\cite{yosh} of $\pi N$ scattering in $S_{11}$ partial wave,
aiming at investigating how the quark-quark interaction in
the constituent quark model can be determined directly by using the reaction data.
Eqs.(52)-(59) are also the basis of
examining the $N^*$ effects\cite{otl} and one-loop coupled-channel effects\cite{ohlee}
on $\omega$ meson photoproduction and the
coupled-channel effects on $K$ photoproduction\cite{chitab}.
                                                                                
The coupled-channel study of both $\pi N$ scattering and $\gamma N \rightarrow \pi N$
in  $S_{11}$ channel by Chen et al\cite{chen} includes
$\pi N$, $\eta N$, and $\gamma N$ channels.
Their $\pi N$ scattering calculation is performed by using Eq.(47), which
is of course equivalent to  Eqs.(52)-(59).
In their $\gamma N \rightarrow \pi N$ calculation, they neglect
the $\gamma N \rightarrow \eta N \rightarrow \pi N $ coupled-channel effect,
and follow the procedure of the DMT model to
 evaluate the resonant term in terms of
the Walker's parameterization (Eq.(64)).
  They find that four
$N^*$  are needed to fit the
empirical amplitudes in $S_{11}$ channel up to $W = 2$ GeV.
                                                                                
A coupled-channel calculation based on Eq.(47) has been carried out
by J\"ulich group\cite{julich} for $\pi N$ scattering.
They are able to describe the $\pi N$ phase shifts
up to $W=1.9$ GeV by including $\pi N$, $\eta N$,
$\pi \Delta$, $\rho N$ and $\sigma N$ channels and 5 $N^*$ resonances :
$P_{33}(1232)$, $S_{11}(1535)$,
$S_{11}(1530)$, $S_{11}(1650)$ and $D_{13}(1520)$. They find that the Roper resonance
$P_{11}(1440)$ is completely due to the meson-exchange coupled-channel effects.
                                                                                
A coupled
channel calculation based on Eq.(47) for both $\pi N$ scattering
and $\gamma N \rightarrow \pi N$ up to $W=1.5$ GeV has
been reported by Fuda and Alarbi\cite{fuda}.
 They include $\pi N$, $\gamma N$, $\eta N$, and $\pi \Delta$
channels and 4 $N^*$ resonances :
$P_{33}(1232)$, $P_{11}(1440)$, $S_{11}(1535)$,
 and $D_{13}(1520)$. The parameters are adjusted to fit the empirical multipole
amplitudes in
a few low partial waves.
                                                                                
Much simpler coupled-channel calculations have been  performed by using
separable interactions. In the model of Gross and Surya\cite{gros}, such separable
interactions are from simplifying the meson-exchange mechanisms
in Figs~\ref{fig:mechanism_1}.(a)-c)
as a contact term
like Fig.~\ref{fig:mechanism_1}(d). They include only $\pi N$ and
$\gamma N$ channels  and 3 resonances: $P_{33}(1232)$, $P_{11}(1440)$ and $D_{13}(1520)$,
and restrict their investigation up to
$W < 1.5$ GeV. To account for the inelasticities in $P_{11}$ and $D_{13}$, the $N^*\rightarrow
\pi\Delta$ coupling is introduced in these two partial waves. The inelasticities in other
partial waves are neglected.

A similar separable simplification is also used in
the chiral coupled-channel models\cite{kais1,oset1} for strange
particle production. There the separable interactions
 are directly determined from
the leading contact terms of SU(3) effective chiral Lagrangian and hence
only act on s-wave partial waves.
They are able to fit the total cross section data for various strange
particle production reaction channels without introducing resonance
states. It remains to be seen whether these models can be further improved to account for
higher partial waves which are definitely needed to give an accurate description of
the data even at energies near production thresholds.

\subsection{Dispersion-relation approaches}

Historically, the approach based on dispersion relations is defined within the
S-matrix theory which was introduced as an alternative to
the relativistic quantum field theory in investigating
 non-perturbative hadron interactions. This approach was first applied
to investigate pion photoproduction by Chew, Goldberger, Low, and 
Nambu\cite{cgln} (CGLN) and electroproduction by
Amaldi, Fubini and Furlan\cite{amaldi}. It was fully developed\cite{donn1,schw}
in the years around 1970 to analyze the data at energies near the $\Delta$
resonance. In recent years, it has been revived by Aznauryan\cite{azn,azn1}, and 
by Hanstein, Drechsel, and Tiator\cite{hans} 
for investigating pion photoproduction 
and $\eta$ photoproduction\cite{azn2}.

The dispersion relation approach assumes that the
scattering amplitude is unitary and
possesses various established symmetry properties such as Lorentz invariance
and gauge invariance. The dynamics is defined by the assumed analytical
property and crossing symmetry. 
For $\pi$ and $\eta$ production, the starting point is the fixed $t$ 
dispersion relation\cite{bell} for the invariant amplitudes $A_i$ defined in
Eq.(24)
\begin{eqnarray}
Re [A^I_k(s,t)] = A^{I,pole}_k(s,t) + \frac{1}{\pi} {\it P}\int_{s_{thr}}^{\infty}
ds^\prime [\frac{1}{s^\prime - s} +\frac{\epsilon^I\xi_k}{s^\prime - u}]
Im[ A^I_k(s^\prime,t)] \,,
\end{eqnarray}
where 
$A^{I,pole}_k(s,t)$ is calculated from pseudo-scalar Born term,
$I=0,+,-$ denote the isospin component, $\xi_1=\xi_2=-\xi_3=-\xi_4=1$
and $\epsilon^+=\epsilon^0=-\epsilon^-=1$ are defined such that the
crossing symmetry relation 
$A^I_k(s,t,u)=\xi_k\epsilon^I A^I_k(u,t,s)$ is satisfied. 
With the definitions Eqs.(23) and (27) and the
 multipole expansion defined by Eqs.(35)-(40), the above fixed-t
 dispersion relation
leads to the following set of
coupled equations relating the real part and imaginary
parts of multipole amplitudes
\begin{eqnarray}
Re [M^I_\ell (W)] =M_\ell ^{I,pole}(W) 
+\frac{P}{\pi} \int_{W_{thr}}^{\infty} d W^\prime \sum_{\ell ^\prime}
K^I_{\ell \ell ^\prime}(W,W^\prime) Im [M^I_{\ell ^\prime}(W)] \,,
\end{eqnarray}
where $M^I_\ell$ is the multipole amplitude, 
$M_\ell^{I,pole}(W)$ is calculated from pseudo-scalar Born term, and
$K^I_{\ell, \ell^\prime}$ contains various kinematic factors.
In the recent work of Ref.\cite{hans}, the procedures of
Ref.\cite{schw} are used to solve the
above equations 
by using the  method of Omnes\cite{omne}. It assumes that
the multipole amplitude can be written as
\begin{eqnarray}
M^I_\ell (W) =exp^{i\phi_\ell (W)} \frac{1}{r_{\ell I}} {\sc M}^I_\ell (W)\,, 
\end{eqnarray}
where ${\sc M}^I_\ell$ is a real function and $r_{\ell I}$ is some kinematic
factor, and hence
\begin{eqnarray}
Im [M^I_\ell (W)] = h^{I*}_\ell(W)   M^I_\ell (W) \,,
\end{eqnarray}
with $h^I_\ell(w)=sin(\phi^I_\ell) exp^{i\phi^I_\ell (W)}$. The phase 
$\phi^I_\ell$ is assummed to be
\begin{eqnarray}
\phi^I_\ell(W) =arctan(\frac{1-\eta^I_\ell (W) cos2\delta^I_\ell(W)}
{\eta^I_\ell(W) sin 2\delta^I_\ell (W)}) \,,
\end{eqnarray}
where $\delta^I_\ell$ and $\eta^I_\ell$ are the phase and 
inelasticity of $\pi N$
scattering in the partial-wave with quantum numbers ($\ell, I)$.

The next approximation is to  limit the  sum over $\ell^\prime$ in
the right-hand side of Eq.(105) to a cutoff $\ell_{max}$. 
For investigating production below $E_\gamma = 500$ MeV, 
$\ell_{max}=1$ is taken. 
Another approximation is needed to
handle the integration over $W$ in Eq.(105). In Ref.\cite{hans}, the integration
is cutoff at $W= \Lambda= 2 $ GeV such that all needed phase $\Phi_\ell$ can be 
determined by the empirical  $\pi N$ phase shifts. The neglected contribution
from $ W > 2 $ GeV is then acounted for by adding vector meson exchange
terms. Eq.(105) then becomes 
\begin{eqnarray}
{\sc M}^I_\ell(W) &=& {\sc M}_\ell ^{I,pole}(W)
+\frac{1}{\pi} \int_{W_{thr}}^{\Lambda}  
\frac{h^{I*}_\ell(W^\prime){\sc M}^I_\ell(W^\prime) d W^\prime}
{W^\prime - W - i\epsilon} \nonumber \\
&+&\frac{1}{\pi} \sum_{\ell^\prime, I^\prime}
 \int_{W_{thr}}^{\Lambda} d W^\prime 
{\sc K}^{I,I^\prime}_{\ell\ell^\prime}(WW^\prime) 
h^{I^\prime *}_{\ell^\prime}(W^\prime)
+{\sc M}^{I,V}_\ell(W)
\end{eqnarray}

The method for solving Eqs.(109) is given in Ref.\cite{hans}.
With the above procedures, the model contains 10 adjustable parameters.
Excellent fit to all $\gamma N \rightarrow \pi N$ data up
to $E_\gamma = 500$ MeV has been obtained in Ref.\cite{hans}. 

The calculation in Ref.\cite{azn} follow the same approach with additional
simplification that the coupling between different multipoles
and the contribution from $ W > \Lambda$ to the integration are neglected;
setting ${\sc K}^{I,I^\prime}_{\ell,\ell^\prime}=0$ and 
${\sc M}^{I,V}_\ell=0$
in solving Eq.(109). These simplifications are justified in calculating the
dominant $\Delta$ excitation amplitude $M_{1+}^{(3/2)}$. But it is questionable
if they can be applied for calculating weaker amplitudes. Thus no attempt was made in
Refs.\cite{azn,azn1} to fit the data directly using dispersion relations.
Rather, the emphasis was in the interpretation of the empirical
amplitudes $M^{(3/2}_{1+}$, $E^{(3/2}_{1+}$ in terms of
rescattering effects and constituent quark model prediction.
By assuming the multipole expansion is also valid in electroproduction,
the $Q^2$-dependence of these $\Delta$ excitation amplitudes are then 
predicted. There are questions regarding the validity of multipole expansion at 
$E_\gamma > $ 500 MeV and large $Q^2$~\cite{craw}.

The dispersion relation approach is also used in Ref.\cite{azn1} to
analyze the pion photoproduction and electroproduction data in the second 
and third resonance region.
It is assumed that the imaginary parts of the amplitudes in 
$M_N+m_\pi < W < 2 $ GeV are
from the resonant amplitudes
parameterized as the Walker's Breit-Wigner form Eq.(64), and
in $ 2.5$ GeV $ < \infty$ from Regge-pole model. The imaginary part of
the amplitude in 2 GeV $ < W < 2.5 $ GeV is obtained by interpolation.
The
real part of the amplitude 
 is then calculated from the dispersion relation described above.
The empirical amplitudes are then fitted by adjusting the resonant parameters.
It turns out that the resulting parameters are close to what were
determined in the single channel K-matrix model described in subsection 4.3.2.

With appropriate modifications, the dispersion relation approach can be
applied to investigate the production of other pseudo-scalar mesons.
This has been achieved in Ref.\cite{azn2} in analyzing the data of
$\eta$ production reactions.

\section{Data and Results of Analyses}

A large volume of data of electromagnetic meson production reactions
is needed to extract the fundamental physics on 
resonance transition form factors or discover new baryon states.
Efforts in this direction in 
the 1970's and 80's at various laboratories were hampered by the low 
duty cycle synchrotrons that were available for these studies, and by 
the use of magnetic spectrometers with relatively small acceptance. 
For a discussion on these results see the excellent review by 
F. Foster and G. Hughes~\cite{foster83}. 
The construction of CW electron accelerators, and the advances in 
detector technologies have made it possible to use detector system 
with nearly 4$\pi$ solid angle coverage, and the ability to operate at
high luminosity. Moreover, the detection of multiple 
photons from $\pi^{\circ}$ or $\eta$ decays with high resolution 
has become feasible with the development 
of high density crystals with sufficient light output, such as 
BGO, CsI, and PbF$_2$. These detectors have become powerful tools in 
the study of baryon spectroscopy and structure.
In this section we will highlight the data and review the results from
the analyses.  We will only consider meson productions from 
nucleon targets. An extensive review of meson production from nuclei has been 
published recently~\cite{krusche03}.
     
\subsection{Single pion production.}

Single pion photoproduction and electroproduction have been the main processes
in the study of the electromagnetic transition amplitudes of the lower
mass nucleon resonances such as $\Delta(1232)$, $P_{11}(1440)$, $D_{13}(1520)$,
$S_{11}(1535)$, and $F_{15}(1680)$.
Data on pion photoproduction now exist from LEGS~\cite{legs_pi0} 
and MAMI~\cite{mami_pi0}, and from GRAAL~\cite{graal_pi+,graal_nstar}, 
including results from measurements of 
polarized beam asymmetries and  beam-target double 
polarization observables~\cite{mami_pi+_double_pol,mami_pi0_double_pol}.


During the past few years high statistics data of pion electroproduction
have been collected at JLab. 
The  $ep \rightarrow ep\pi^{\circ}$ data cover a large range in             
invariant mass W from threshold to  2.5 GeV,  a wide range
in momentum-transfer $Q^2 = 0.1 - 6$~GeV$^2$, and the full range in azimthal and polar angles in the $p\pi^{\circ}$ center-of-mass.  
                                                                                
In the past there have been very limited data on
$n\pi^+$, mostly at forward center of mass 
angles\cite{bonn1,desy1,nina1}, some at
backward angles\cite{bonn2}. Even fewer data exist in $\pi^-$ production from deuterium~\cite{nina2}. 
This has limited our ability to extract reliable resonance
transition amplitudes in the high mass region where many isospin ${1\over 2}$
states exist, which couple more strongly to $n\pi^+$ than 
to $p\pi^{\circ}$.                                                                     New $n\pi^+$ data from CLAS \cite{clas_npip02} have nearly full angular coverage and
span the range W = 1.1 - 1.6 GeV and $Q^2 = 0.3 - 0.6$~GeV$^2$. They vastly
increase the covered kinematics with high statistics, and will be extended to 
$W \le 2.5$~GeV$^2$ and $Q^2 = 0.1 - 6$~GeV$^2$.  

For the first time there are also significant amounts of polarized
beam asymmetries data and data on the beam helicity response function $\sigma_{LT^{\prime}}$
available, both for $p\pi^{\circ}$~\cite{joo02,joo03_1,mami02,bates_oops03}
and for $n\pi^+$~\cite{joo03_2}.
The most complete data sets will come from JLab for both the
$\vec{e}p \rightarrow ep\pi^{\circ}$, and the $\vec{e}p \rightarrow en\pi^+$
channels.
Some response functions at few low $Q^2$ have also been measured
MIT-Bates and Mainz. In particular, the experiments using the
OOPS of MIT-Bates yield rather precise data of
$A_{TT}$ and $A_{TL}$  at $Q^2=0.126$ (GeV/c)$^2$ covering a limited
angular range. In table ~\ref{tab:pion_data}, we summarize these new data.
\begin{table}[htbp]
\tcaption{Summary of the single pion electroproduction data}
\centerline{\footnotesize\smalllineskip
\begin{tabular}{l c l l l }\\
\hline
{Reaction} &{Observable} &W range &$Q^2$ range & Lab/experiment \\
{} &\phantom0 &(GeV) & (GeV$^2$)&\phantom0 \\
\hline\\
{$ep \rightarrow ep\pi^{\circ}$}       &$d\sigma / d\Omega$               &$< 1.8$             &0.4 - 1.8  &JLab-CLAS~\cite{joo02}\\
				       &$d\sigma / d\Omega$               &$\Delta(1232)$      &0.1 - 0.9 &ELSA-Elan~\cite{elsa_elan}\\
				       &$d\sigma / d\Omega$               &$\Delta(1232)$      &2.8, 4.0  &JLab-Hall C~\cite{frolov99}\\
				       &$d\sigma / d\Omega$               &$ < 2.5~ GeV$          &2 - 6     &JLab-CLAS~\cite{jlab-e99-107}\\
				       &$d\sigma / d\Omega$               &$\Delta(1232)$      &7.5       &JLab-CLAS~\cite{jlab-e01-002}\\
    	   			       &$d\sigma / d\Omega$               &$< 2.0$             &1.0       &JLab-Hall A~\cite{laveissiere03}\\
{$ep \rightarrow en\pi^+$}             &$d\sigma / d\Omega$               &$< 1.6$             &0.3 - 0.65 &JLab-CLAS~\cite{clas_npip02}\\
			               &$d\sigma / d\Omega$               &$\Delta(1232)$      &0.1 - 0.9 &ELSA-Elan~\cite{elsa_elan}\\
				       &$d\sigma / d\Omega$               &$ < 2.5~ GeV$          &2 - 6     &JLab-CLAS~\cite{jlab-e99-107}\\
{$\vec{e}p \rightarrow ep\pi^{\circ}$} &$A_e,~A_{TT},~A_{TL}$                             &$\Delta(1232)$      &0.2        &MAMI-A1~\cite{mami02}\\
				       &$\sigma_{LT^{\prime}}$            &$\Delta(1232)$      &0.3 - 0.65 &JLab-CLAS~\cite{joo03_1}\\
				       &$A_e$                             &$\Delta(1232)$      &0.126       &Bates-OOPS~\cite{bates_oops03}\\
{$\vec{e}p \rightarrow en\pi^+$}       &$\sigma_{LT^{\prime}}$            &$< 1.6$             &0.3 - 0.65 &JLab-CLAS~\cite{joo03_2}\\
{$\vec{e}p \rightarrow e\vec{p}\pi^{\circ}$} & pol. resp. fct.		  &$\Delta(1232)$      &1.0       &JLab-Hall A~\cite{jlab-e91-011}\\
{$\vec{e}\vec{p} \rightarrow ep\pi^{\circ}$} &$A_p, A_{ep}$               &$\Delta(1232)$      &0.5 - 1.5      &JLab-CLAS~\cite{biselli03}\\
{$\vec{e}\vec{p} \rightarrow en\pi^+$} &$A_{ep}$                          &$< 1.85$            &0.4, 0.65, 1.1 &JLab-CLAS~\cite{devita02}\\
\hline\\
\end{tabular}}
\label{tab:pion_data}
\end{table}
                                                                                
One of the main outcomes from the analyses of these single pion production data
 is a more detailed understanding of
the $\Delta(1232)$ resonance.
The focus has been
on the determination of the magnetic $M1$, electric $E2$,
and Coulomb $C2$ form factors of the $\gamma N \rightarrow \Delta$ transition.
This development will be discussed in detail  in this subsection.
The single pion production in the second and third resonance 
regions will be covered mainly in 
section 5.3 where some  $N^*$ parameters extracted from a combined analysis 
including the data of $\eta$  production will be discussed.

\subsubsection{Pion photoproduction}
                                                                                                                                                                
\begin{figure}[htbn{figure}[tbh]
\vspace{8.5cm}
\centering{\includegraphics{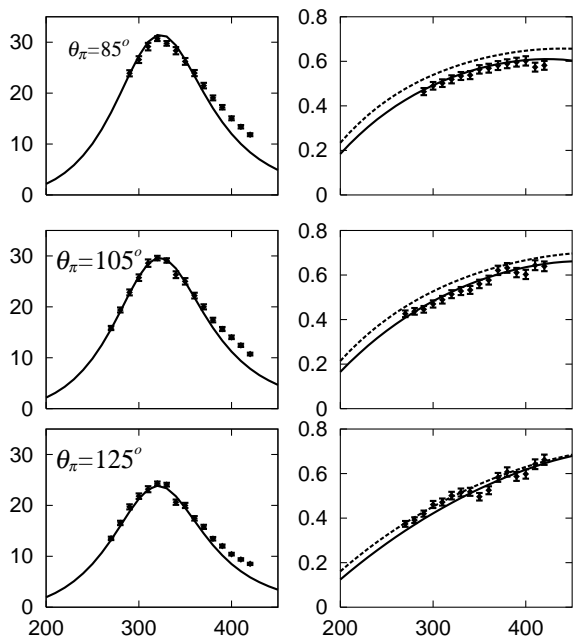}}
 \caption[]{The differential cross section($d\sigma/d\Omega$) and
           photon asymmetry($\Sigma$) of the $p(\gamma,\pi^0)p$ reaction,
           calculated from the Sato-Lee model,
          are compared with the  data of Mainz\cite{mami_pi0}. The photon
          asymmetry data of LEGS\cite{legs_pi0} agree  with those shown
          here and hence are not displayed. 
          The dashed curves are obtained from setting $R_{EM}=0$.
          The dashed and solid curves for $d\sigma/d\Omega$ are 
          indistingushable.}
\label{fig:legs}
\end{figure}

The high statistics of
the photon asymmetry data is essential in
determining the small $E_{1+}$ amplitude
of the $\gamma N \rightarrow \pi N$ reaction, which determines the electric $E2$
strength of the $\gamma N \rightarrow \Delta$ transition
through Eq.(100). Fig.~\ref{fig:legs} shows the comparison of the
results from the Sato-Lee (SL) model\cite{sl1} and the
$\gamma p \rightarrow p\pi^{\circ}$ data from Mainz and LEGS. 
When the  $E_{1+}$ amplitude is
turned off in the SL model,
the predicted photon asymmetries (dotted curves) deviate from the
data. 
By performing the amplitude analyses of these new data by several groups,
we now have a world averaged value of the $R_{EM}$ ratio, defined by
Eq.(102), $R_{EM} = (- 2.38 \pm 0.27) \%$\cite{e2m1} at photon point.
The magnetic $M1$ transition strength, defined by Eq.(99),
has also been determined as $G^*_M(0) =
3.18 \pm 0.04$.

\begin{table}[hbpt]
\caption{\small Helicity amplitude $A_{3/2}$ and $E2/M1$ ratio $R_{EM}$ for
the $\gamma N \rightarrow \Delta$ transition at $Q^2=0$ photon point.
 $A_{3/2}$ is in unit of $10^{-3} GeV^{-1/2}$ and $R_{EM}$ in \%.
The references are:
a(68), b(71), c(55),
 d(50), e(101), f(19).
 $^*$ This value differs from the value $226$
listed in Ref.69, because of a kinematic factor is not included correctly.}
\centerline{\footnotesize\smalllineskip
\begin{tabular}{lccccl}\\
\hline
              & \multicolumn{2}{c}{$A_{3/2}$}
              & \multicolumn{2}{c}{$R_{EM}$}      & Refs.\\
                &  Dressed  & Bare & Dressed  & Bare & \\ \hline
Dynamical Model & -258$^*$      & -153 & -2.7     & -1.3 & a\\
                & -256      & -136 & -2.4     & 0.25 & b\\
K-Matrix        & -255      &  $-$    & -2.1     &  $-$    & c\\
Dispersion      & -252      &   $-$   & -2.5     &  $-$    & d\\
Quark Model     &  $-$      & -186 &   $-$       & $\sim 0$ & e \\
                &  $-$      & -157 &    $-$     & $\sim 0$ & f\\
                \hline
\end{tabular}}
\label{tab:Delta_multipoles}
\end{table}
For the dynamical models, it is
possible to also get the bare transition strengths $G_M(0)$ and
$G_E(0)$ which are obtained by
separating the pion cloud effects from the full(dressed)
transition strengths, as defined by Eq.(84) and illustrated 
in Fig.\ref{fig:dressgm}.
In Table ~\ref{tab:Delta_multipoles}, we show the importance of the pion cloud.
We see that 
 the helicity amplitude $A_{3/2}$
 extracted from three different  analyses are
very close to each other and are about 40 $\%$ larger than
the bare strengths extracted
within the dynamical models of Refs.\cite{sl1,kamyan}.
We now note that these bare values are
within the ranges predicted by two constituent quark models.
This suggests that
the bare parameters of the dynamical model are more likely to be identified
 with the current hadron structure calculations.
In Table ~\ref{tab:Delta_multipoles} we also see that the
 differences between dressed and bare values of
$R_{EM}$ are even larger.
The bare values from two dynamical model analyses are quite
different, indicating some significant differences in their formulations
as discussed in section 4.

\subsubsection{Pion electroproduction}

\begin{figure}[thbp]
\vspace{80mm}
\centering{\includegraphics{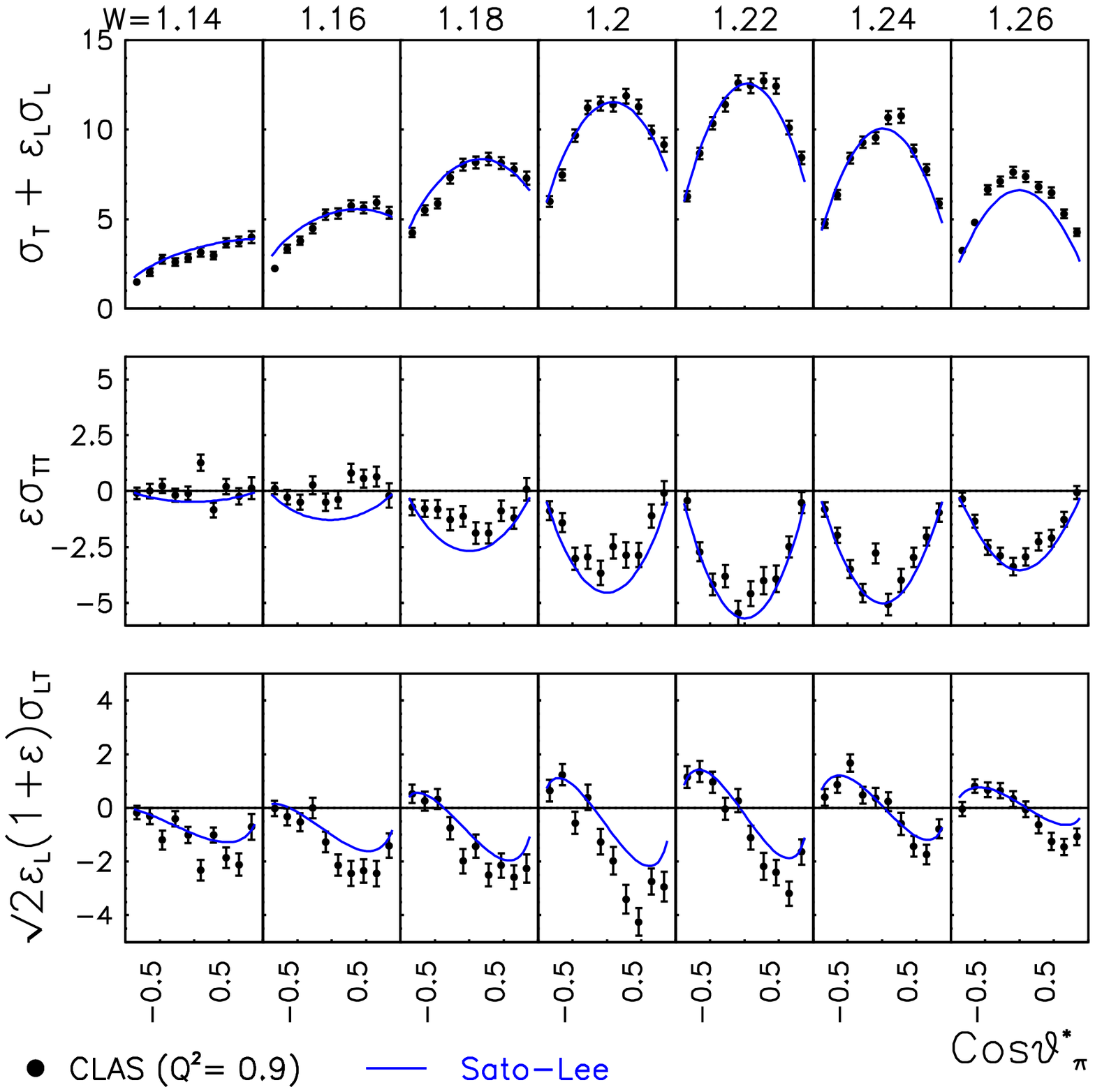}}
\caption{$p(e,e^\prime \pi^0)$ cross section data from CLAS at JLab are
compared with the predictions from the SL Model.}
\label{fig:jlab_pi0_1}
\end{figure}

\begin{figure}[tphb]
\vspace{8cm}
\includegraphics{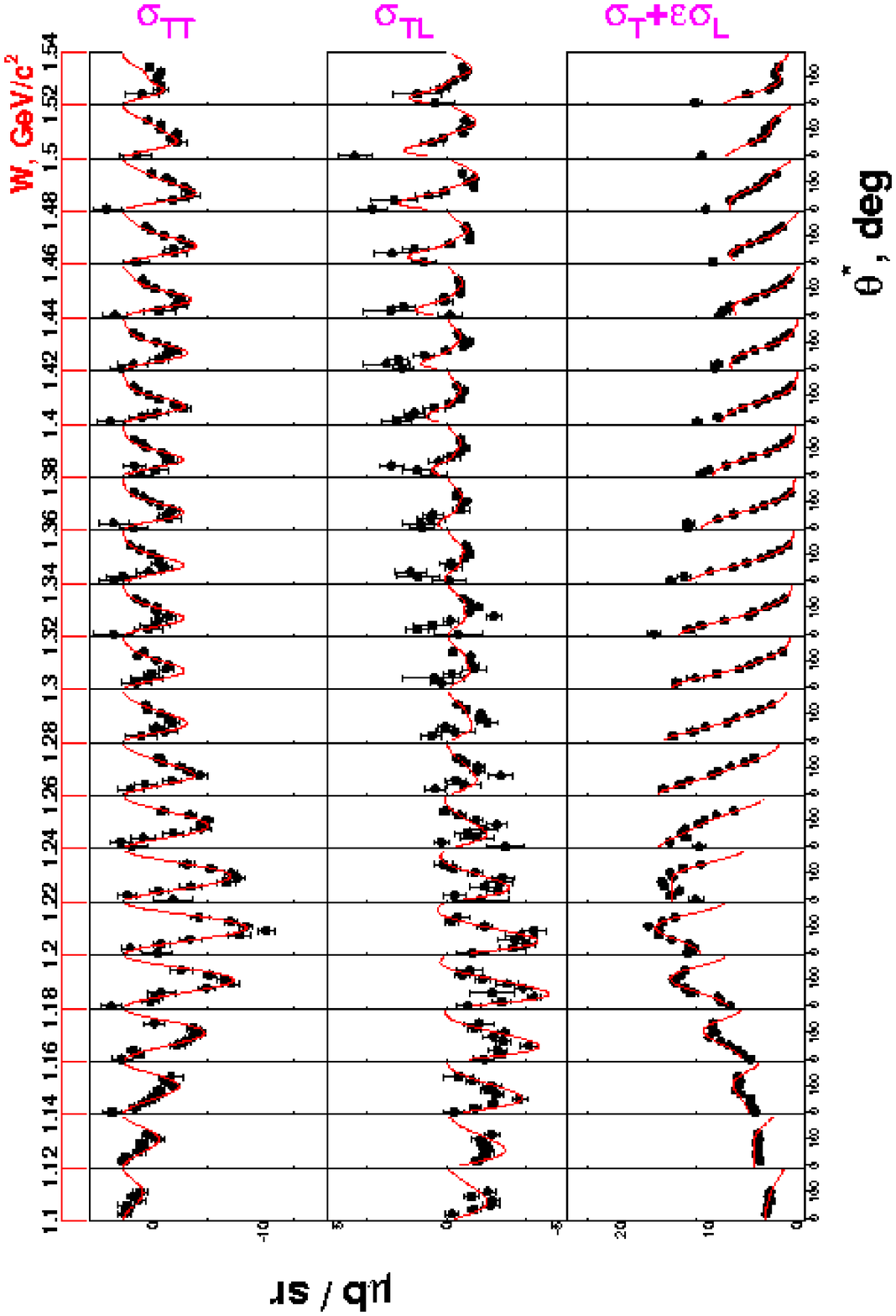}
\caption{\small $ep \rightarrow e n\pi^+$ data from  CLAS. 
Separated response functions $\sigma_T + \epsilon \sigma_L$, $\sigma_{TT}$, and $\sigma_{LT}$ are
shown. The curve represents a fit to the data using the 
JLab/Yeveran unitary isobar model .}
\label{fig:clas_electro_npipl}
\end{figure}

\begin{figure}[thbp]
\vspace{70mm}
\centering{\includegraphics{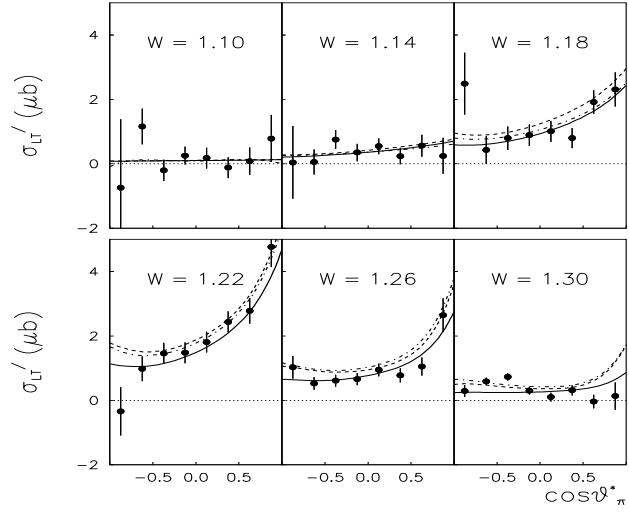}}
\caption{CLAS data on $\sigma_{LT^{\prime}}$ of $p(e,e^\prime \pi^+)n$
reaction in the $\Delta(1232)$ region are compared 
with predictions from SL(solid), MAID(dashed), and
 DMT(dashed-dotted) models.}
\label{fig:rltp_pip_c1_q1}
\end{figure}

\begin{figure}[thbp]
\vspace{80mm}
\centering{\includegraphics{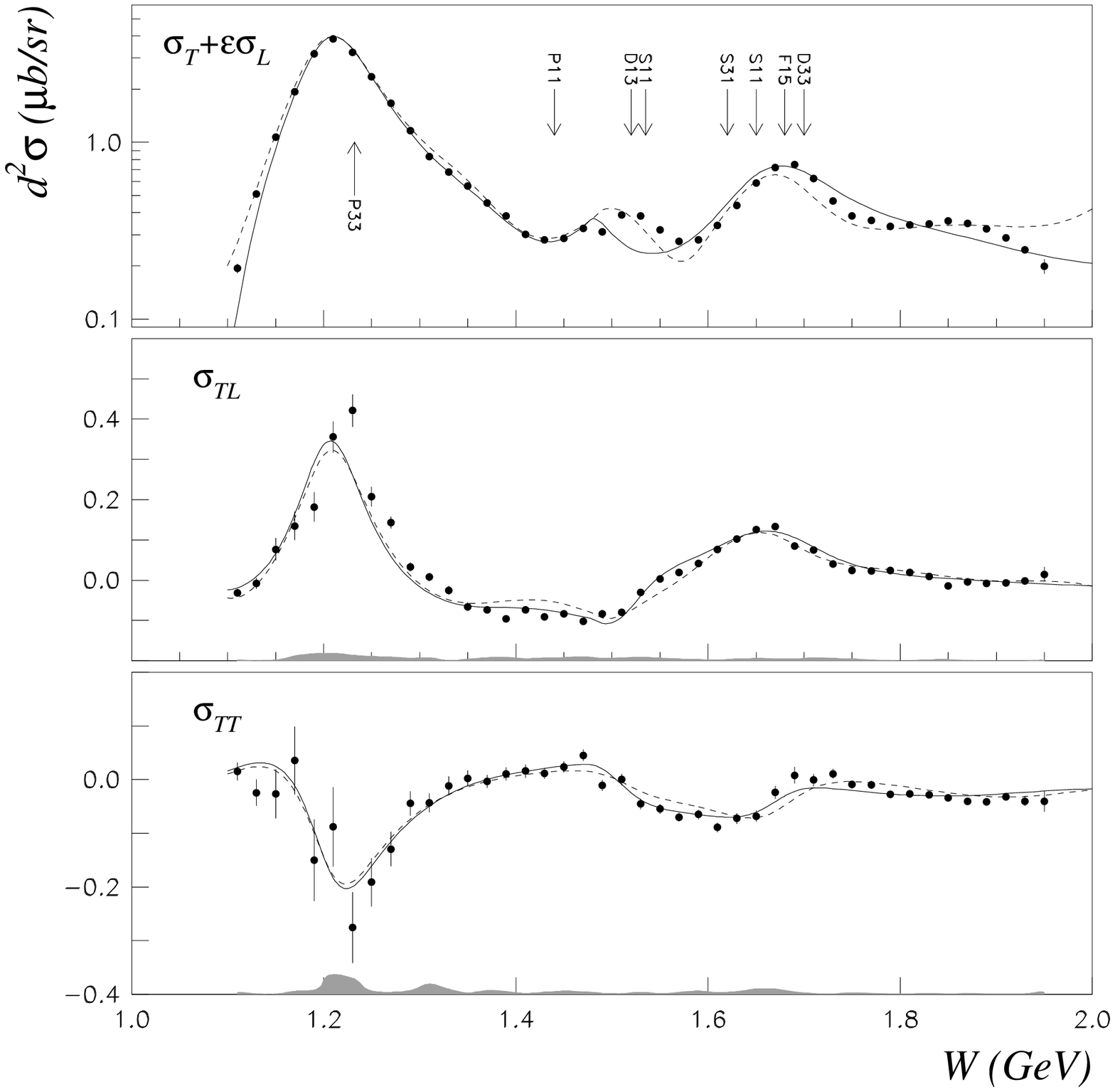}}
\caption{Unpolarized response functions for $p\pi^{\circ}$ at $\theta^{cm}_{\pi^{\circ}}=170^{\circ}$ 
from JLab Hall A. The solid line 
corresponds to a fit to the data using the MAID03 implementation. 
The dashed line corresponds to the SAID solution.}
\label{fig:halla_pi0_rf}
\end{figure}

As can be seen in Table~\ref{tab:pion_data}, pion electroproduction data
are now very extensive and of high quality. 
In  Figs.~\ref{fig:jlab_pi0_1},~\ref{fig:clas_electro_npipl}, and \ref{fig:rltp_pip_c1_q1},
we show some sample data from CLAS at JLab.
As an example for a spectrum with high statistics data on $\pi^{\circ}$ electroproduction
at a fixed backward angle of $\theta^{cm}_{\pi^{\circ}}=170^{\circ}$ we show
in Fig.~\ref{fig:halla_pi0_rf} response functions recently obtained from
JLab Hall A~\cite{laveissiere03}.

In Fig.~\ref{fig:jlab_pi0_1} and Fig.~\ref{fig:rltp_pip_c1_q1},
the predictions from the SL, MAID, and DMT models
are also displayed to illustrate the status of current reaction 
models.
The analyses of these new data in the past few years have led
to rather accurate determinations of the
 $\gamma N \rightarrow \Delta$ transition form factors.
We now discuss this advance in more detail.

\subsubsection{The $\gamma N\rightarrow \Delta(1232)$ transition form factors}

With the fairly extensive coverage over angles and energies, the data
 from JLab have allowed
nearly model-independent determinations of $\gamma N\rightarrow \Delta(1232)$ 
form factors.  Theses analyses by the CLAS collaboration are  based on the
following considerations. At the $\Delta$ peak, the dominant amplitude is
$M_{1+}$ and the small $E_{1+}$ and $S_{1+}$ can become
accessible through their interference with the dominant $M_{1+}$ amplitude. 
One thus can start the analysis by
using a truncation, in which only terms involving $M_{1+}$ are retained. 
With the partial-wave decomposition defined by
 Eqs.(35)-(40), the differential cross section in Eq.(17) can then be
written as
\begin{eqnarray}
\frac{d\sigma}{d\Omega^*} = \sum_{\ell=0}^{2}A_\ell P_\ell(cos \theta^*)
+[\sum_{\ell=1}^{2} B_\ell P^{\prime}_\ell(cos \theta^*)] cos  \phi_M 
+[C_2 P^{\prime}_2(cos \theta^*)] cos 2 \phi_M
\label{equ:diff_pwa}
\end{eqnarray}
The coefficients of the above equation are related to
$|M_{1+}|^2$ and its projection onto the other s- and p-wave multipoles
$E_{1+}$, $S_{1+}$, $M_{1-}$, $E_{0+}$, $S_{0+}$ :
\begin{eqnarray}
|M_{1+}|^2 &=& A_0/2,  \\
Re(E_{1+}M^*_{1+}) &=& (A_2 -2C_2/3)/8,  \\
Re(M_{1-}M^*_{1+}) &=& -[A_2 + 2(A_0 + C_2/3)]/8,  \\
Re(E_{0+}M^*_{1+}) &=& A_1/2, \\
Re(S_{0+}M^*_{1+}) &=& B_1,  \\
Re(S_{1+}M^*_{1+}) &=& B_2/6.
\end{eqnarray}
The partial wave coefficents of Eq.(110) are determined by 
fitting the differential cross
sections data such as those displayed in 
Figs.~\ref{fig:jlab_pi0_1} -~\ref{fig:rltp_pip_c1_q1}.
From the relations Eqs.(111)-(116), one then obtains the $M_{1+}$, $E_{1+}$ and
$S_{1+}$ amplitudes for determining the $\gamma N \rightarrow \Delta$ 
form factors through Eqs.(99)-(101).

The results from using the above procedure must be corrected for the systematic errors due to
the truncation of higher multipoles. This can be accomplished by calculating the
effects of higher partial waves using a realistic
parametrization of the higher mass resonances and a realistic model for the
background amplitudes. 
For not too large $Q^2$ values, this method results in
reliable multipoles. 

\begin{figure}[hbtp]
\vspace{80mm}
\centering{\includegraphics{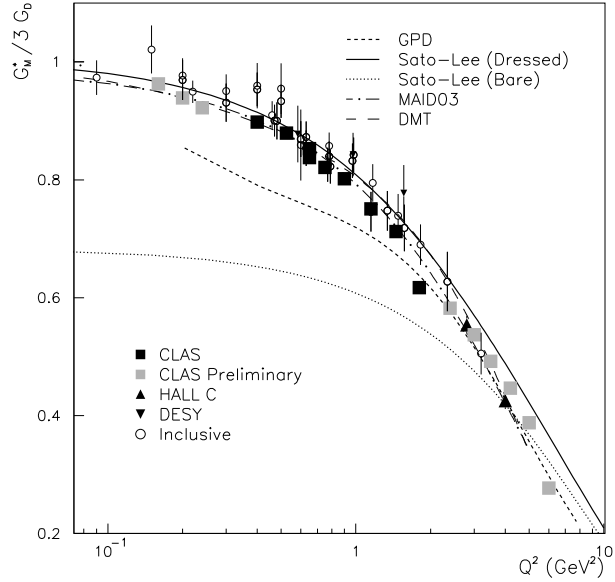}}
\caption{\small The data of magnetic form factor $G^*_M$ for 
the $\gamma N \rightarrow \Delta(1232)$ transition are
compared to various models. Results from old single arm electron scattering experiments are
labeled ``inclusive''. All other results have been obtained from a multipole expansion
of exclusive $\pi^{\circ}$ production from protons.}
\label{fig:gm_Delta}
\end{figure}

\begin{figure}[thbp]
\vspace{90mm}
\centering{\includegraphics{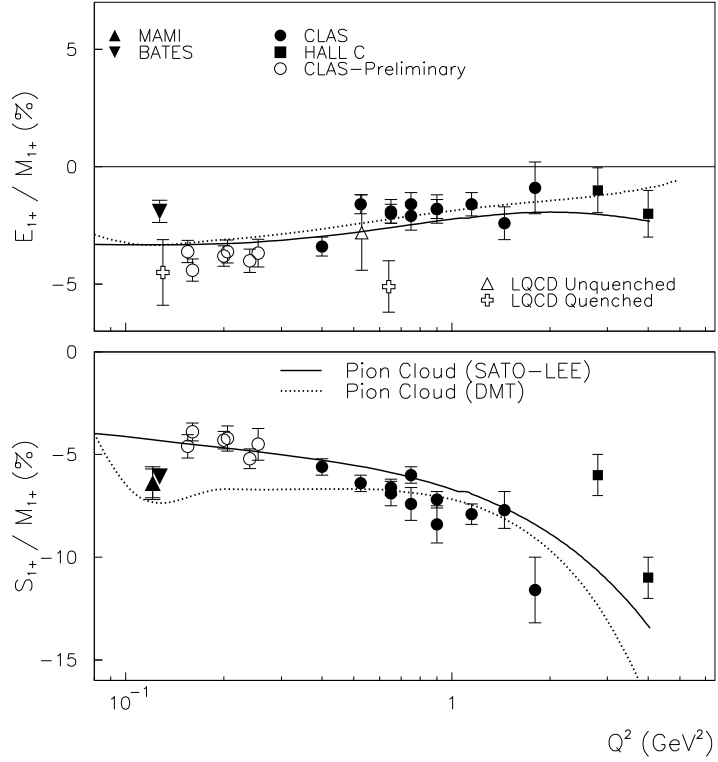}}
\caption{\small Ratios $R_{EM}$(denoted as $E_{1+}/M_{1+})$ 
and $R_{SM}$(denoted as $S_{1+}/M_{1+})$ for the 
$\gamma N \rightarrow \Delta(1232)$ transition. 
These two ratios are related to the 
$E^{3/2}_{1+}$, $S^{3/2}_{1+}$, and $M^{3/2}_{1+}$ multipole
amplituds of $\gamma^* N \rightarrow \pi N$, as
defined in Eqs.(102)-(103).
Preliminary data from CLAS at low $Q^2$ are also included.}
\label{fig:mrat_full}
\end{figure}

With the above largely model-independent procedure, results 
for $G^*_M$  up to
$Q^2=$ 6 GeV$^2$ and the ratios
$R_{EM}$ and $R_{SM}$, defined in Eqs.(102) and (103), up to
$Q^2 = 4$~GeV$^2$ have been obtained at JLab and are compared with various
 theoretical predictions in Figs.~\ref{fig:gm_Delta} and  
 ~\ref{fig:mrat_full}.
We now explain how the displayed results from SL, MAID, and DMT
models are obtained.
Within the MAID model, 
the $Q^2$-dependence of the $\gamma N \rightarrow N^*$ transition strengths
$\bar{A}^\alpha$ of Walker's parameterization Eq.(64)
is determined from fitting the differential cross section
data. The resulting multipole amplitudes are then used to
extract the $\gamma N \rightarrow \Delta$ form factors by using Eqs.(99)-(101).
On the other hand, within the  SL and
DMT dynamical models, the parameters of the bare quantities 
$G_M(Q^2)$, $G_E(Q^2)$ and $G_C(Q^2)$, defined by Eq.(93),
are adjusted to fit the data.
The dressed form factors of the SL model
 are then predicted by using Eq.(84) 
to calculate the meson cloud effect.
As shown in Ref.\cite{sl1}, at the $\Delta$ mass $W=1.232$ GeV  
this procedure is 
equivalent to that based on Eqs.(99)-(101). The parameters of these three
models have been determined by using the data up to $Q=4$ GeV$^2$.
The results at $Q^2 > 4$ GeV$^2$ are their predictions.

In  Fig.~\ref{fig:gm_Delta},
we see that  
the theoretical predictions of $G^*_M(Q^2)$
at $Q^2 >$ 4 (GeV/c)$^2$ from SL , MAID,  and DMT models agree
well with the new  data from Jlab. The prediction by Stoler\cite{stoler03}, 
which is based on a PQCD-motivated model,
is also displayed there for comparison. It also agrees well with the data at 
relatively high $Q^2$. 

The dotted curve in Fig.~\ref{fig:gm_Delta} is obtained from setting the
pion cloud effect, defined by Eq.(84) and illutrated in 
Fig.~\ref{fig:dressgm},  to zero within the Sato-Lee model.
We see  that the pion cloud effect
is very large at low $Q^2$, but becomes
much smaller at high $Q^2$.
Clearly this $Q^2$ dependence plays an important
role in getting the agreement with the data up to $Q^2$=6 GeV$^2$. It will be
interesting to see whether the predicted pion cloud effect will agree with
the data at even higher $Q^2$. 

\begin{figure}[thbp]
\vspace{30mm}
\centering{\includegraphics{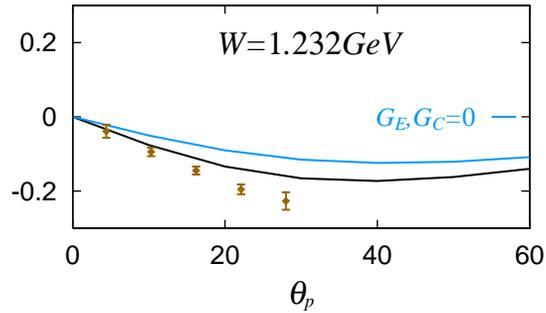}}
\caption{ The data of $A_{LT}$ from MIT-Bates are compared with the results 
from Sato-Lee model. The dashed curve is obtained from setting the 
$G_C$ and $G_E$ of the $\gamma N \rightarrow \Delta$ form factor to zero.  }
\label{fig:bates}
\end{figure}
\begin{figure}[hpbt]
\vspace{40mm}
\includegraphics{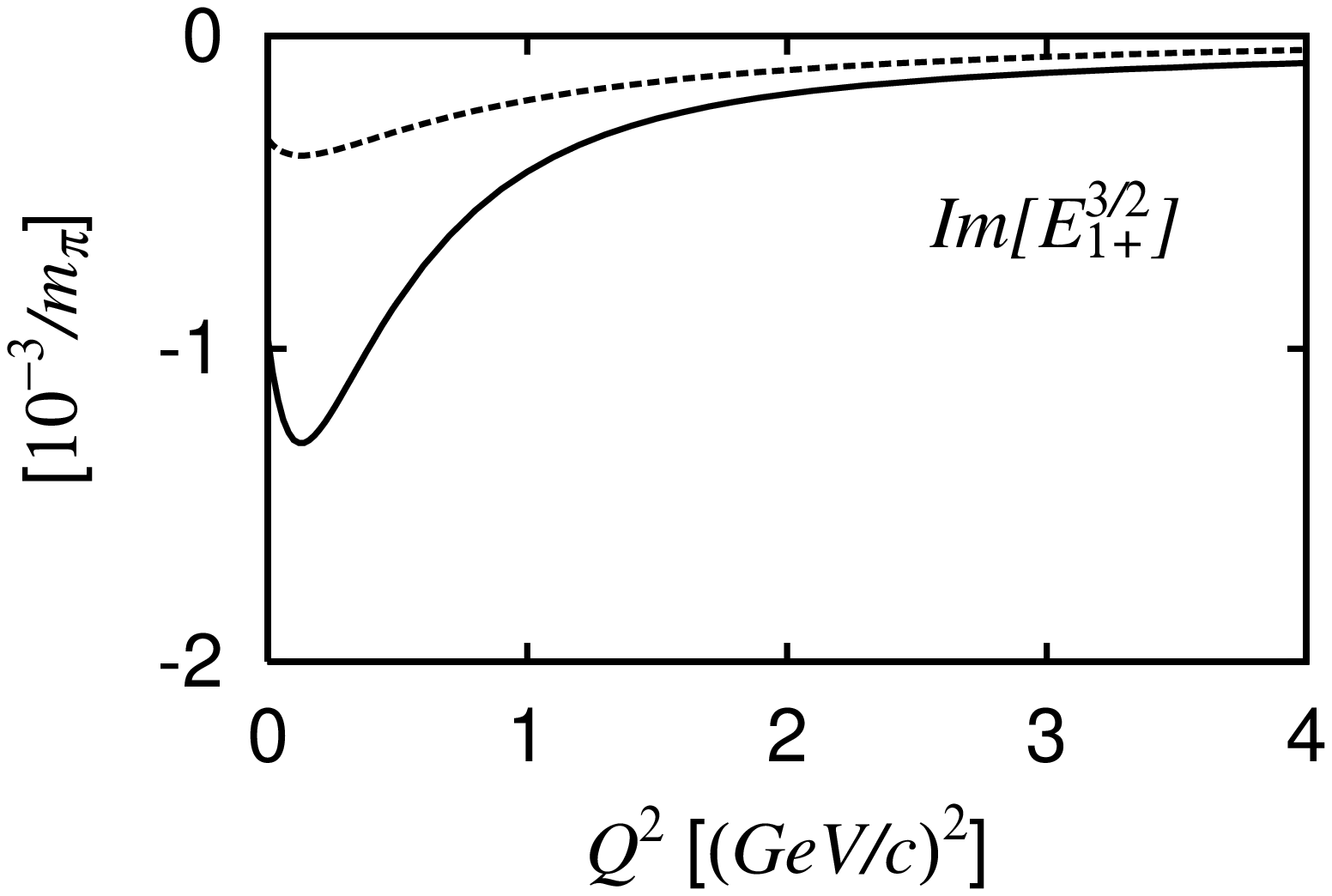}
\includegraphics{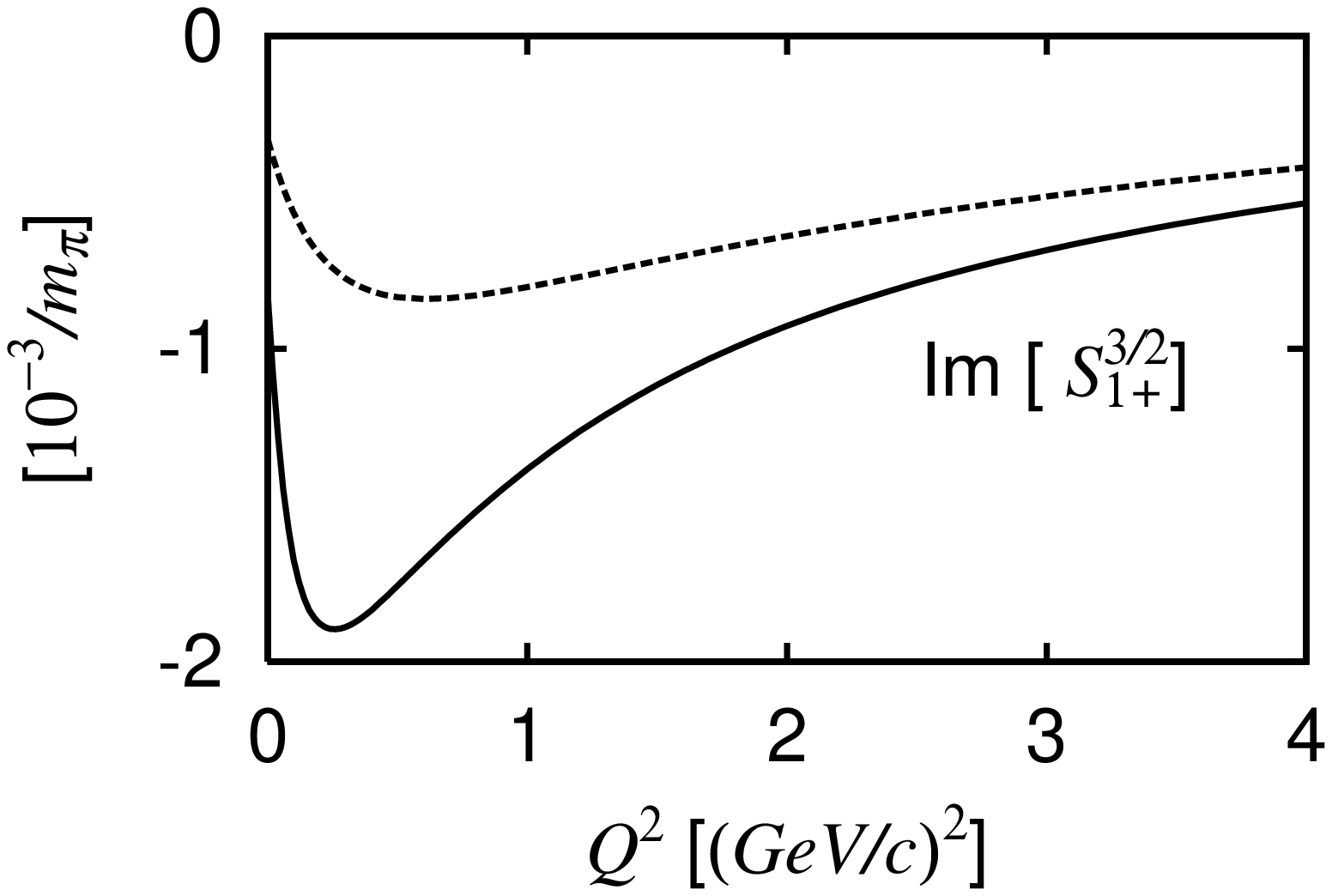}
\caption{The imaginary parts ($Im$) of the $E_{1+}^{3/2}$ and 
$S_{1+}^{3/2}$ calculated from the Sato-Lee model. The dotted curves are 
obtained from 
setting the pion cloud effect to zero.}
\label{fig:e1s1}
\end{figure}

In the upper part of Fig.~\ref{fig:mrat_full}, we see that
the preliminary CLAS data for $R_{EM}$
at low $Q^2 < 0.2 $ (GeV)$^2$~\cite{lcsmith04} 
are in good agreement with the predictions from the SL and DMT
 models. On the other hand, the new Jlab data for the  
ratio $R_{SM}$ (lower part of Fig.~\ref{fig:mrat_full}) in the
low $Q^2 < 0.2 $ GeV$^2$ region prefers the prediction from
the SL model.
The data points at $Q^2 = 0.127$~GeV$^2$ from MAMI and Bates have a larger 
magnitude for $R_{SM}$. These data points were used by DMT in fixing their
parameterization for $G_C(Q^2)$.
It should be noted that the points from MAMI and Bates are not the result
of an independent multipole fit, but are from data sets with more limited 
angle coverage fitted to the MAID parametrization. 
One of the data sets from MIT-Bates is shown in Fig.~\ref{fig:bates}.
Clearly, only very limited angles are covered.
Nevertheless, these data are  very useful in revealing the
non-zero values of the 
$G_E$ and $G_C$ form factors within the dynamical model, 
as illustrated in the difference
between the solid and dotted curves.

We now note that the ratios $R_{EM}$ 
and $R_{SM}$  calculated from the dynamical models are 
very much related to the predicted pion cloud effects. These are illustrated in
Fig.~\ref{fig:e1s1} from the SL model. We see that the pion cloud effect can 
strongly enhence the the $E^{3/2}_{1+}$ and $S^{3/2}_{1+}$
 amplitudes of $\gamma^* N \rightarrow \pi N$ at low $Q^2$.
As defined in Eqs.(99)-(101), these two amplitudes are related to the
$G_E(Q^2)$ and $G_C(Q^2)$ of $\gamma N \rightarrow \Delta$ transition.
The non-trivial pion cloud effects shown  in Fig.~\ref{fig:e1s1}
 are clearly verified by the JLab data,
as seen in Fig.~\ref{fig:mrat_full}.

\begin{figure}[thbp]
\vspace{50mm}
\centering{\includegraphics{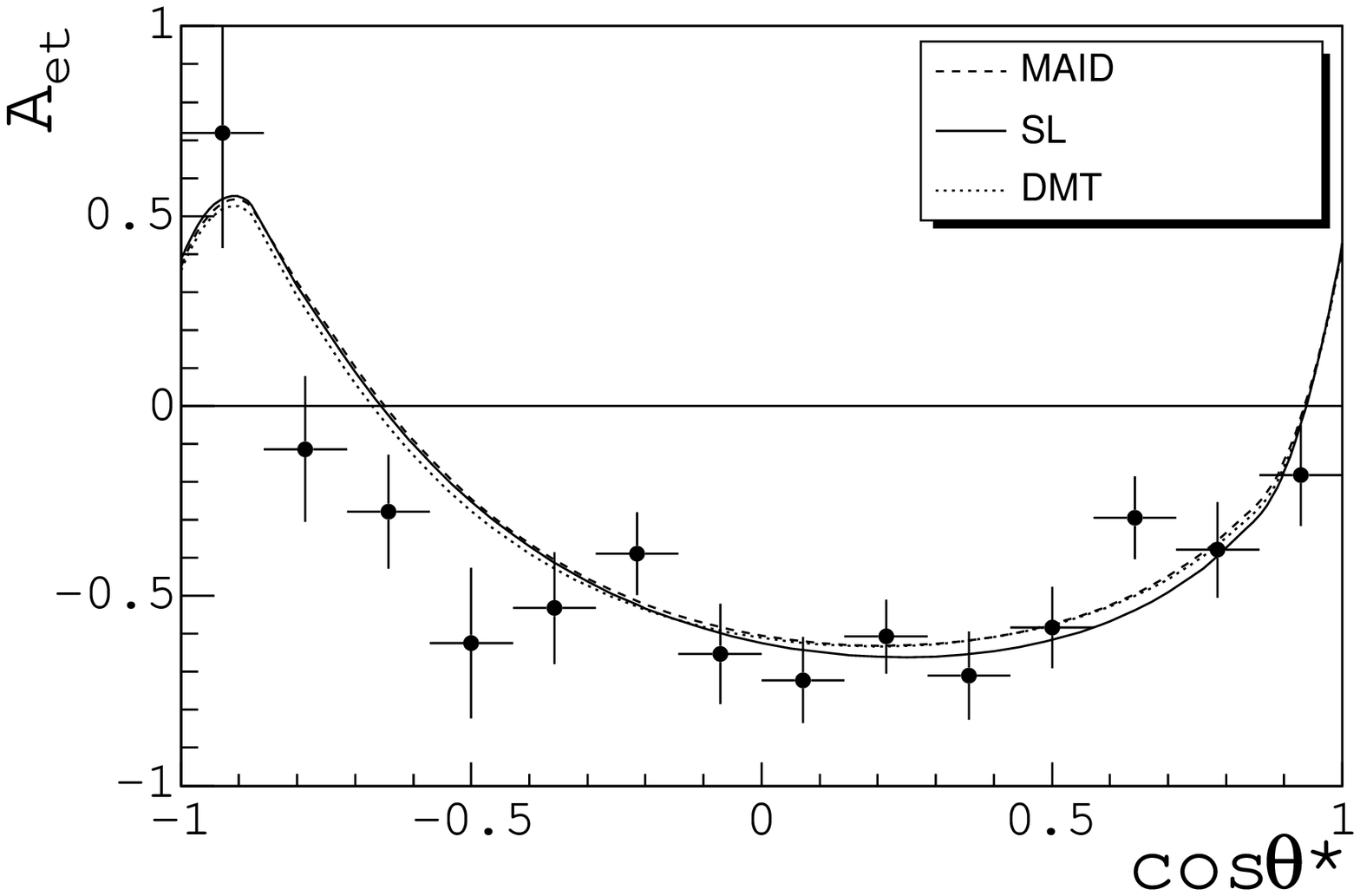}}
\centering{\includegraphics{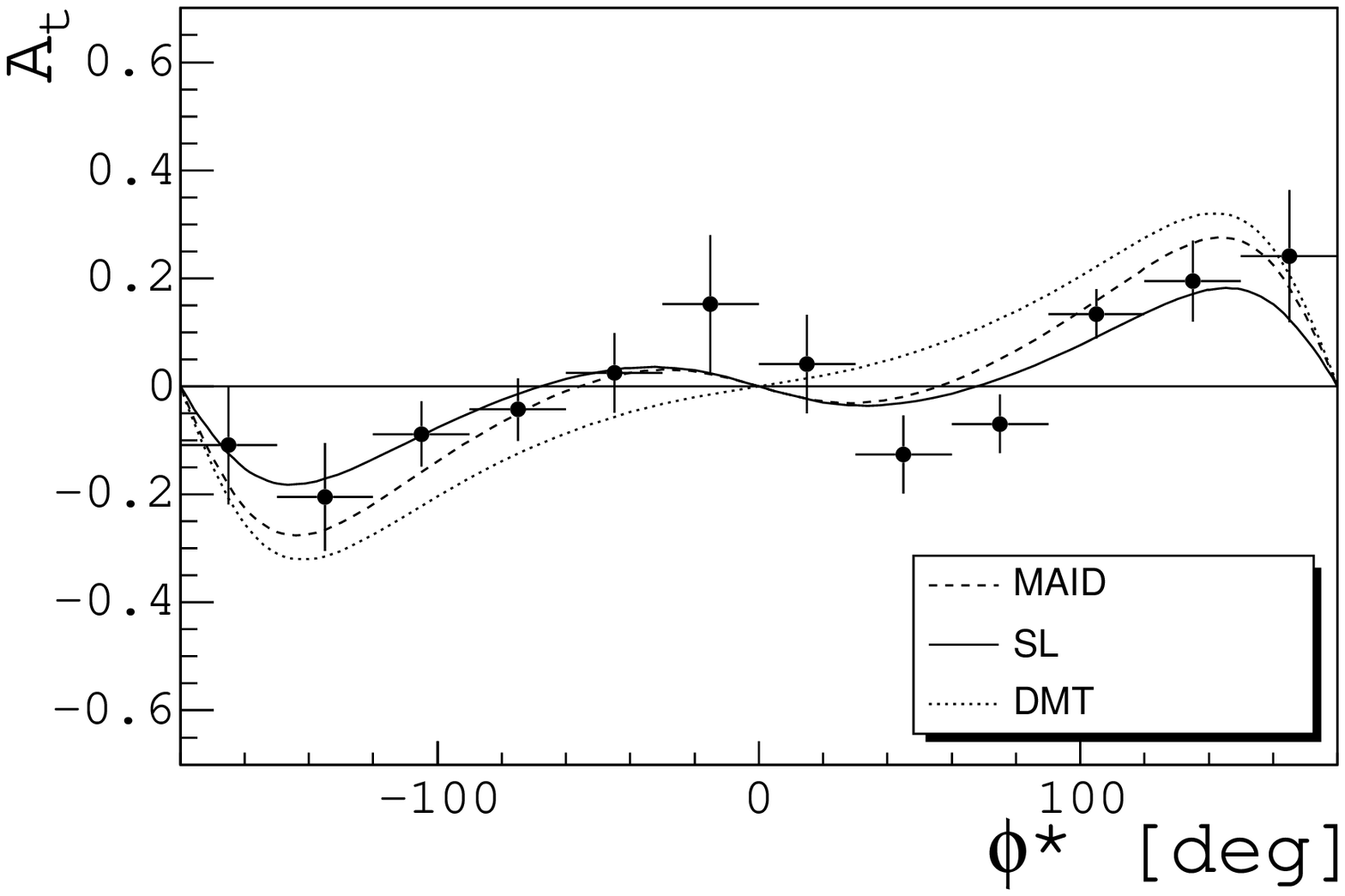}}
\caption{$A_{et}$ and $A_{t}$ from JLab at W=1.225 GeV, $Q^2 = 0.46 GeV^2$
are compared with the predictions from SL, MAID, and
 DMT models. The left panel is integrated in a range $\phi = (144^{\circ}, 160^{\circ})$.
The right panel is integrated in a range $\cos{\theta^*} = (-0.6, -0.8)$.}
\label{fig:clas_beam_target_asy}
\end{figure}

To further improve the determination of the $\gamma N \rightarrow
\Delta$ form factors, data of polarization observables must be 
included in the theoretical analyses. 
Measurements  using polarized electron beam and/or a
longitudinally polarized hydrogen target~\cite{biselli04}
have yielded data of
double spin beam-target asymmetry $A_{et}$ and 
target asymmetry $A_t$. Samples of asymmetry data from CLAS are
shown in Fig.~\ref{fig:clas_beam_target_asy}.
The double polarization asymmetry $A_{et}$ is largely given by the 
well determined  $M_{1+}$
multipole and is well described by all models. 
However, significant differences can be seen in
the $A_t$ asymmetry which is sensitive to interferences
 between the non-resonant and resonant amplitudes. The discrepancy in the model descriptions 
can be attributed to their different treatments of the non-resonant 
amplitudes, as discussed in section 4.

\begin{figure}[hbtp]
\vspace{8.5cm}
\includegraphics{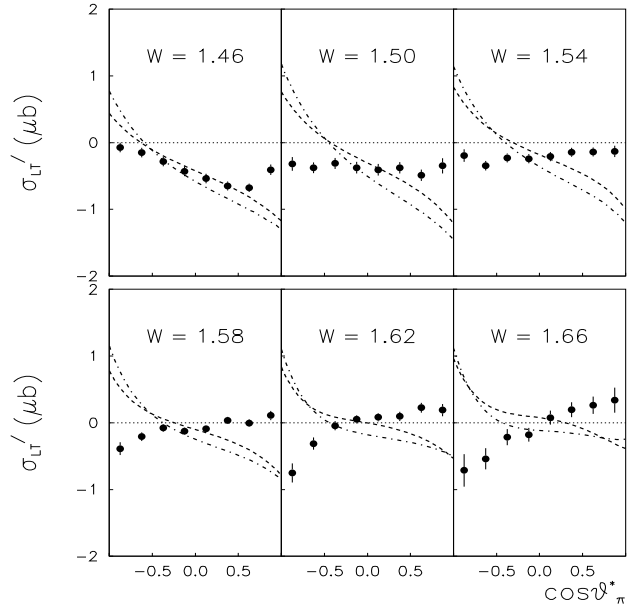}
\caption{\small{The data of $\sigma_{LT^{\prime}}$ for 
 $\gamma^* p \rightarrow p\pi^{\circ}$ 
in the $2^{nd}$ resonance region are compared with 
MAID (dashed) and the DMT (dashed-dotted) predictions.}}     
\label{fig:sigmaltp_pi0_c3}
\end{figure}

Extensive pion electroproduction data in the second and third resonance regions 
have also been obtained using CLAS.
Some typical results are shown in Figs.~\ref{fig:sigmaltp_pi0_c3}.
Here we see that the displayed theoretical predictions 
do not agree well with the data
at $W > 1.4$ GeV. This is not surprising since the parameters
of these single-channel models are fixed by mainly fitting the data at
$W < 1.4 $ GeV.
Recently, fits to these higher $W$ data have been achieved by
using a single-channel K-matrix model and fixed-t dispersion relations. In these
analyses $p\pi^{\circ}$ and $n\pi^+$ data are fitted simultaneously using 
unpolarized cross section data as well as beam spin response function results.
It has been found that these very different approaches 
give consistent results, e.g. in the analyses of 
Aznauryan et al.~\cite{azn2}. 
It indicates that the model-dependence  may be relatively small. 
Nevertheless, the extracted resonance parameters must be taken with
caution before a rigorous investigation of 
the coupled-channel effect has been carried out. Progress in this direction
is being made~\cite{lms}. The results of these fits are discussed in section 5.3. 
\newpage
\subsection{Photoproduction and electroproduction of $\eta$ mesons.}
\label{sect:eta_production}

\begin{figure}[tbh]
\vspace{80mm}
\centering{\includegraphics{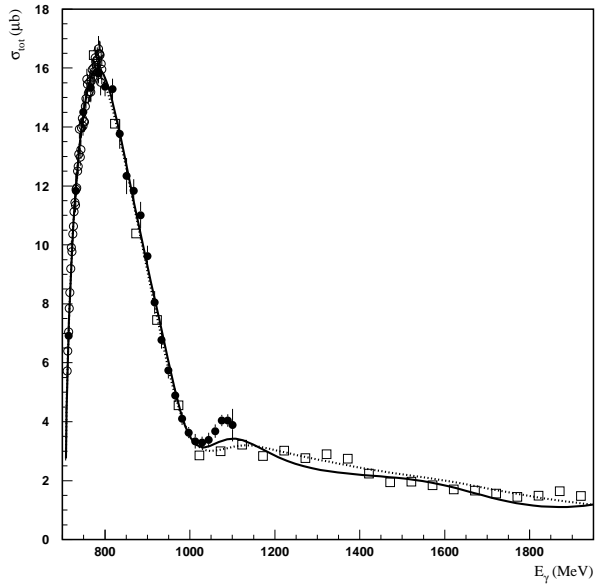}}
\caption{\small Total cross section for $\eta$ photoproduction from protons. Data from MAMI (open circles),
GRAAL (full circles), and CLAS (open squares) are shown. 
The curves are the fits explained in section 5.2.4.}
\label{fig:eta_photo_tot}
\end{figure}

In contrast to the pion with isospin I = 1, the eta is an isoscalar meson with
no charged partners. As such it can only couple with nucleons to form I = 1/2 
resonances.
This makes the production of $\eta$'s from nucleon targets an ideal tool to 
separate isospin $1\over 2$ $N^*$ resonances from isospin $3\over 2$ $\Delta^*$ resonances.
The total photoproduction cross section, shown in 
Fig. \ref{fig:eta_photo_tot}, exhibits a rapid rise just above 
threshold, indicative of a strong s-wave contribution near threshold. 
This behavior is known to be due to the first negative parity nucleon resonance,
the $S_{11}(1535)N_{1/2^-}$, which couples with approximately 55\% to the 
$N\eta$ channel~\cite{armstrong99}. The nearby $D_{13}(1520)$ has a 
branching ratio of much less than 1\% to this channel\footnote{Even though the 
$D_{13}(1520)$ coupling to $N\eta$ is very small, its close proximity to the
$S_{11}(1535)$ causes large interferences with the dominant $E_{0+}$ transition 
amplitude of the $S_{11}$. This in turn allows a precise determination of the 
$D_{13}(1520) \rightarrow N\eta$ branching ratio.}. The next higher mass nucleon 
resonance with a significant $N\eta$ coupling is the $P_{11}(1710)$, nearly 200~MeV/c$^2$
higher in mass. This fact makes the production of $\eta$'s from nucleon targets the reaction
of choice for detailed studies of the electromagnetic transition from the ground state to 
the $S_{11}(1535)$. The $N\eta$ channel effectively isolates this state from other nearby
resonances, similar to the $\Delta(1232)$ which is well separated from higher 
mass resonances 
in the $p\pi^{\circ}$ channel. In distinction to the $\Delta(1232)$, whose 
electromagnetic transition form fcators drop rapidly with increasing photon virtuality $Q^2$, the $S_{11}(1535)$ 
remains a prominent resonance even at the highest $Q^2$ that are currently accessible. 

In the following subsections we discuss the status of the electromagnetic production of
$\eta$'s from nucleons, and analyses to extract the photocoupling helicity amplitudes for 
the $\gamma N\rightarrow S_{11}(1535)$ transition, and their $Q^2$ evolution. We finally compare the
results with model predictions.  
Table \ref{tab:eta_data} gives an overview of the kinematics covered in recent $\eta$ production 
measurements~\cite{dugger02,crede03,rebreyend02,kuznetsov04,ahrens03,krusche95,thompson01,renard02,ajaka98,denizli04}.

\begin{table}[tbh]
\tcaption{Summary of $\eta$ production data}
\centerline{\footnotesize\smalllineskip
\begin{tabular}{l l l c l}\\
\hline\\
{Reaction} &{Observable} &W range &$Q^2$ range &Lab. \\
{} &\phantom0 &(GeV) & (GeV$^2$) & \phantom0 \\ 
\hline\\
{$\gamma p \rightarrow p\eta$} & $d\sigma / d\Omega $ & $< 2.0$  & &JLab-CLAS\\
				 & $d\sigma / d\Omega$ &$< 1.7$  & & GRAAL\\
				& $d\sigma /d\Omega$ & $< 2.3$  &  &ELSA-CB\\
				& $d\sigma / d\Omega$ &$< 1.53$  &  &MAMI-TAPS\\
{$\gamma (n/p) \rightarrow (n/p) \eta$} & $d\sigma_n /d\sigma_p$ &$< 2.3$  &  &GRAAL\\
{$\vec{\gamma} p \rightarrow p\eta$} & $\Sigma$ &$< 2.3$  & & GRAAL\\
{$\gamma \vec{p} \rightarrow p \eta$} & $T$  & $< 2.3$  & & ELSA\\
{$\vec{\gamma} \vec{p} \rightarrow p\eta$} & $E$ & $< 1.53$  & &MAMI-A2\\
$ep \rightarrow ep\eta$ & $\sigma_{LT}$, $\sigma_{TT}$, $\sigma_T+\epsilon\sigma_L$ &$< 2.2$  &2.8 - 4.0 &JLab-Hall-C\\
& $\sigma_{LT}$, $\sigma_{TT}$, $\sigma_T+\epsilon\sigma_L$ &$< 2.2$  &0.3 - 4.0 &JLab-CLAS\\
\hline\\
\end{tabular}}
\label{tab:eta_data}
\end{table}
\newpage
\subsubsection{$\eta$ photoproduction from protons}

\begin{figure}[tbh]
\vspace{100mm} 
\centering{\includegraphics{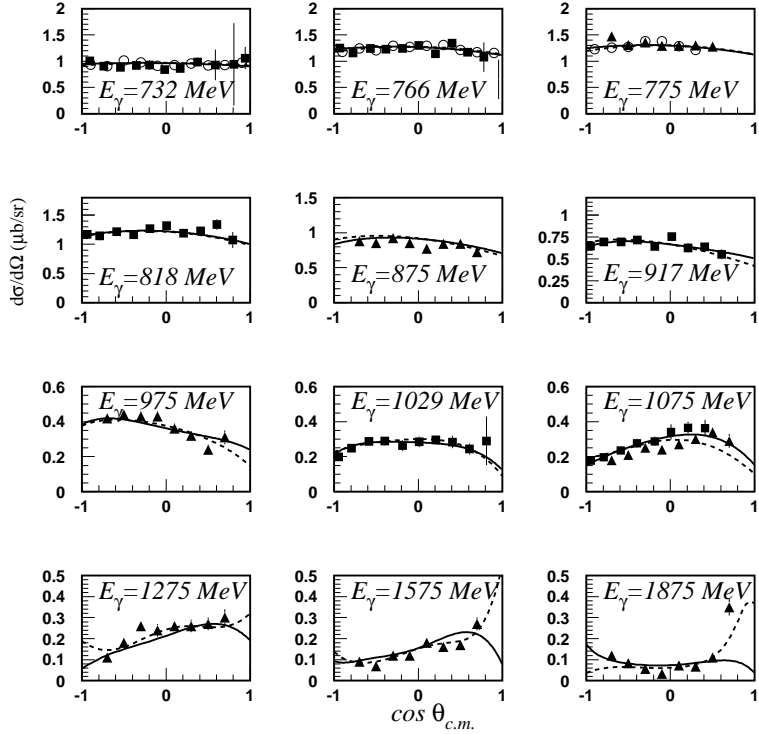}}
\caption{\small Differential cross section for $\gamma p \rightarrow p \eta$ 
from TAPS, GRAAL, and CLAS are shown for photon energies 
near threshold to 1.875 GeV. The curves are the fits explained in
 section 5.2.4. The solid (dotted) line represents the UIM (DR) analysis. }
\label{fig:all_eta_photo_cs}
\end{figure}

With the new measurements in recent years, 
the data base for $\eta$ photoproduction reaction
has been improved tremendously. 
The differential cross section data 
now cover the $p\eta$ mass range   
up to W = 2.3 GeV, and are available for
most of the angular range 
in the hadronic center-of-mass system. Some of these data are shown
Figure \ref{fig:all_eta_photo_cs}.  
The data from all three experiments
agree well.
In the mass region of the $S_{11}(1535)$ 
resonance the angular distributions are nearly flat, indicating dominant s-wave components 
with only slight indications of higher partial wave contributions. 
In the mass region above 1.750 GeV, the angular distributions become increasingly 
forward-peaked, indicating significant non-resonant behavior presumably due to t-channel processes.

\begin{figure}[t]
\vspace{70mm} 
\centering{\includegraphics{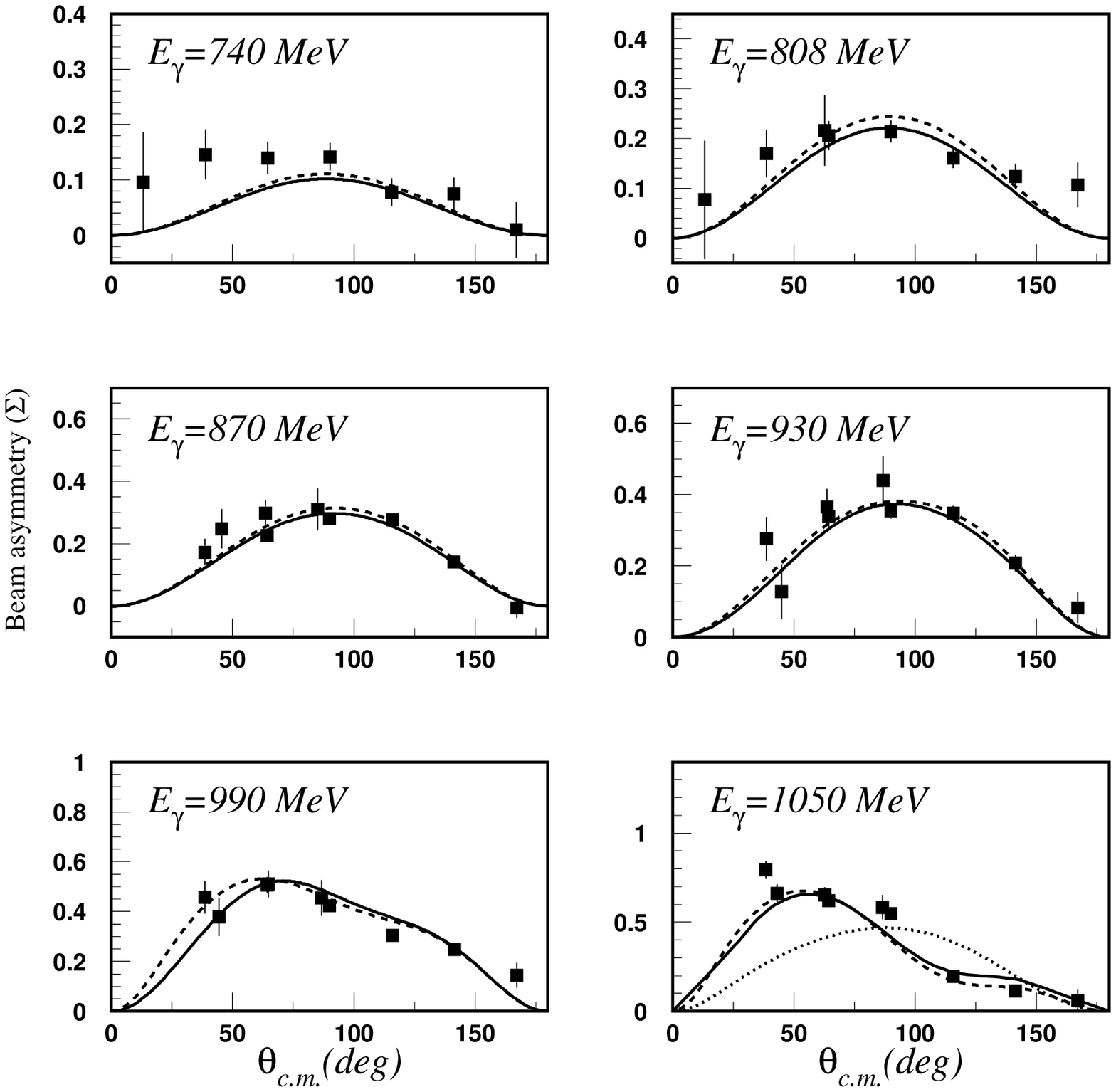}}
\centering{\includegraphics{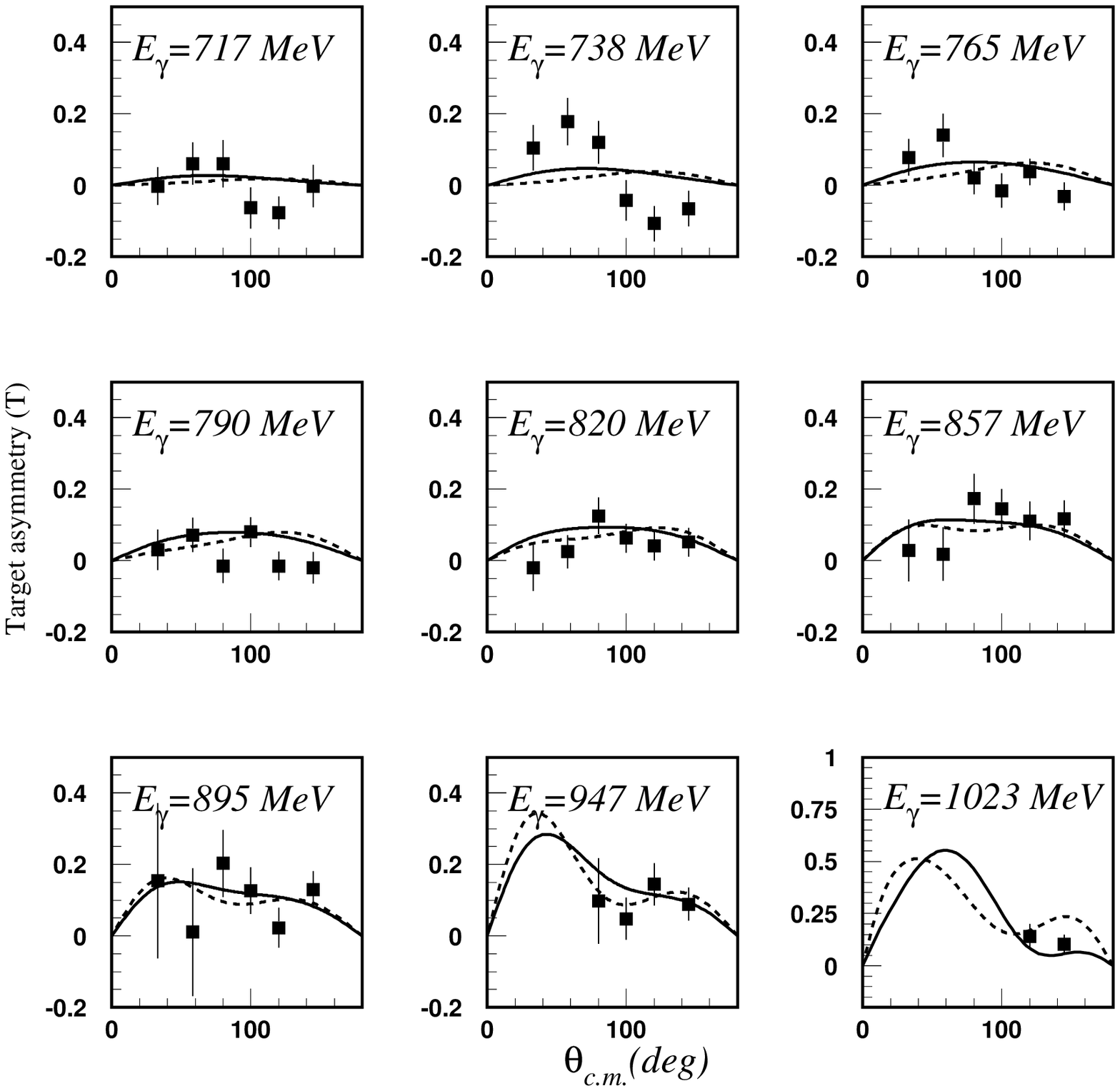}}
\caption{\small  The fits to the 
$\gamma p \rightarrow p \eta$ data using the Unitary Isobar Model (solid curves)
 and Dispersion Relation (short dashed curves)approaches. 
Left panel: Beam asymmetry measured at GRAAL. The dotted curve in the 1050 MeV panel
shows the fit without the $F_{15}(1680)$ resonance.  
Right panel: Target asymmetry measured at ELSA.  } 
\label{fig:asymmetry_eta}

\end{figure}

GRAAL has measured the beam asymmetries using laser light backscattered from the 6 GeV 
electrons to generate high-energy linearly polarized photons. The 
beam asymmetry $\Sigma_{\gamma}$ is
defined as 
\begin{eqnarray}
\Sigma_{\gamma} = {1\over P_{\gamma}} {{{d\sigma \over d\Omega} (\phi=0^{\circ}) - {d\sigma \over d\Omega}(\phi=90^{\circ})} \over
{{d\sigma \over d\Omega} (\phi=0^{\circ}) + {d\sigma \over d\Omega}(\phi=90^{\circ})}}
\end{eqnarray}
where $P_{\gamma}$ is the photon polarization, $\phi$ is the azimuthal angle between the 
plane defined by the linear photon polarization and the hadronic plane defined by the photon beam and 
the $p\eta$ final state. The measured beam asymmetries are shown in Fig. \ref{fig:asymmetry_eta}. 
Just above $\eta$ threshold and at the resonance position $\Sigma_{\gamma}$ shows a symmetric angular distribution, 
approximately following a $\sin^2\theta^*$ behavior, while at higher energies
 the asymmetry is 
more forward peaked.  The $\sin^2\theta^*_{\eta}$ behavior 
near the resonance pole at the lower energies is prominent in the data, and is also reflected in the 
model descriptions included in Fig. \ref{fig:asymmetry_eta}.

Asymmetries have also been measured with a transversely polarized proton target
at ELSA. The target asymmetry $T$ is given by:
\begin{eqnarray}
T = {1 \over P_T}{{{d\sigma \over d\Omega}(\uparrow) - {d\sigma \over d\Omega}(\downarrow)} \over  
{{d\sigma \over d\Omega}(\uparrow) + {d\sigma \over d\Omega}(\downarrow)}}.
\end{eqnarray}
The arrows indicate the direction of the proton polarization relative to the hadronic plane. 
Results are shown in the right panel in Fig. \ref{fig:asymmetry_eta}. 

\subsubsection{$\gamma p\rightarrow S_{11}(1535)$ parameters extracted from 
gobal fit to $\eta$
photoproduction} 

The $S_{11}(1535)$ has long been known as a strong nucleon resonance with a large branching 
ratio to the $N\eta$ and $N\pi$ channels. However, there have been assertions that the strong
enhancement near this mass is not due to the excitation of a resonance~\cite{hoehler98}. One model implies 
that the state is dominantly a dynamically generated $\Sigma\bar{K}$ resonance \cite{weise95}. 
On the other hand, recent 
Lattice QCD calculations \cite{dong03,lee_nstar02} show that there is a strong 3-quark state 
at this mass with the spin-parity  $J^P = {1\over 2}^-$, indicating that the $S_{11}(1535)$ 
is indeed an excited state of the nucleon.  

The photocoupling amplitudes and their $Q^2$ dependence are powerful tools in determining 
the internal resonance structure and will help solve this controversy. 
For the purpose of this article we consider the $S_{11}(1535)$ as a baryon resonance 
with well defined quantum numbers. 
In the following we discuss results obtained in a global analysis of all observables 
in $\eta$ photoproduction with the goal to extract the photocouplings amplitudes for resonances 
coupling to the $p\eta$ channel, especially the $S_{11}(1535)$.


A number of analyses have been performed on the $p\eta$ photoproduction 
channel~\cite{azn2,chiang03,saghai01}.
Here, we describe the recent global analysis by Aznauryan\cite{azn2} 
as it allows to
also assess the model-dependence of the results.
Differential cross sections~\cite{renard02,dugger02} were included  as well as  
polarized beam asymmetries~\cite{ajaka98},
and polarized target asymmetries~\cite{bock98}. 
All established $N^*$  resonances above the $N\eta$ threshold were included,  
i.e. $S_{11}(1535)$, 
$D_{13}(1520)$, $S_{11}(1650)$, $D_{15}(1675)$, $F_{15}(1680)$, 
$D_{13}(1700)$, $P_{11}(1710)$, $P_{13}(1720)$. 
The cross section data are fitted for photon energies up to 2 GeV, corresponding to 
invariant masses in the range W = 1.49 - 2.15 GeV, i.e. covering the entire 
resonance region. 
The polarization data cover only the range up to W = 1.7 GeV. 
Figure \ref{fig:all_eta_photo_cs} and Fig. \ref{fig:asymmetry_eta} show
samples of the fit to the cross sections and asymmetry data.
The Unitary Isobar Model (UIM) and Dispersion-relations (DR)
 approaches give consistent results for 
the $S_{11}(1535)$, $D_{13}(1520)$, and $F_{15}(1680)$ resonances. 
The first result is the confirmation of a large photocoupling amplitude 
for the $S_{11}(1535)$ which is determined with good precision. The results are 
summarized in table \ref{tab:eta_photo_s11}.
\begin{table}[htbp]
\tcaption{$S_{11}(1535)$ photocoupling from gobal fit in units ($10^{-3}~GeV^{-1/2}$).}
\centerline{\footnotesize\smalllineskip
\begin{tabular}{l l l l l l}\\
\hline
Resonance  &Mass (MeV) & $\Gamma$ (MeV) &$A^p_{1/2}$ & Model & \\ 
\hline\\
$S_{11}(1535)$ &1527  &142 &  96  & Isobar model& \\
  & 1542 &195 &  119 & Dispersion relations& \\
  &1520-1555 &100-200 &  60 - 120 & PDG estimate & \\
\hline\\
\end{tabular}}
\label{tab:eta_photo_s11}
\end{table}
They are compared with the range given by the PDG~\cite{pdg04}. The new results are 
within the upper part of the range given by the PDG.  
The lower range in the PDG value comes from an analysis of pion production data by the George Washington University (GWU) 
group\cite{gwu96}. We note here that the results from the
 global fits are also in good agreement 
with a combined analysis of $\pi$ and $\eta$ electroproduction data. 
We will discuss this in section 5.3.

From the fit to the differential cross sections one can then also extract the total 
photo absorption cross section for $\eta$ production. The fit results are compared with the 
experimental data in Fig. \ref{fig:eta_photo_tot}. All three  experiments
agree well in the region where the $S_{11}(1535)$ resonance dominates, while there is a 
discrepancy near 1100 MeV photon energy. 
Since the angular distributions agree well, this discrepancy must 
be entirely due to different models used for the extrapolation into the unmeasured angular 
regions. This emphasizes the importance of measureing complete angular 
distributions which are now available~\cite{crede03}.      

The global analysis incorporates also the beam asymmetry in the fit. 
To illustrate the sensitivity of $\Sigma$ to contributions from the $D_{13}(1520)$
we express $\Sigma_{\gamma}$ in the approximation that only S-waves, P-waves, and D-waves 
with spin $J \le {3 \over 2}$ contribute as \cite{chiang02}:
\begin{eqnarray}
\Sigma_{\gamma} \approx {3 \sin^2\theta Re[E^*_{0+}(E_{2-} + M_{2-})] \over |E_{0+}|^2} 
\end{eqnarray}

\begin{table}[thbp]
\tcaption{Summary of $\eta$ photoproduction gobal fit results. 
The uncertainty in $\beta_{\eta N}$ reflects the model dependence in
using the unitary isobar model and dispersion relation approach.}
\centerline{\footnotesize\smalllineskip
\begin{tabular}{l c c c c c}\\
\hline
Resonance & Mass(MeV)  &$\Gamma(MeV)$ &$\beta_{\eta N}$(\%) & $\beta_{\pi N}$(\%) & \\ 
\hline\\
$D_{13}(1520)$ & 1520 & 120  &  0.05 $\pm$ 0.02& 50 - 60& \\
$F_{15}(1680)$ & 1675 & 130  &  0.15 $\pm$ 0.03& 60 - 70& \\
\hline\\
\end{tabular}}
\label{tab:eta_photo_table1}
\end{table}

This expression can be fitted to the measured beam asymmetry $\Sigma_{\gamma}$. Using $E^*_{0+}$ 
from fits to the cross section data, the multipoles $E^{\eta}_{2-},~ M^{\eta}_{2-}$ for the 
$D_{13}(1520)$ can then be determined. 
Since the corresponding pion multipoles  $E^{\pi}_{2-},~ M^{\pi}_{2-}$ are known 
with high precision from pion
production, the branching ratio $\beta_{\eta N}$ can be extracted.
The analysis also allows to extract the $N\eta$ branching ratio for the $F_{15}(1680)$ by analyzing
the forward-backward asymmetry in $\Sigma_{\gamma}$ seen in Fig.~\ref{fig:asymmetry_eta} at $E_{\gamma} \approx  1$~GeV. The dotted curve 
in the figure for 1050 MeV (left panel) 
shows the fit when the small $F_{15}(1680)$ amplitudes are turned off. Clearly, the 
(
interference effects strongly enhance this contribution.   
The results for the $D_{13}(1520)$ and $F_{15}(1680)$ are summarized in 
table \ref{tab:eta_photo_table1}. Both results represent significantly 
improved values for the branching ratios.

%
\begin{figure}[htbp]
\vspace{75mm} 
\centering{\includegraphics{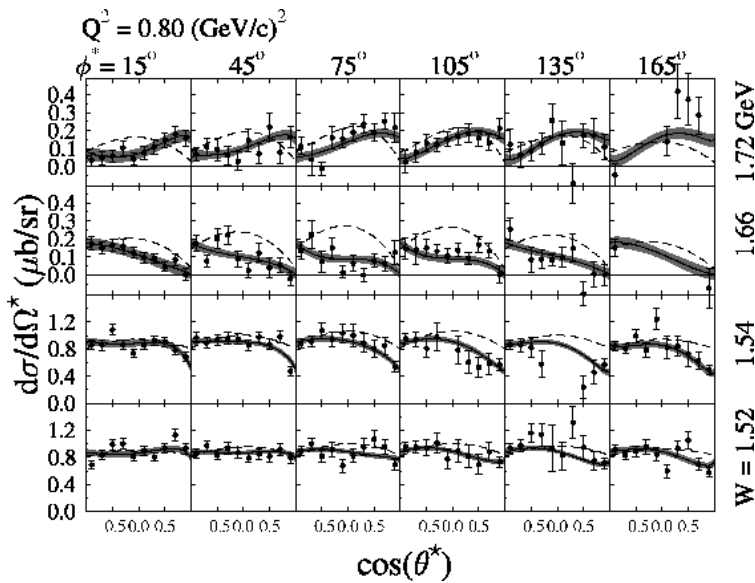}}
\caption{\small Samples of differential cross section for $\gamma^* p \rightarrow p \eta$ from CLAS
at fixed $Q^2=0.8$~GeV$^2$ and different W and $\phi$ values. The shaded bands indicate the fit results 
using 9 Legendre polynomials, as explained in section 5.2.5. The dashed curve represents the 
predictions of $\eta$MAID.} 
\label{fig:clas_eta_electro1}
\end{figure}

\subsubsection{Eta electroproduction}

Eta electroproduction experiments have focussed on the 
$Q^2$ evolution of the $S_{11}(1535)$ transverse photocouplings amplitude 
$A_{1/2}(Q^2)$. Experiments at  DESY~\cite{desy_eta1,desy_eta2} and
Bonn~\cite{bonn_eta_1,bonn_eta2}  found a very slow falloff with $Q^2$. 
Recent experiments at Jefferson Lab~\cite{armstrong99,thompson01,denizli04} have studied this behavior 
in detail with high statistics, and also extended 
the kinematics range. Figure \ref{fig:clas_eta_electro1}
shows samples of differential cross sections measured with CLAS~\cite{denizli04}. Even at the peak of 
the $S_{11}(1535)$ resonance the angular distributions are not completely 
flat indicating that higher partial waves are present in addition to the dominant S-wave.
\begin{figure}[tbph]
\vspace{75mm} 
\centering{\includegraphics{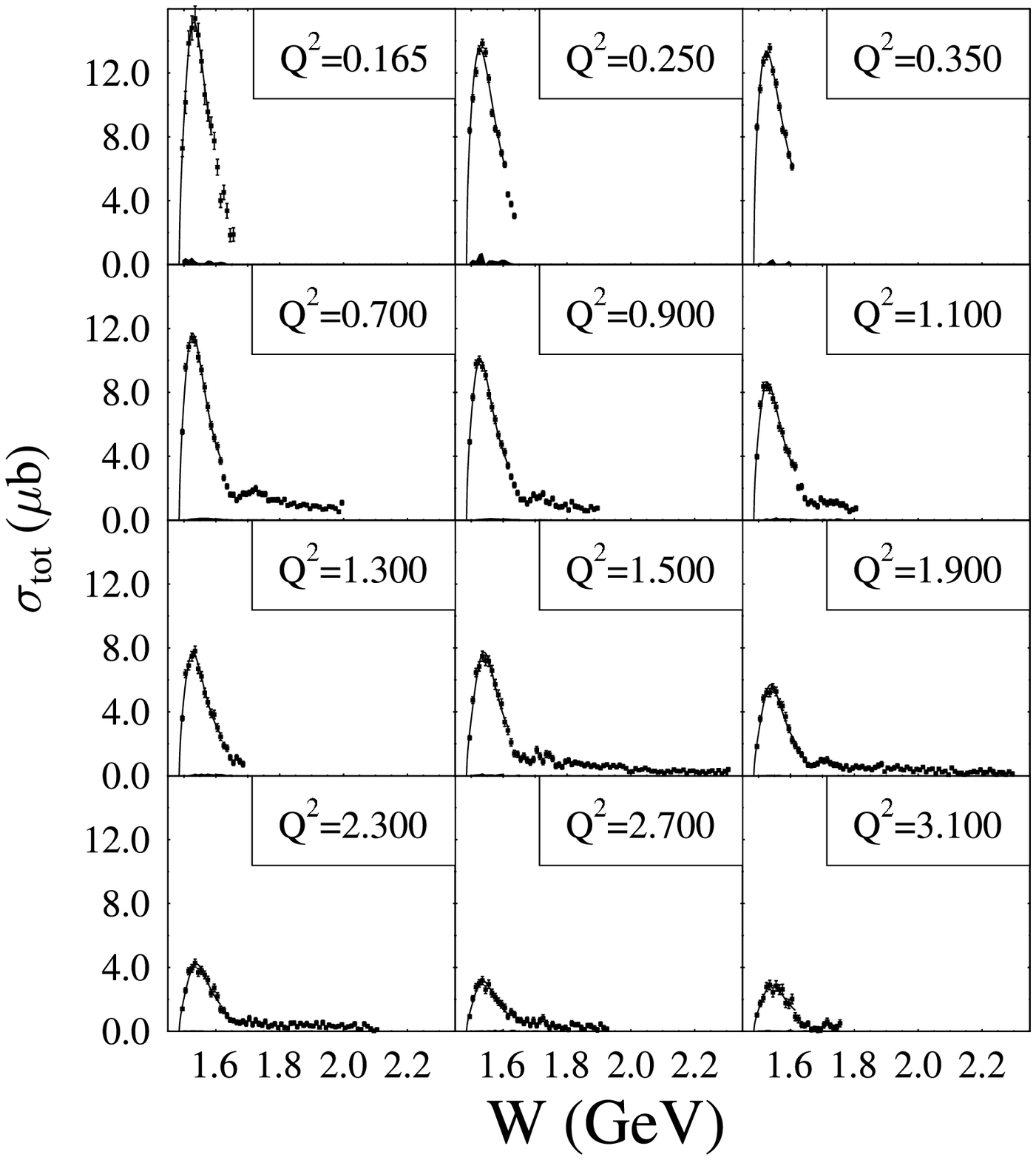}}
\centering{\includegraphics{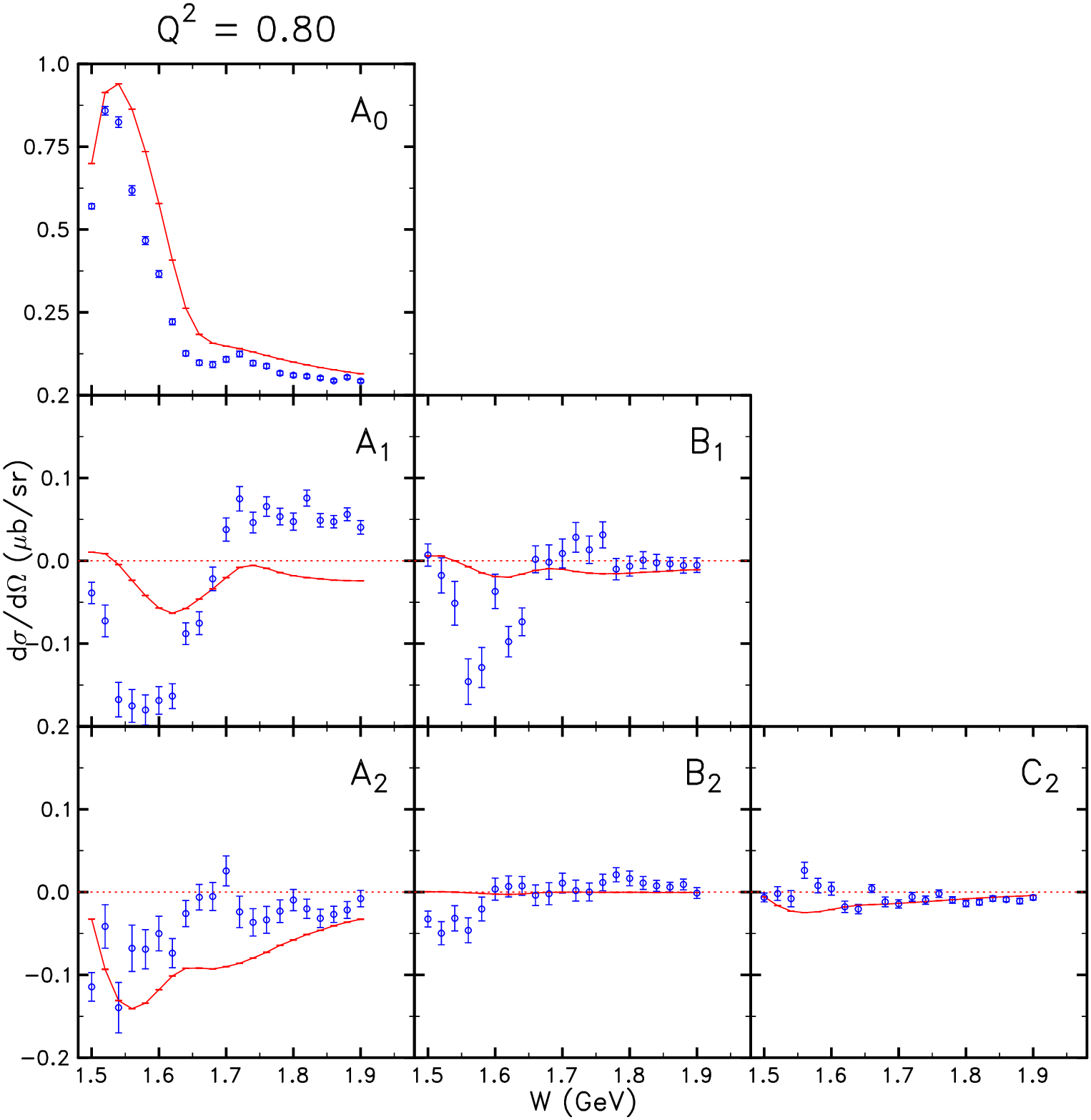}}
\caption{\small Left panel: Total cross section for $\gamma^* p \rightarrow p \eta$ vs W,  
at various photon virtualities $Q^2$. The right panel shows partial wave fit parameters vs W
compared to predictions of $\eta$MAID. All data are from CLAS.}
\label{fig:clas_eta_electro_fit1}
\label{fig:clas_eta_tot_electro1}
\end{figure}
Figure \ref{fig:clas_eta_tot_electro1} shows samples of total cross sections at fixed $Q^2$. 
In contrast to the $\Delta(1232)$ which rapidly drops with $Q^2$, the $S_{11}(1535)$ remains 
prominent even at the highest $Q^2$.       

\begin{figure}[pbth]
\vspace{80mm} 
\centering{\includegraphics{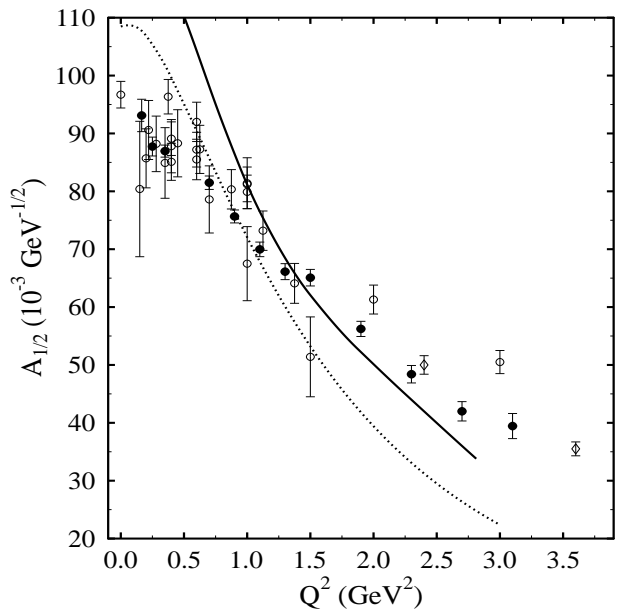}}
\caption{\small  $Q^2$ evolution of the $\gamma p \rightarrow
 S_{11}(1535)$ photocouplings amplitude $A_{1/2}$. 
The full symbols are the most recent CLAS results. The open diamonds are the JLab data from Hall C, 
and the open circles are from previous experiments. The solid curve is
from Ref.(177), and the dotted curve is from Ref.(175). }
\label{fig:A12_S11_Q2}
\end{figure}

Most of the published results on the $Q^2$ dependence 
of the $S_{11}(1535)$ transition amplitude have been obtained in
single resonance fits. This has been justified with the dominant contributions of the $S_{11}(1535)$
to the $\gamma^* p \rightarrow p\eta$ cross section. It is, however, not a fully satisfactory solution, as higher mass states that couple to 
$N\eta$ may also contribute in the lower mass region. The results have to be taken with caution.
The differential cross sections are fitted to the 
expression Eq.(17).
The dependence on the $\eta$ scattering angle ($\theta^*_{\eta}$)
can be examined by expanding each component of the
differential cross section in terms of Legendre polynomials: 
\begin{eqnarray}
{d\sigma_T \over {d\Omega}^*_{\eta}} + \epsilon {d\sigma_L \over {d\Omega}^*_{\eta}} 
&=& {\sum_{\ell=0}^{\infty}} A_{\ell}P_{\ell}(\cos\theta^*_{\eta})\\
\sqrt{2\epsilon(\epsilon+1)} {d\sigma_{LT}\over {d\Omega}^*_{\eta}} &=&  {\sum_{\ell = 1}^{\infty}} B_{\ell}P_{\ell}^{\prime}(\cos\theta^*_{\eta})\\
 \epsilon {d\sigma_{TT}\over {d\Omega}^*_{\eta}} &=&  {\sum_{\ell = 2}^{\infty}} 
C_{\ell}P_{\ell}^{\prime\prime}(\cos\theta^*_{\eta})
\end{eqnarray}
If the expansion is limited to $\ell = 2$, only the coefficients $A_0,~A_1,~A_2,~B_1,~B_2$ and $C_2$ are retained.  
Results from the fit at fixed $Q^2$ are shown in the right panel of Fig. \ref{fig:clas_eta_electro_fit1}.
Strong variations of $A_1$ and $B_1$ are seen in the W range from 1.6 to 1.7 GeV, indicating large interference 
effects involving s- and p-waves. Possible p-wave candidates are the $P_{11}(1710)$ and $P_{13}(1720)$ states.   
$A_0$ is mostly due to the $S_{11}(1535)$ resonance, and is the by far largest amplitude. The longitudinal 
and transverse response functions cannot be separated in this analysis. In earlier experiments~\cite{bonn_eta2,desy_eta2} 
the longitudinal and transverse 
cross sections were separately determined at some fixed $Q^2$ values, showing that $A_0$ is dominated by the transverse 
amplitude $A_{1/2}^2$ in the $Q^2$ range of this study. The combined 
analysis of $\pi$ and $\eta$ electroproduction data, to be discussed in
the next section, 
also finds small longitudinal contribution to $A_0$. Assuming $\sigma_L = 0$, $|A_{1/2}|$ can be 
computed from $A_0$.    
Figure \ref{fig:A12_S11_Q2} shows a compilation of results for the 
$\gamma p\rightarrow S_{11}(1535)$ photocouplings helicity amplitude $A_{1/2}(Q^2)$. The slow falloff 
with $Q^2$ confirms the unusually hard transition form factor that persists to the highest measured 
values of $Q^2$.   The solid and dotted curves
are the prediction of Close and Li~\cite{close90}, and of Giannini, Santopinto 
and Vassallo~\cite{giannini03}. 
It should be noted that the absolute normalization of the data displayed in
Fig.\ref{fig:A12_S11_Q2} is uncertain to the extent that the branching ratio
$\beta_{N\eta}(S_{11}) = 0.55$ and a total widths of 150 MeV have been used in 
extracting $A_{1/2}$. The Review of Particle Properties 2002 allows a large range of 0.30 - 0.55 for the branching ratio. 
However, the recent analysis of Armstrong et al.~\cite{armstrong99}, gives 
a value of $\beta_{N\eta}(S_{11}) \approx 0.55$. 
The use of this value is consistent with 
the values $\beta_{N\eta}(S_{11})=0.55$ and $\beta_{N\pi}(S_{11})=0.4$
used in the combined analysis of $\pi$ and $\eta$ electroproduction
which is the subject of next section.

\subsection{Combined analysis of $\pi$ and $\eta$ electroproduction data }
\label{sect:combined_analysis}

\begin{figure}
\vspace{7cm}
\centering{\includegraphics{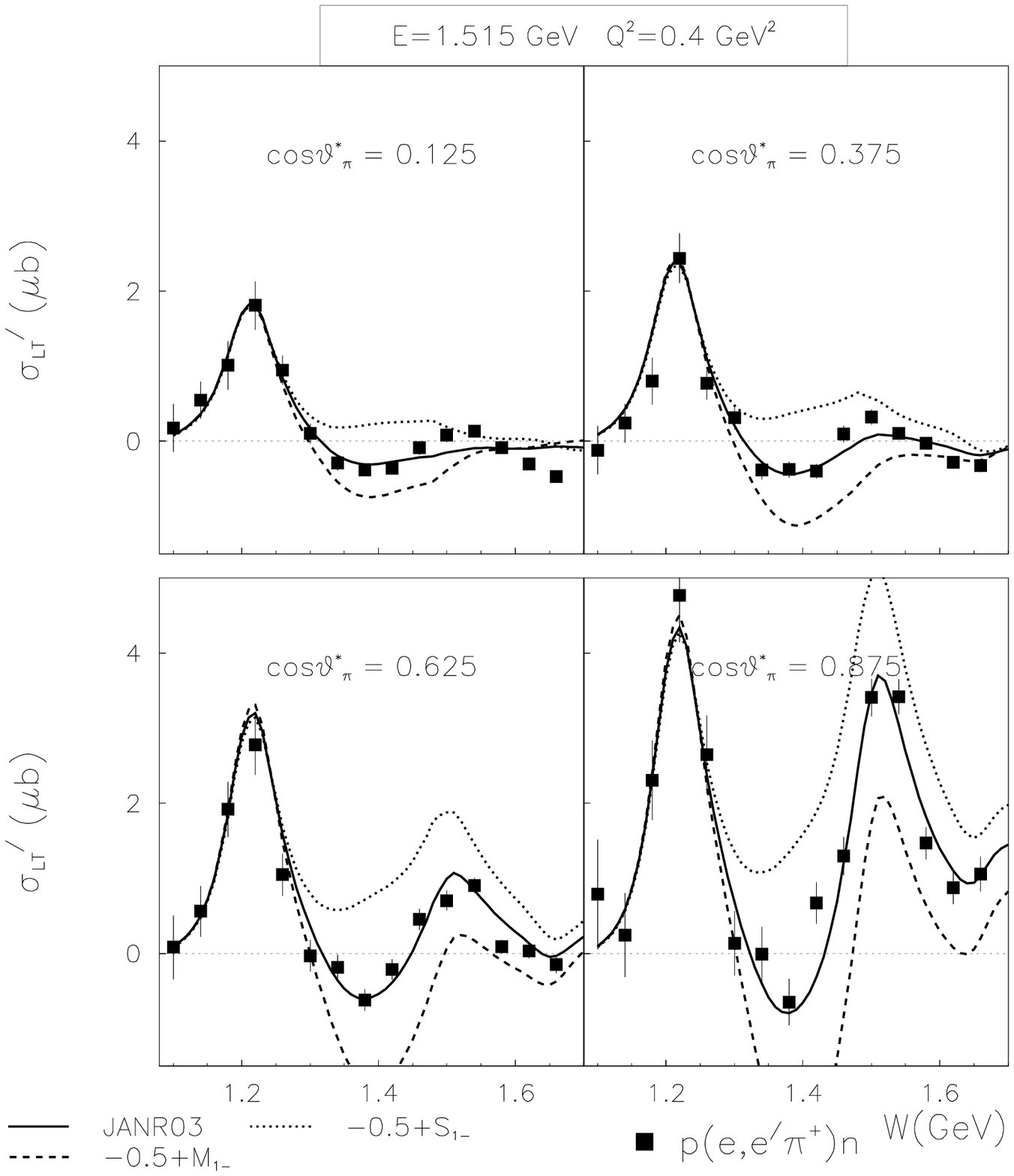}}
\centering{\includegraphics{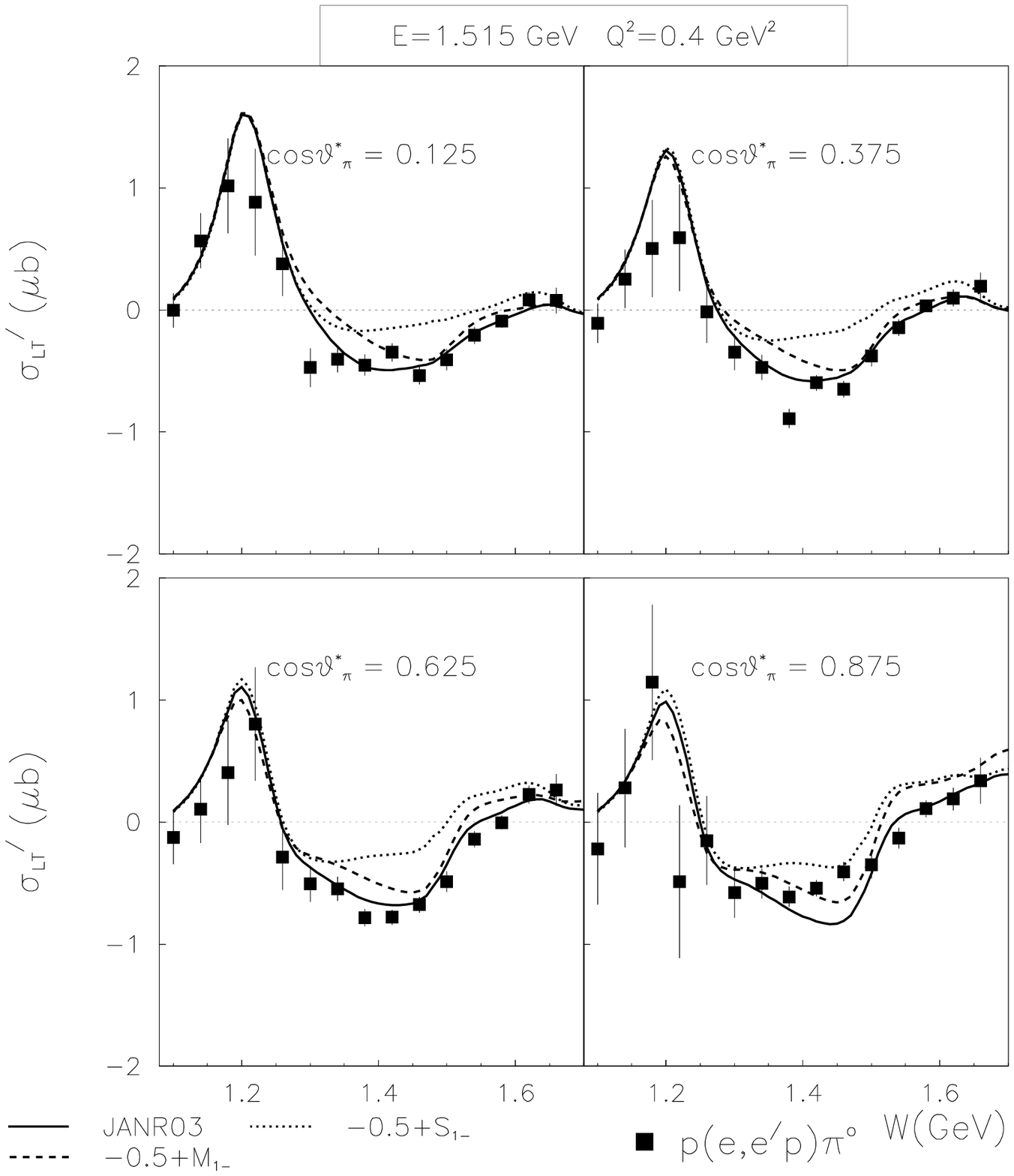}}
\caption{\small Sensitivity of CLAS data of $\sigma_{LT^{\prime}}$ to changes in multipoles 
$M^{1/2}_{1-}$ (dashed)
and $S_{1-}^{1/2}$ (dotted) for $n\pi^+$ channel (left) and $p\pi^{\circ}$ channel (right). 
Solid line shows best fit 
using the Unitary Isobar Model. The -0.5 in the legend refers to shifting the Breit-Wigner 
amplitude by -0.5 $\mu b^{1/2}$. }
\label{fig:p11_sensitivity}
\end{figure}

\begin{figure}[pbth]
\vspace{60mm} 
\centering{\includegraphics{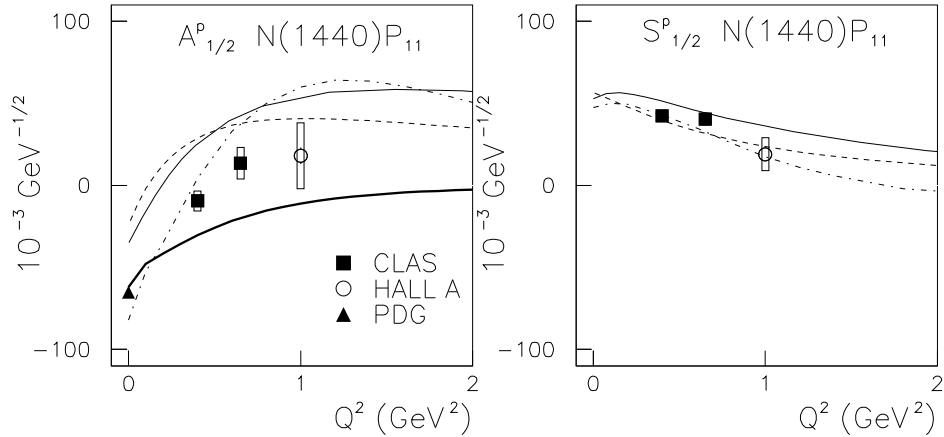}}
\caption{\small Results for the $P_{11}(1440)$ amplitudes $A^p_{1/2}$ and $S^p_{1/2}$. CLAS points 
include model error. The Hall A point shows MAID03 fit model error. The curves are described in the text.} 
\label{fig:p11_global_fit}
\end{figure}

The large amount of data taken by the CLAS detector allows simultaneous measurements of cross sections and polarization 
observables for several channels, e.g. $p\pi^{\circ},~n\pi^+,~p\eta$. Also, the large acceptance 
provides complete angular distributions, including the full azimuthal dependence. Use of a highly polarized 
electron beam provides data on the helicity-dependent response function $\sigma_{LT^{\prime}}$ covering the 
full angle range. Results of a MAID and DMT analysis of the channel $p\pi^{\circ}$ have recently been 
reported~\cite{tiator03}. Combined analyses of these 
data in a multi-channel global fit provides much more stringent constraints on resonance parameters than 
single-channel analyses can. The full set of data taken with a hydrogen target have been analyzed within the 
unitary isobar model~\cite{azn1} and the dispersion relation approach~\cite{azn2} described in section 4. 
The data on 
$\sigma_{LT^{\prime}}$ are especially sensitive to small resonance contributions in a large non-resonant background.
The sensitivity is the result of the interference term that mixes real and imaginary amplitudes 
\begin{eqnarray}
\sigma_{LT^{\prime}} \sim Im(L)\cdot Re(T) + Im(T)\cdot Re(L)~, 
\end{eqnarray}
where L and T represent the longitudinal and transverse amplitudes, respectively. Figure ~\ref{fig:p11_sensitivity} shows the 
the sensivity of $\sigma_{LT^{\prime}}$ to the $P_{11}(1440)$ multipoles $M_{1-}$ and $S_{1-}$.
Both channels show sensitivity to changes in the multipoles, however, the effect is much larger in
the $n\pi^+$ channel. This is due to a combination of two factors, the stronger coupling of $I=1/2$ 
nucleon states to the 
$n\pi^+$ channel, and the larger background terms contributing to this channel. Using pion and eta 
production to study the same resonances, e.g. the $S_{11}(1535)$ allows tests of possible flavor-dependence of the 
results which could be significant if rescattering effects,  
such as present in the $\Delta(1232)$ region, play an important 
role also for higher mass states.  The use of two
conceptually very different approaches also allow to 
estimate the model-dependence of the resulting amplitudes.
The results for the $P_{11}(1440)$ are shown in Fig.~\ref{fig:p11_global_fit}. For the first
time a consistent trend is emerging: The magnitude of $A_{1/2}(Q^2)$ drops rapidly for $Q^2 > 0$, with 
a sign change near $Q^2 = 0.5$~GeV$^2$. There is also a strong longitudinal coupling.
Bold, solid, dashed and dot-dashed lines are from various 
calculations\cite{libuli,capstick95,cardarelli00,cano98}.
Nonrelativistic quark models\cite{warns90,close90} predict large negative $A_{1/2}(Q^2)$ in the entire 
$Q^2$ range, and do not describe the data. The hybrid model\cite{libuli} describes 
the fast drop of $|A_{1/2}|$ qualitatively, but has no sign change, and predicts 
$S_{1/2}(Q^2)=0$, while the data show a 
sizeable $S_{1/2}$ amplitude. The relativistic models of Capstick and 
Keister~\cite{capstick95} and of Cardarelli and Simula~\cite{cardarelli00} predict the sign change for $A_{1/2}(Q^2)$ 
but show a much faster rise  than is observed. The magnitudes and trends of both amplitudes are well described by a model 
that describes the $P_{11}(1440)$ with a small quark core and a $q\bar{q}$ cloud~\cite{cano98}. In this model, 
the low $Q^2$ behavior is entirely due to the $q\bar{q}$ contribution while the $qqq$ core defines 
the high $Q^2$ behavior. 
\begin{figure}[bhpt]
\vspace{50mm} 
\centering{\includegraphics{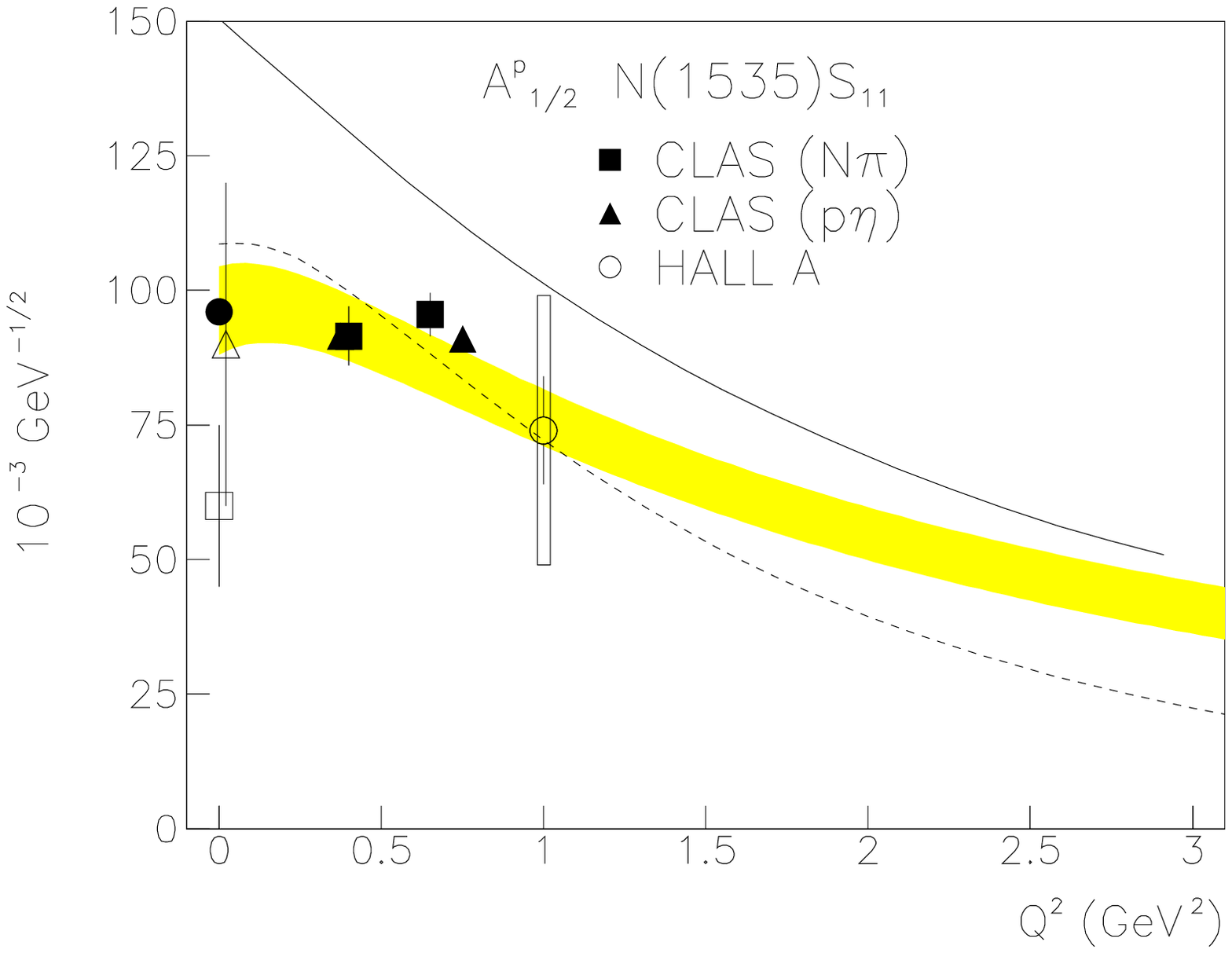}}
\centering{\includegraphics{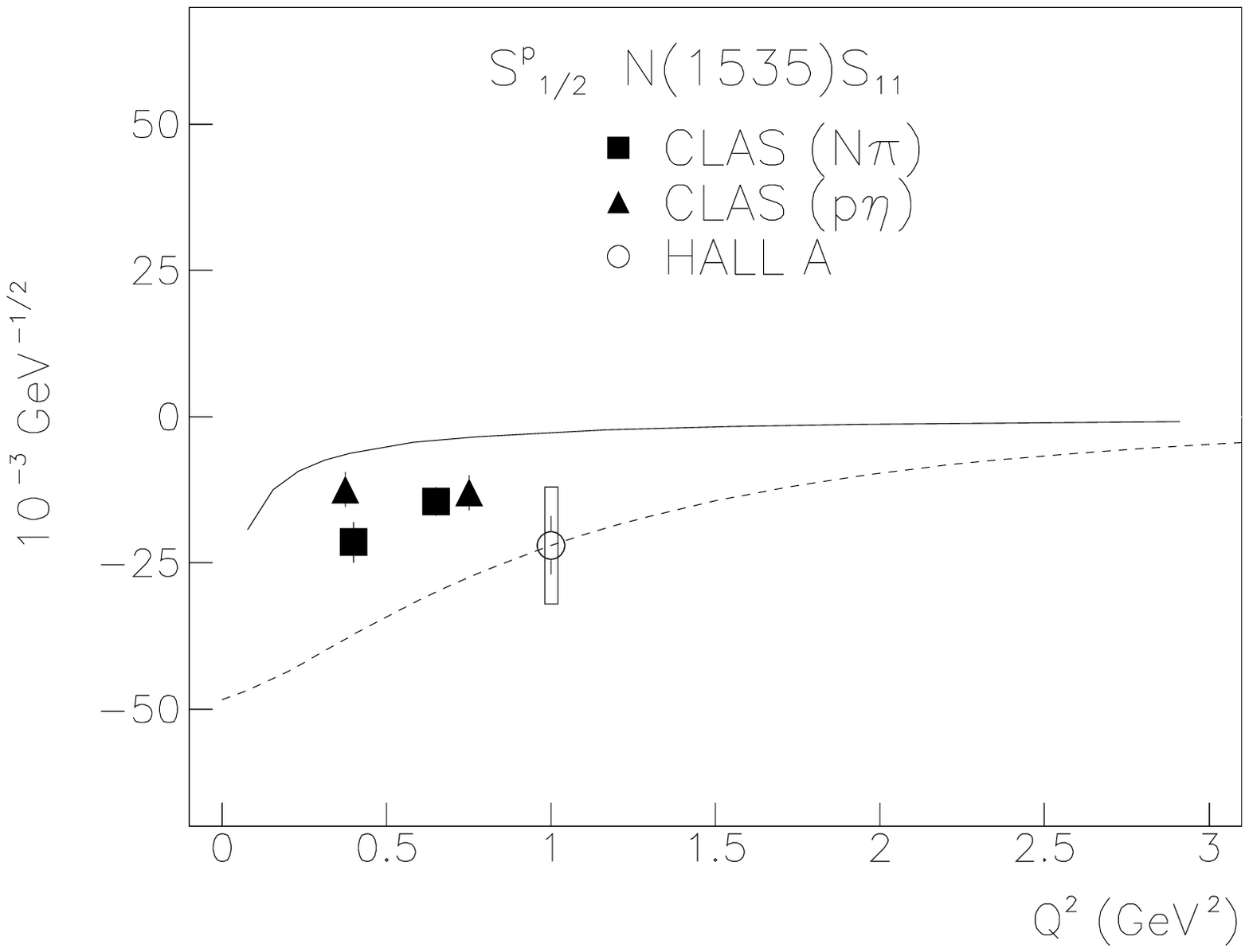}}
\caption{\small  $Q^2$ evolution of the $S_{11}(1535)$ photocoupling amplitudes $A_{1/2}(Q^2)$ (left), 
and $S_{1/2}(Q^2)$ (right). Cross section data from $p\eta$, $p\pi^{\circ}$, and $n\pi^+$ have been 
used, as well as polarized beam response function $\sigma^{\pi^+}_{LT^{\prime}}$ and  
$\sigma^{\pi^{\circ}}_{LT^{\prime}}$. The shaded band indicates the uncertainties seen in previous 
analysis using $p\eta$ cross sections data. Solid and dotted lines are from quark model 
calculations described in the text.} 
\label{fig:s11_global_fit}
\end{figure}
\begin{figure}[pbth]
\vspace{80mm} 
\centering{\includegraphics{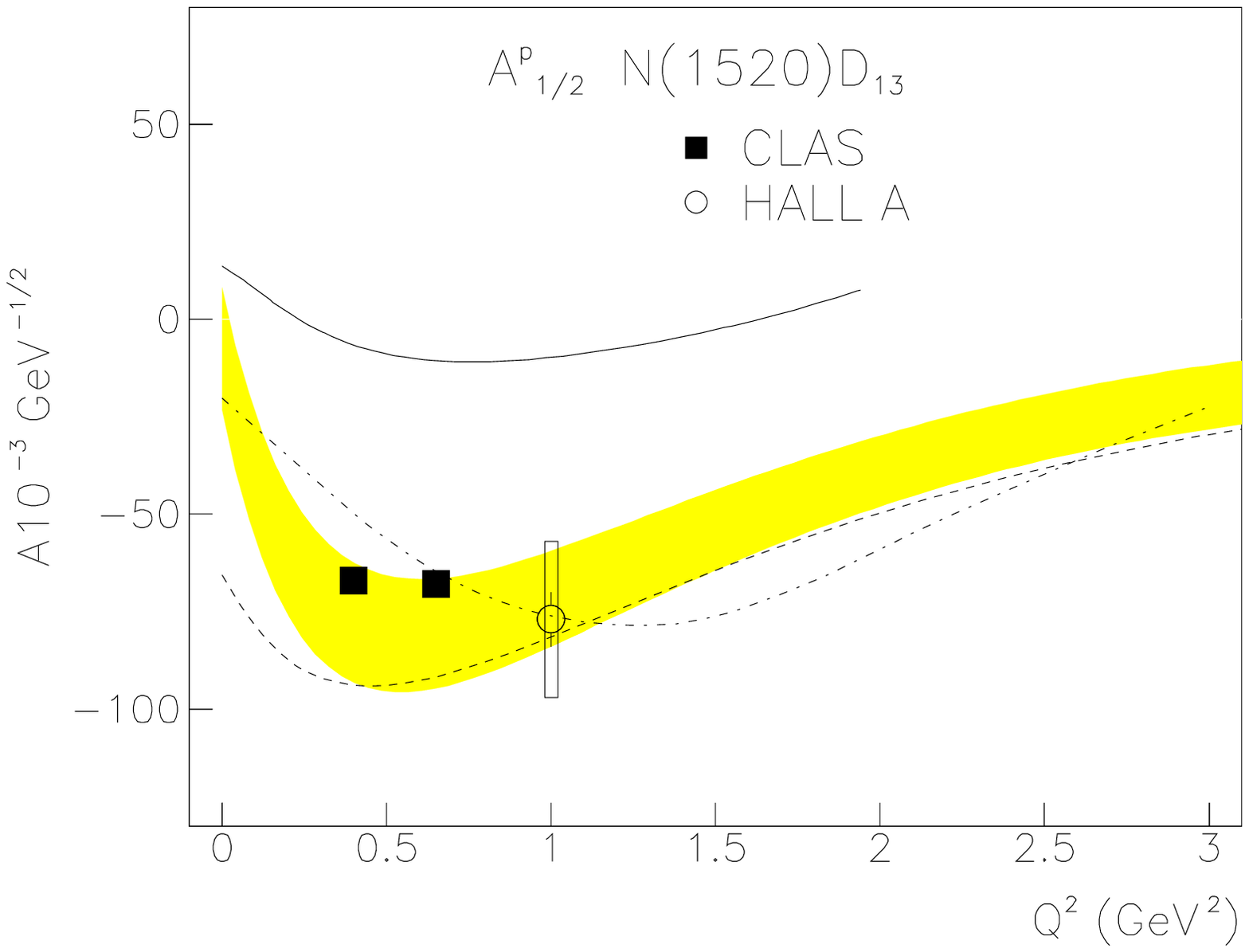}}
\centering{\includegraphics{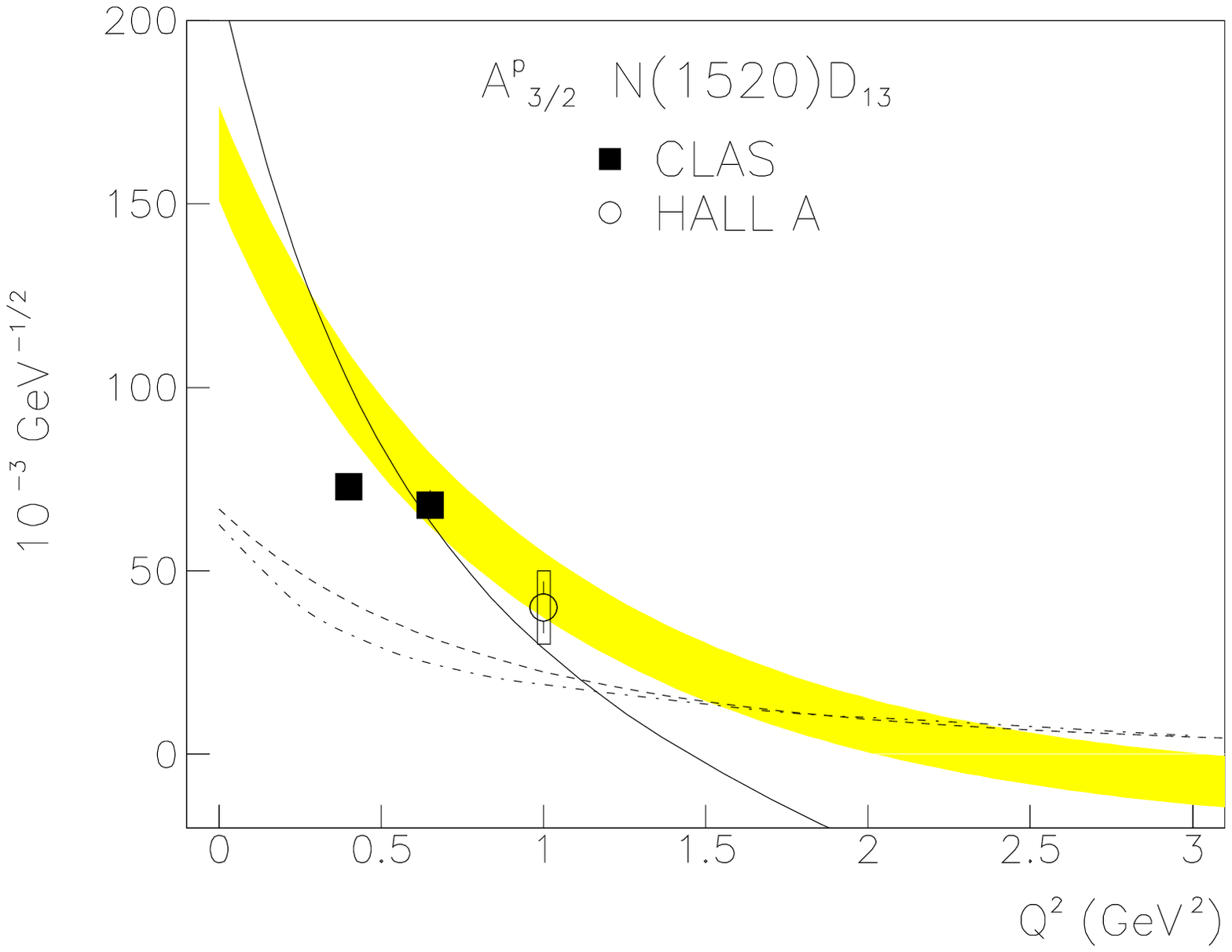}}
\centering{\includegraphics{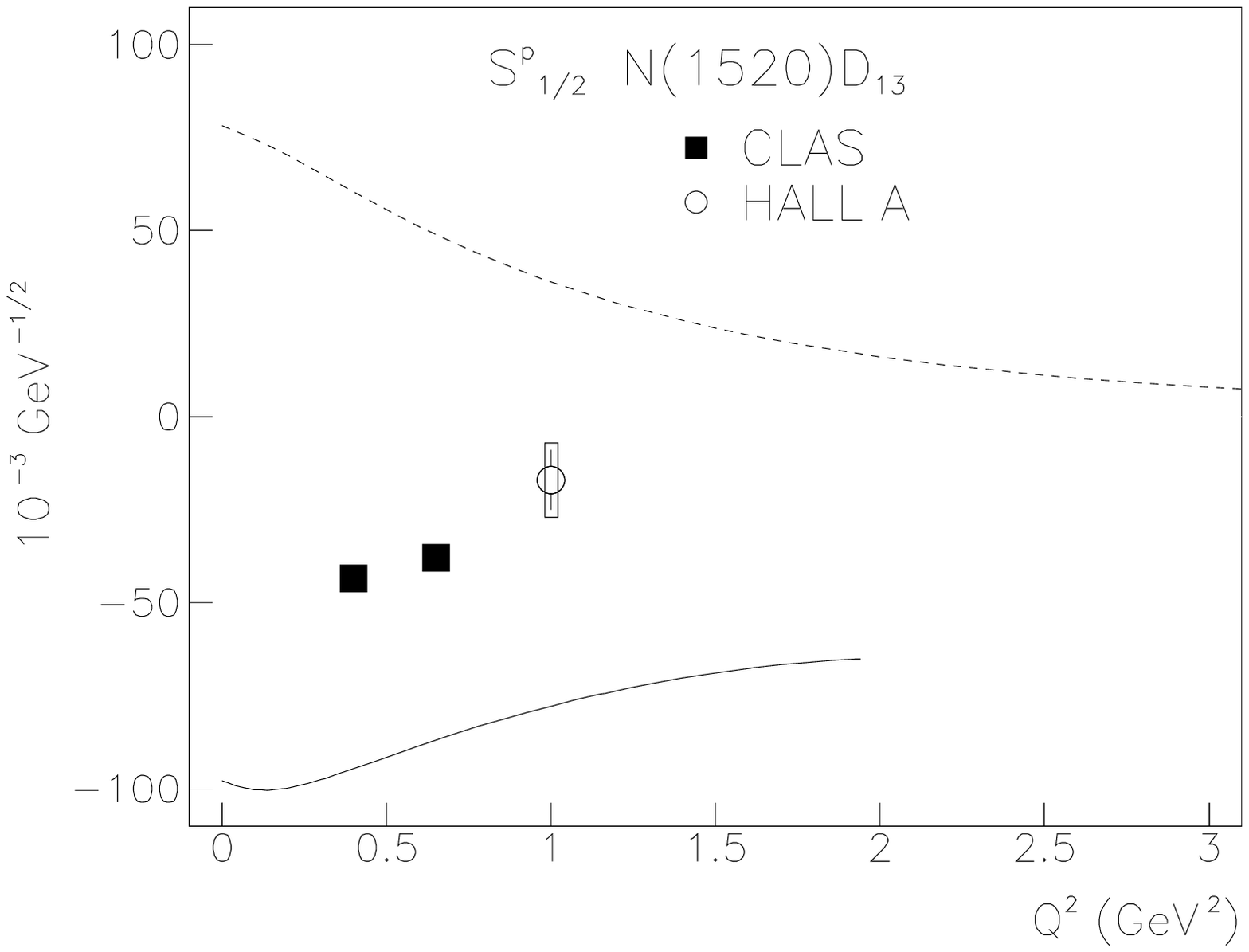}}
\caption{\small  $Q^2$ evolution of the $D_{13}(1520)$ photocoupling amplitudes $A_{1/2}$, $A_{3/2}$, and
$S_{1/2}$. 
Cross section data from $p\pi^{\circ}$, and $n\pi^+$ have been 
used, as well as polarized beam response function $\sigma^{\pi^+}_{LT^{\prime}}$ and  
$\sigma^{\pi^{\circ}}_{LT^{\prime}}$. The shaded bands indicate the uncertainties seen in previous 
analysis using mostly $p\pi^{\circ}$ cross section data. The theoretical
curves are explained in the text.} 
\label{fig:d13_global_fit}
\end{figure}
The results for the $S_{11}(1535)$ are depicted in Fig. \ref{fig:s11_global_fit}. They show consistent results
for $A_{1/2}$ in the $p\eta$ and the $N\pi$ channels for the 
Unitary Isobar Model (UIM) and Dispersion-Relation (DR) analyses. Also, for the first time, 
stable results for the longitudinal coupling $S_{1/2}$ have been obtained. 
The solid and dotted curves in Fig.\ref{fig:s11_global_fit}
represent quark model calculations using a harmonic oscillator potential~\cite{close90} and a 
hypercentral potential~\cite{giannini03}, respectively.

The results of the global fit for $A_{1/2},~A_{3/2}$ and $S_{1/2}$ for the $D_{13}(1520)$ are 
shown in Fig. \ref{fig:d13_global_fit}. Both the UIM and DR analyses give consistent results. 
\noindent
To summarize, the inclusion of polarization observables in addition to the differential cross section 
into a global analysis results in a less model-dependent description of 
$\pi$ and $\eta$ 
photoproduction and electroproduction processes in the resonance region. There are several noteworthy results:
First, consistent results are obtained for the mass of the $S_{11}(1535)$ for 
the $\eta$ and $\pi$ channel, both for photo- and electroproduction. The mass is in the range 
$M(S_{11})=1531 \pm 5$~MeV.  
Second, the discrepancy between $\eta$ and $\pi$ photoproduction results for 
$A_{1/2}$ amplitude seems to have been resolved. The analysis of electroproduction data
 gives also good agreement for the $p\eta$ and $N\pi$ channels and in both UIM and DR approaches. 
Third, the $Q^2$ evolution of the $A_{1/2}$ and $S_{1/2}$ amplitudes for the $P_{11}(1440)$ are 
consistent with the predictions of a meson cloud model.      
This is in line with what has been found earlier for the $N\Delta(1232)$ transition, that
meson cloud effects can be sizeable for some of the resonance transitions.   

\subsubsection{Analysis of Resonance Transitions in the Single Quark Transition Model}
Properties of nucleon resonances such as mass, spin-parity, and flavor
fit well 
into the representation of the $SU(6)\otimes O(3)$ symmetry group, which 
describes the spin-flavor and orbital wave functions of the 3-quark 
system. This symmetry group leads to supermultiplets 
of baryon states with the same orbital angular momentum $\vec{L}$ of the 
3-quark system, and  degenerate energy levels. Within a supermultiplet 
the quark spins are aligned to form a total quark spin
$\vec{s}$, with $s = {1\over 2}$, $3 \over 2$, which combines with the orbital 
angular momentum $L$ to form the total angular momentum $\vec{J} = \vec{L} + \vec{s}$. 
A large number of explicit dynamical quark models have been developed 
to describe the electromagnetic transitions between the nucleon 
ground state and its excited states~\cite{koniuk80,warns90,capstick95,giannini03}. 
Measurement of resonance
transitions and the dependence on the distance scales, given by the 
virtuality of 
$Q^2$ of the photon, provides information on the nucleon wave function. 
In order to compute the transition, assumptions on the 3-quark potential and 
the quark-quark interactions have to be made. These are then tested by predicting 
photocoupling helicity amplitudes which can then be confronted with experimental data. 
\begin{figure}
\vspace{7.0cm}
\includegraphics{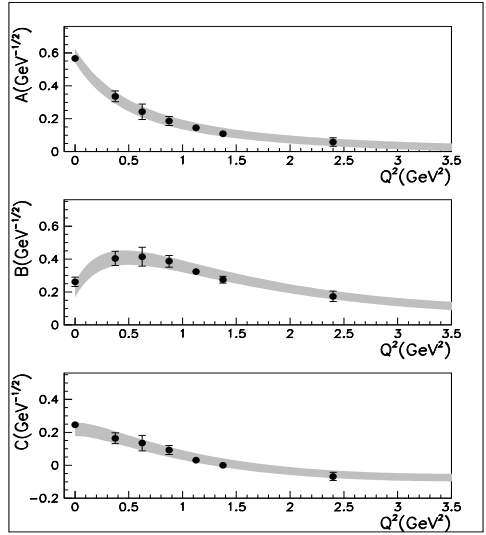}
\caption[]{Single quark transition amplitudes A, B, C as functions of
 $Q^2$. The amplitudes have been 
extracted from the recent JLab, MAMI, and GRAAL data on the $S_{11}(1535)$ 
and world data on $D_{13}(1520)$ photocoupling 
amplitudes. The shaded band is a parameterization of the data from Ref. 178. }
\label{fig:abc_70}
\end{figure}
\begin{table}
\caption{\label{tab:amp_70_proton} Helicity amplitudes for the electromagnetic transition from the ground state 
$[56,0^+]$ to the  $[70,1^-]$ multiplet as a function of the 
SQTM amplitudes. $\theta$ is the mixing angle relating two $J^P = {1\over 2}^-$ states with 
$s_{3q}= {3\over 2}$ and $s_{3q} = { 1\over 2}$. There is also a small mixing angle for the two 
${3\over 2}^-$ states resulting in the physical states $D_{13}(1520)$ and $D_{13}(1700)$. We have not 
included the latter mixing angle in the table. Note that the excitation of the $D_{13}(1700)$ 
from a proton target is only possible because of $SU(6)$ symmetry breaking leading to the mixing 
with the lower mass $D_{13}(1520)$.}
\centerline{\footnotesize\smalllineskip
\begin{tabular}{c l l}
\\
\hline \\
{State} & {Proton target} & {Neutron target} \\
\hline \\
$S_{11}(1535)$ & $A^+_{1/2}$ =~ ${1\over 6} (A + B - C)\cos\theta$ & $A^{\circ}_{1/2} = - {1\over 6}(A + {1\over 6} B - {1\over 3} C)$ \\
$D_{13}(1520)$ & $A^+_{1/2} = ~{1\over 6\sqrt{2}} (A - 2B - C)$ & $A^{\circ}_{1/2} = - {1\over 18\sqrt{2}}(3A - 2 B -  C) $\\  
	       & $A^+_{3/2} = ~{1\over 2\sqrt{6}} (A + C) $      & $A^{\circ}_{3/2} =~~ {1\over 6\sqrt{6}} (3A - C) $\\
$S_{11}(1650)$ & $A^+_{1/2} = ~{1\over 6} (A + B - C)\sin\theta$ & $A^{\circ}_{1/2} = ~~{1\over 18}( B -  C)$ \\
$D_{13}(1700)$ & $A^+_{1/2} = ~0$&  $A^{\circ}_{1/2} = ~~{1\over 18\sqrt{5}}( B - 4C)$ \\  
               & $A^+_{3/2} = ~0$&  $A^{\circ}_{3/2} = ~~{1\over 6\sqrt{15}}( 3B - 2C)$ \\
$D_{15}(1675)$ & $A^+_{1/2} = ~0$&  $A^{\circ}_{1/2} = -{1\over 6\sqrt{5}}( B  + C)$ \\  
               & $A^+_{3/2} = ~0$&  $A^{\circ}_{3/2} = -{1\over 6}\sqrt{2\over 5}( B + C)$\\
$D_{33}(1700)$ & $A^+_{1/2} = ~{1\over 6\sqrt{2}} (A - 2B - C)$ &~~ same \\  
	       & $A^+_{3/2} = ~{1\over 2\sqrt{6}} (A + C) $      &~~ same \\
$S_{31}(1620)$ & $A^+_{1/2} = ~{1\over 18} (3A - B + C)$ &~~ same \\  
\hline
\end{tabular}}
\label{tab:sqtm_abc_a12a32}
\end{table}
Algebraic relations have been derived for resonance transitions assuming 
the transition only affects a single quark in the nucleon.
The parameters in these algebraic equations can be determined 
from the experimental analysis\cite{burkert_sqtm}. Based on the symmetry properties of the Single Quark Transition Model (SQTM), 
predictions for a large number of resonances belonging to the same $SU(6)\otimes O(3)$supermultiplet can be made. 
The fundamentals of the SQTM are described in 
references \cite{hey,cottingham}, where the symmetry properties 
have been discussed for the transitions from the ground state nucleon 
$[56,0^+]$ to the $[70,1^-]$ and the $[56,2^+]$ supermultiplets. 
The $[70,1^-]$  contains states which are prominent in electromagnetic
excitations,  and it is the only supermultiplet for which sufficient
data on resonance couplings of two states are available to extract the 
SQTM amplitudes and test predictions for other states.
The coupling of the electromagnetic current is considered for the 
transverse photon component, and the quarks in the nucleon are assumed to 
interact freely with the photon. In such a model 
the quark transverse current can be written in general as a sum of four terms~\cite{hey,cottingham,close}:
\begin{equation}
J^+= {AL^+ +B\sigma^+L_z  + C\sigma_zL^+  + D\sigma^-L^+L^+}~,\\
\end{equation}
where $\sigma$ is the quark Pauli spin operator, and the terms with 
$A$, $B$, $C$, $D$ in front operate on the quark spatial wave function 
changing the component of orbital angular momentum along the direction of the 
momentum transfer (z- axis). 
The $A$ term corresponds to a quark orbit flip with  $\Delta L_z = +1$, 
term $B$ to a quark spin flip with $\Delta L_z = 0$, 
the $C$ and $D$ terms correspond to simultaneous quark orbit and quark spin flip
with orbital angular momentum flips of $\Delta{L_z} = +1$ 
and $\Delta{L_z} = +2$, respectively. For the transition  from the $[56,0^+]$
to the $[70,1^-]$ supermultiplet with $L = 1$, only $A$, $B$, and $C$ are 
allowed. The relationship between the A, B, C amplitudes and the usual helicity 
photocoupling amplitudes $A_{1/2}$ and $A_{3/2}$ is listed in table~\ref{tab:sqtm_abc_a12a32}. 
\begin{figure}[hbpt]
\vspace{15.0cm}
\includegraphics{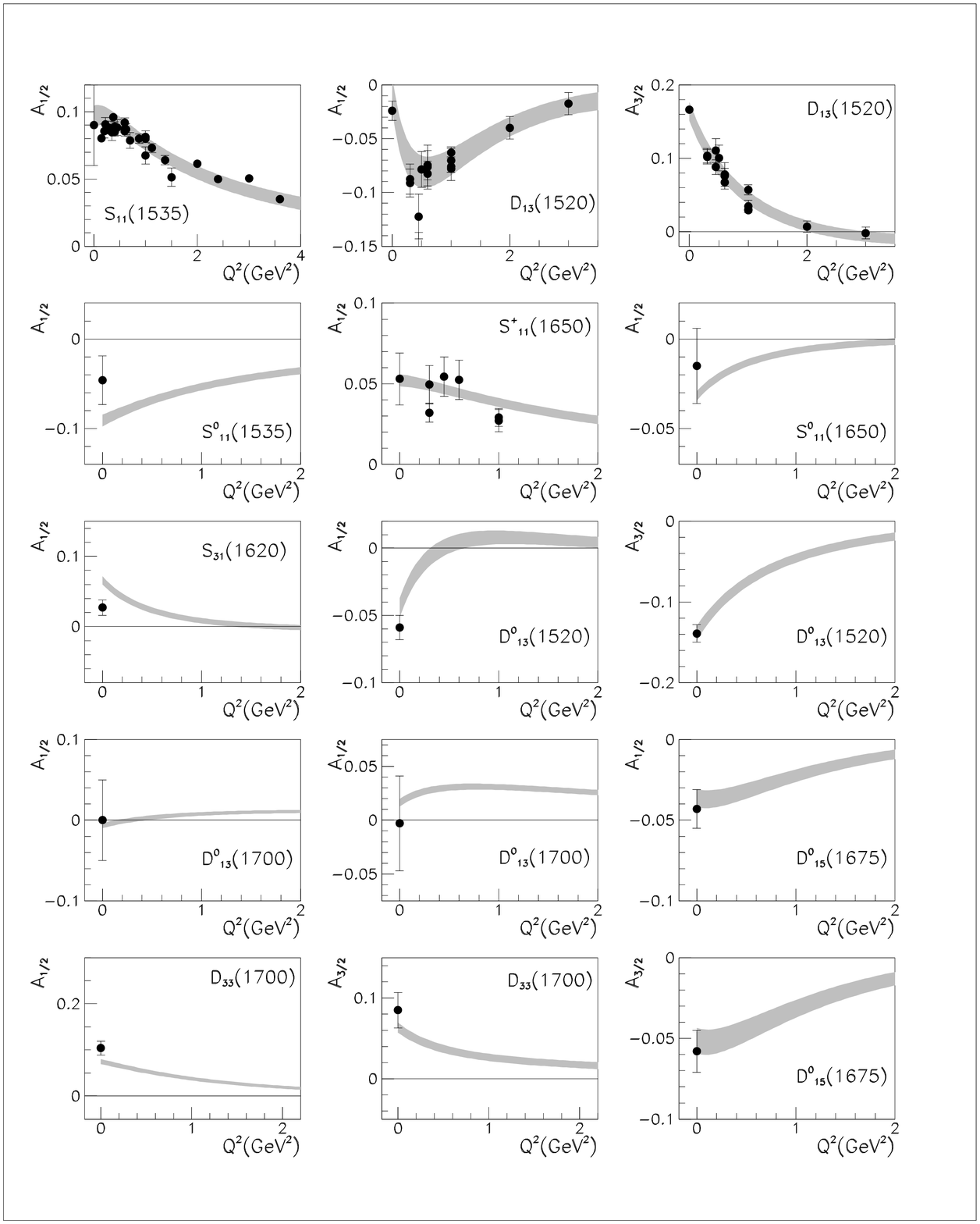}
\caption[]{Single quark model transition prediction for the $[70,1^-]$ multiplet. 
The SQTM predictions are shown by the 
shaded band in comparison with the experimental data. At $Q^2=0$ the full 
circle is the Particle Data Group estimate. For $Q^2>0$, measurements from 
JLab, Bonn, DESY, and NINA in $\eta$ and $\pi$ electroproduction are shown. 
For the $S_{11}(1535)$, the results of an analysis of the world data 
in $\eta$-electroproduction presented in Ref. \cite{thompson01} are also included.
The superscript $^o$ refers to neutron states.}
\label{fig:sqtm_70}
\end{figure}
Using the extracted photocoupling amplitudes from the $S_{11}(1535)$ and the 
$D_{13}(1520)$, the A, B, C amplitudes for the $\gamma + [56,0^+] \rightarrow [70,1^-]$
have been extracted~\cite{burkert_sqtm}. 
The results are shown in Fig. \ref{fig:abc_70}.
Knowledge of the 3 amplitudes and of two mixing angles for the transition 
to the $[70,1^-]$ allows predictions for 16 amplitudes of states 
belonging to the same supermultiplet.
If they can be confirmed for some of the 
amplitudes, one then has a measure of the degree to which electromagnetic 
transitions of nucleon resonance are dominated by single quark transitions at 
the photon point ($Q^2=0)$ and, using electroproduction data, examine if 
and how this is changing as a function of the distance scale at increasing
photon virtuality. The SQTM predictions for the proton  and neutron 
amplitudes are shown in Fig. \ref{fig:sqtm_70}.
There is remarkable agreement between the predictions and the data at 
the photon point. For electroproduction, there is good agreement where consistent data sets 
are available, i.e. for the $S_{11}(1650)$. Much improved electroproduction 
data are needed for more definite conclusions. Most of the states belonging to the $[70,1^-]$ 
supermultiplet with masses near 1700~MeV couple
strongly to $N\pi\pi$ channels. Studies of these channels require use of large acceptance 
detectors and new, sophisticated analysis techniques. Progress made in these areas will be 
discussed in the next section.
There are similar relations for the transition from the nucleon ground state to the 
members of the $[56,2^+]$ supermultiplet. In this case four SQTM amplitudes can contribute.
Unfortunately, the only state for which the two transverse photocoupling amplitudes have been 
measured in electroproduction is the $F_{15}(1680)$. This is insufficient to extract the 
four SQTM amplitudes.

\subsection{Two-pion production.}
Two-pion channels dominante the electromagnetic meson production
cross sections in the second and third resonance regions where we hope to
resolve the {\it missing resonance} problem\cite{koniuk80} and
 ultimately determine what
basic symmetry group\cite{kirchbach} is underlying the baryon spectrum.
Thus, a detailed understanding of two-pion production is very important
in the $N^*$ study, and has been pursued
very actively in recent years. Very extensive  two-pion production
data have now been accumulated at JLab, MAMI and ELSA, but have not
been fully anaylzed and understood  theoretically. Here we will mainly
report on the status of the data and describe some very
preliminary attempts to identify $N^*$ states.

The study of $N\pi\pi$ channels requires the
use of detectors with nearly $4\pi$ solid angle
coverage for charged or neutral particle detection. Several such detectors have been in operation
for a number of years, and 
have generated large data sets for the following reactions
\begin{eqnarray} 
\gamma p &\rightarrow& p \pi^{\circ}\pi^{\circ} \,,\\
\gamma p &\rightarrow& p \pi^+ \pi^- \,, \\
e p &\rightarrow& e p \pi^+ \pi^- \,.
\end{eqnarray}

The two-pion channels in the above processes
 can be projected onto various isobar channels which are more useful in
identifying the nucleon resonances from the data.
The $p\pi^+\pi^-$ final state is sensitive to the 
$\Delta^{++}\pi^-$ isobar channel which could have large sensitivity to resonance decays
but has also very strong contributions from non-resonant mechanisms. 
The $p\pi^{\circ}\pi^{\circ}$ final state has the advantage of high sensitivity 
to resonance contributions with fewer non-resonant contributions. 
The $p\pi^+\pi^-$ channel is sensitive to the $p\rho^{\circ}$ isobar channel, 
while the
$p\pi^0\pi^0$ channel does not couple to $p\rho^{\circ}$.  
Table \ref{tab:ppipi_data} gives an overview of recent 2-pion production data 
obtained at various laboratories. 
\begin{table}[htbp]
\tcaption{Summary of $\gamma p \rightarrow p\pi\pi$ production data}
\centerline{\footnotesize\smalllineskip
\begin{tabular}{l l c c l l}\\
\hline
{Reaction} &{Observable} &W range &$Q^2$ range & Lab. \\
{} &\phantom0 &(GeV) & (GeV$^2$) &\phantom0 & \phantom0 \\ 
\hline\\
{$\gamma^* p \rightarrow p\pi^+\pi^-$} & $\sigma_{tot},~M_{p\pi^+},~M_{\pi^+\pi^-},
~{d\sigma \over dcos_{\theta_{\pi^-}}}$ & $< 2.1$  &0.65 - 1.3  & CLAS~\cite{ripani03}\\
& & $< 2.7 $ & 1.5 - 4.0 & CLAS~\cite{burkert99}\\
{$\gamma p \rightarrow p\pi^+\pi^-$} & $\sigma_{tot},~M_{p\pi^+},~M_{\pi^+\pi^-},
~{d\sigma \over dcos_{\theta_{\pi^-}}}$ & $< 2.0$  & 0  & CLAS~\cite{bellis04}\\
{$\gamma p \rightarrow p\pi^{\circ}\pi^{\circ}$} & $\sigma_{tot},~M_{p\pi^{\circ}},
~M_{\pi^{\circ}\pi^{\circ}}$ & $< 1.9$ & 0 &GRAAL~\cite{assafiri03} \\
					 & $\sigma_{tot},~\Sigma,
~M_{p\pi^{\circ}},~M_{\pi^{\circ}\pi^{\circ}}$  &$< 1.55$  & 0 & MAMI~\cite{wolf00,harter97} \\
$\gamma p \rightarrow p\pi^{\circ}\pi^{\circ}$ & event-by-event analysis & $ < 2.6 $ & 0 & CB-ELSA~\cite{thoma02} \\
\hline\\
\end{tabular}}
\label{tab:ppipi_data}
\end{table}

\subsubsection{Analysis of the data with $p\pi^+\pi^-$ final state}

\begin{figure}[tbph]
\vspace{55mm} 
\centering{\includegraphics{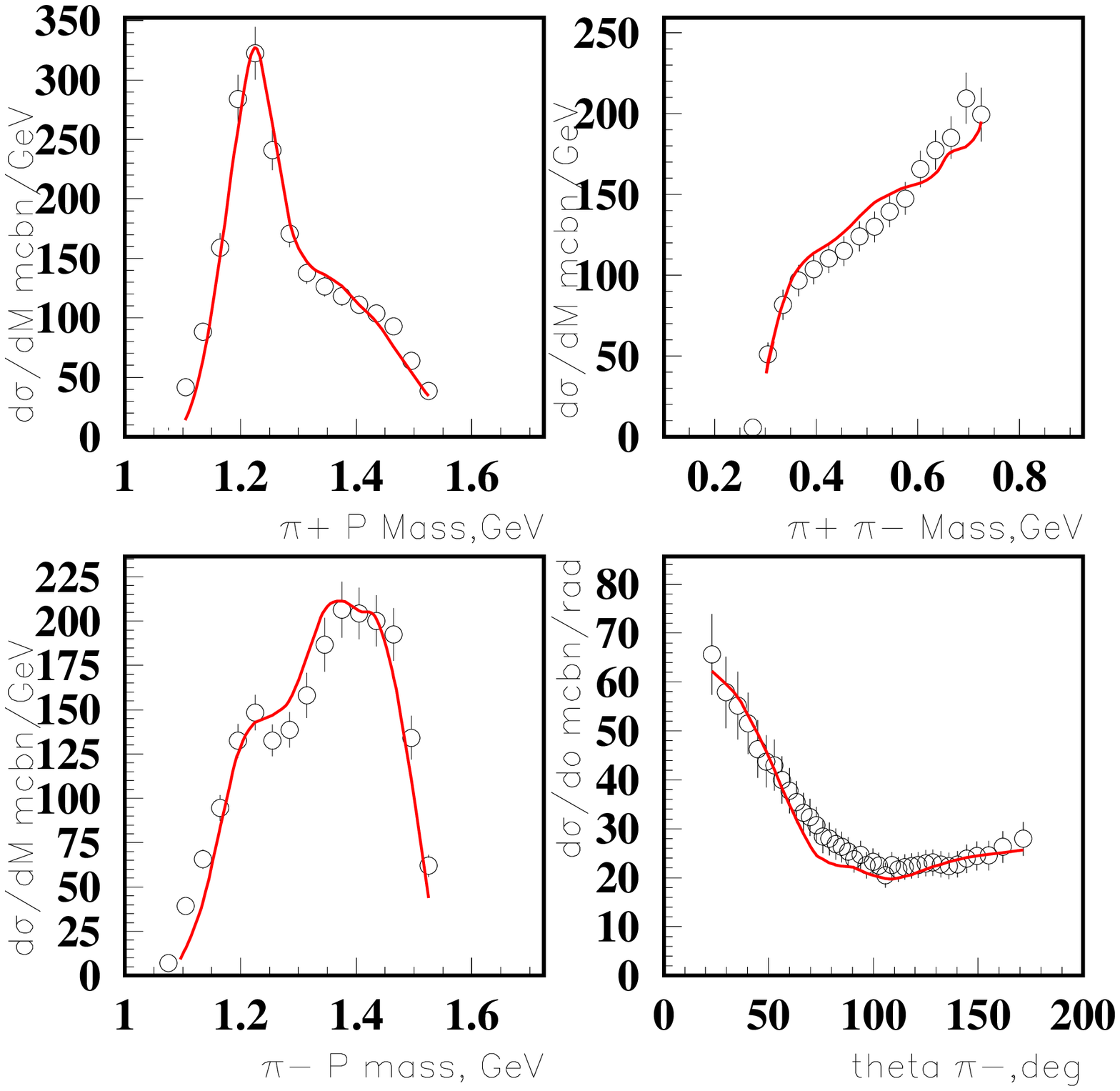}}
\centering{\includegraphics{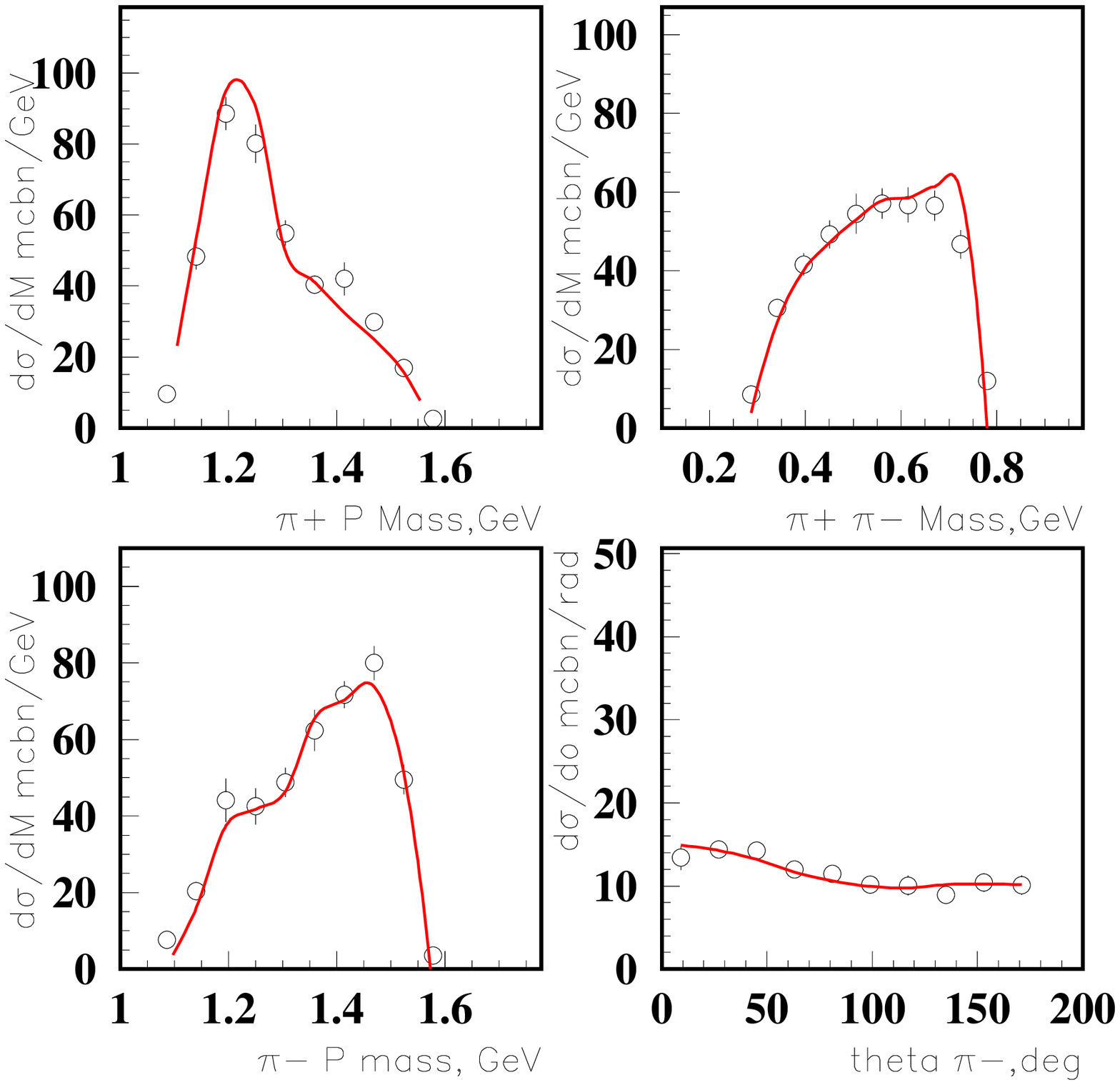}}
\caption{\small One-dimensional projections for the 
$\gamma p \rightarrow p\pi^+\pi^-$ reaction cross sections. 
Left: $Q^2 = 0$. Right: $Q^2 = 0.95$ GeV$^2$. Both data set are from CLAS. Curves are 
explained in the text.}
\label{fig:diff_cs_ppippim}
\end{figure}

The quality of the recent data with $p\pi^+\pi^-$ final state is very high.
An example is shown in Fig.~\ref{fig:diff_cs_ppippim} for
 the differential cross sections in one-dimensional projections.
The data show evidence for the formation of
the $\Delta^{++}\pi^-$,
$\Delta^{\circ}\pi^+$, and $p\rho^{\circ}$ isobar channels.
In the real photon case (left panel) the $p\rho^{\circ}$
contribution  dominates the higher invariant mass (W) region.  This
contribution drops significantly for virtual photons (right panel). 
We also see that  the $\pi^-$ angular distribution is
much more forward-peaked in the case of real photons, and becomes
flatter with increasing $Q^2$. This behavior can be qualitatively understood as a consequence of
vector meson photon duality. With increasing $Q^2$ the probability of the photon to fluctuate into a
virtual vector meson is reduced
leading to reduced diffraction-like scattering.

\begin{figure}[pthb]
\vspace{55mm} 
\centering{\includegraphics{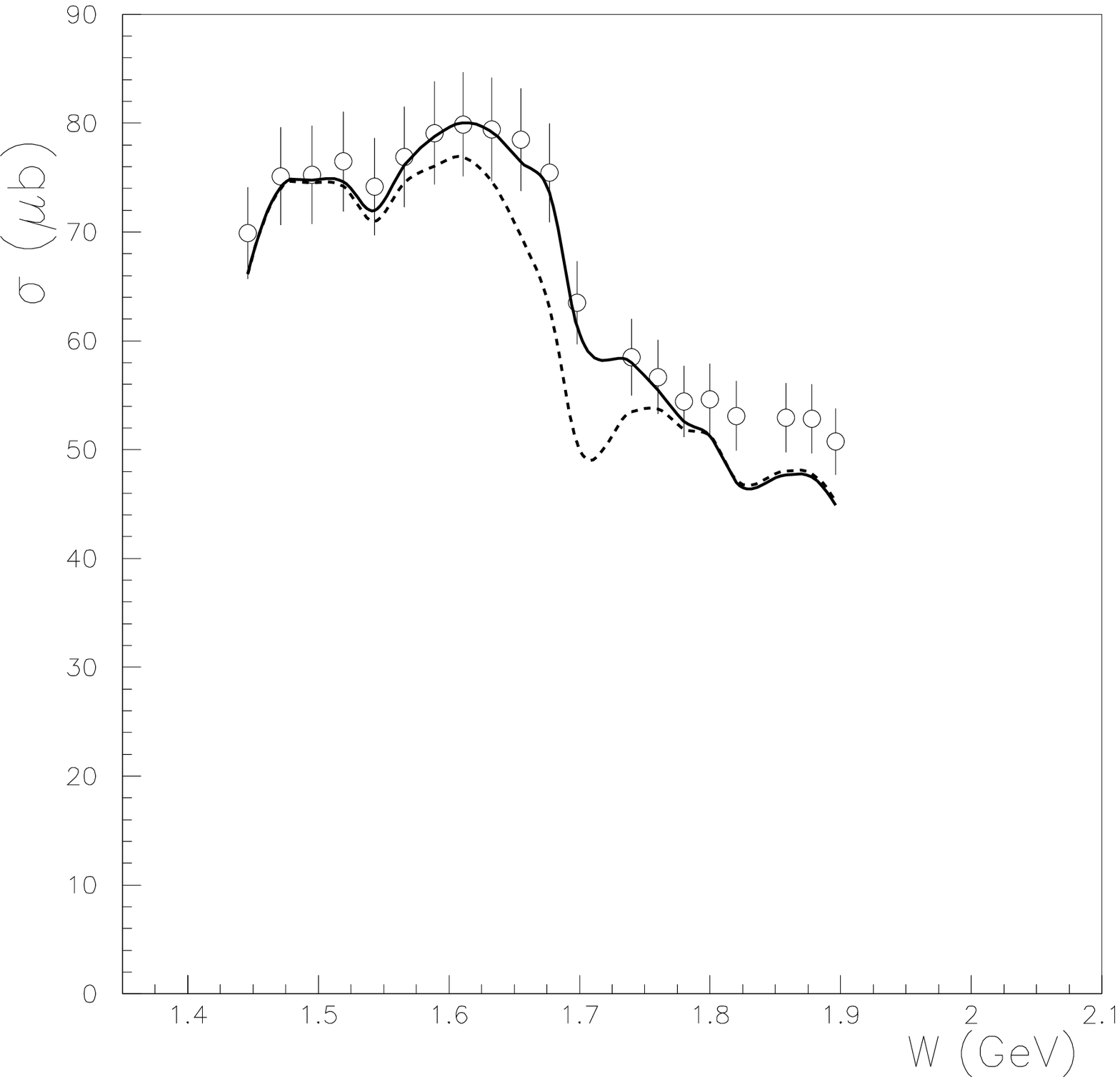}}
\centering{\includegraphics{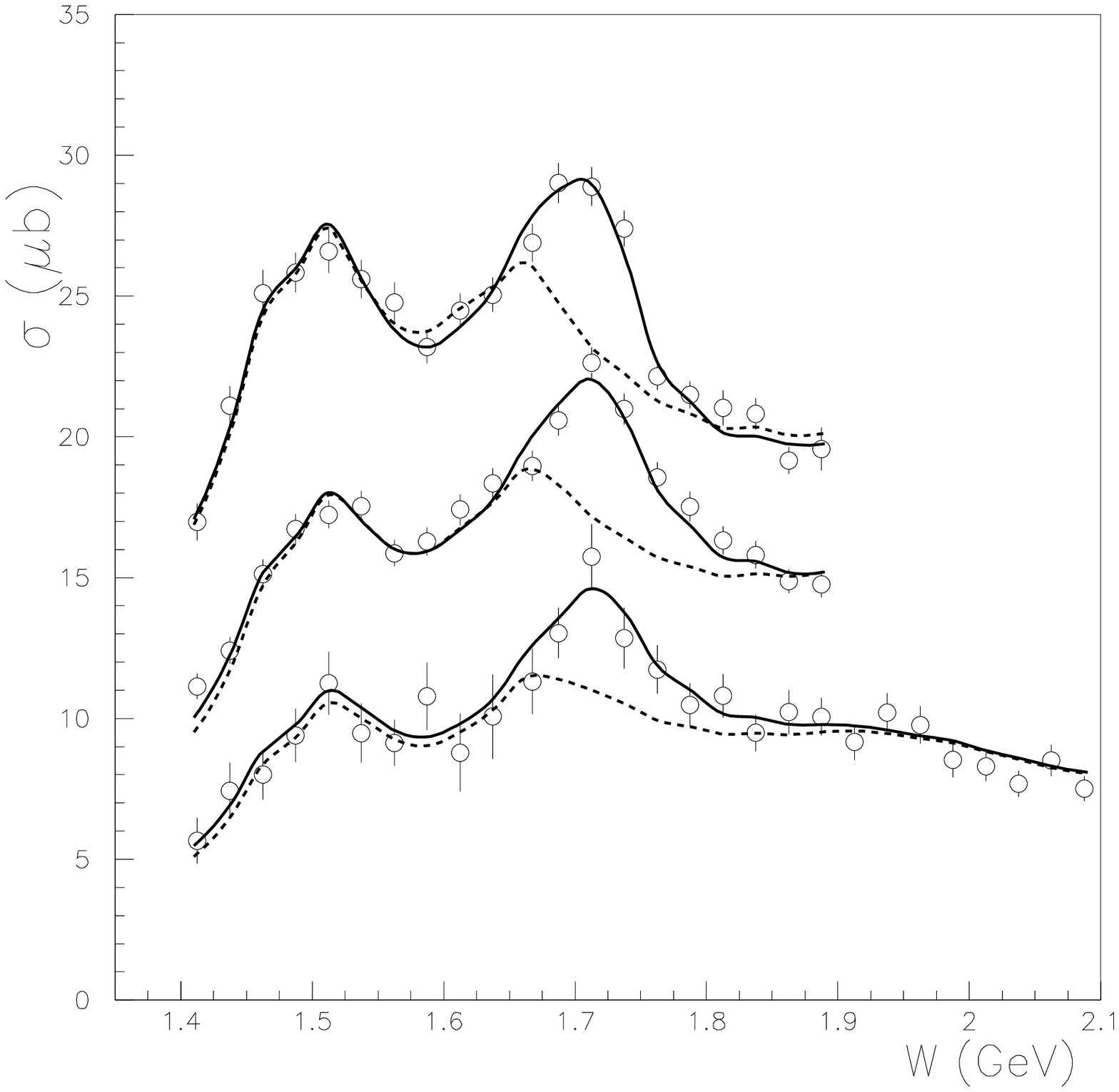}}
\caption{\small Total cross section for photoproduction (left) and electroproduction (right) 
of $p\pi^+\pi^-$ at $Q^2 = 0.65,~0.95,~1.30$GeV$^2$ (from the top). 
Data sets are from CLAS. Curves are explained in the text. }
\label{fig:total_cs_ppippim}
\end{figure}

\begin{figure}[pthb]
\vspace{40mm} 
\centering{\includegraphics{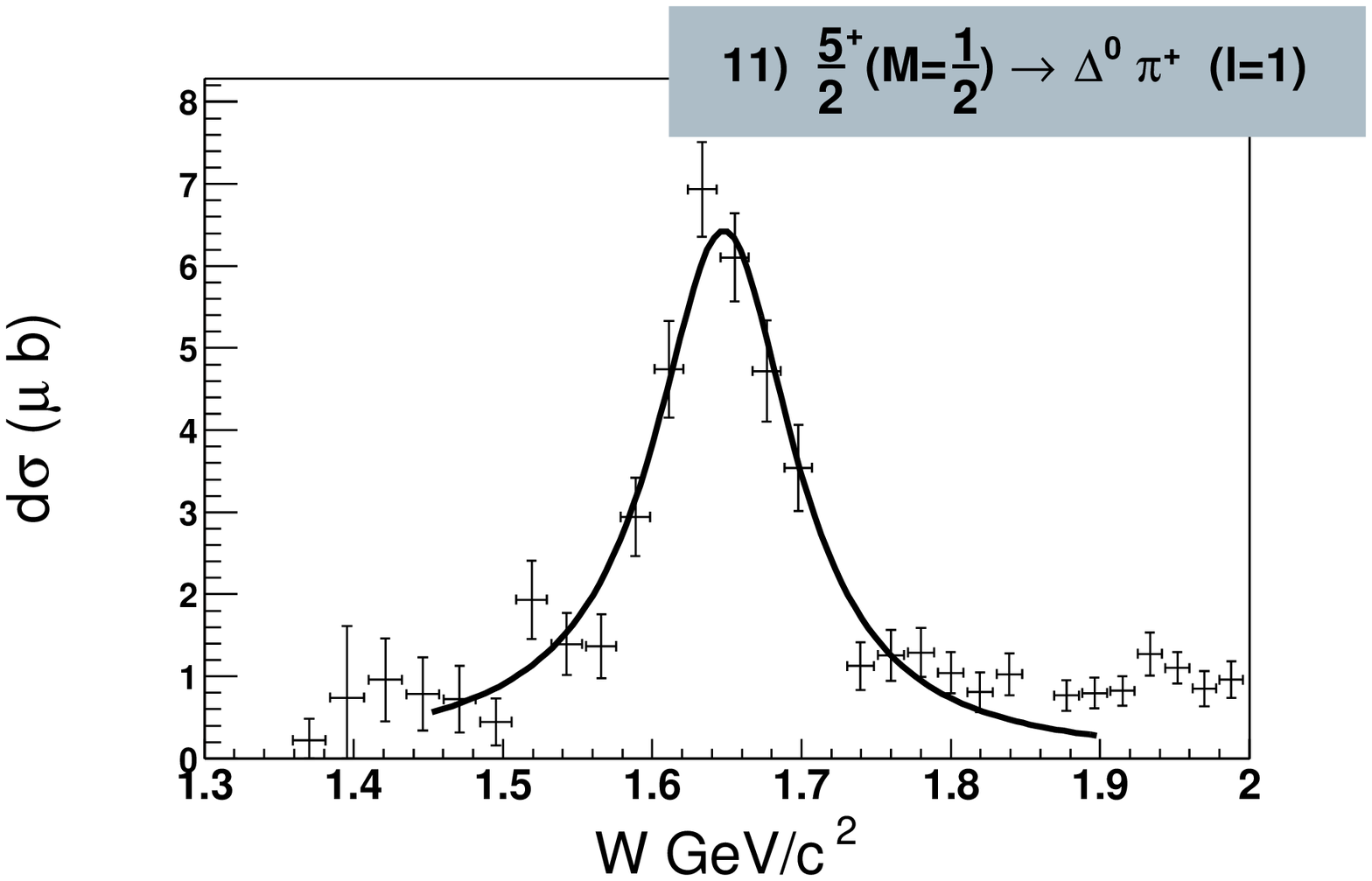}}
\centering{\includegraphics{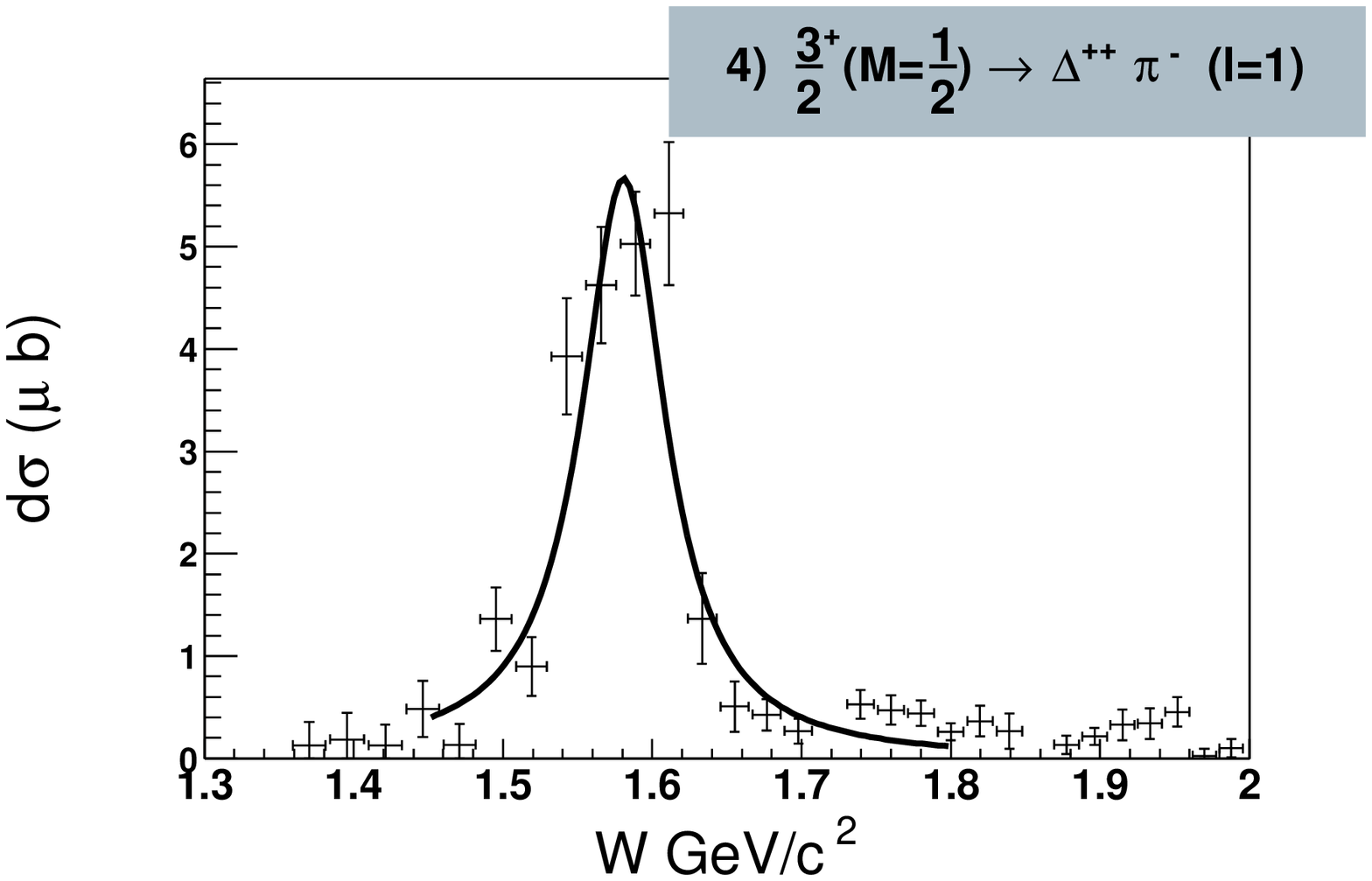}}
\caption{\small Preliminary results of a partial wave analysis in the $\gamma p \rightarrow p\pi^+\pi^-$
showing the $m = {1\over 2}$ spin projections for the well-known $F_{15}(1680)$ (left), and evidence 
for the poorly known $P_{33}(1600)$ (right).}
\label{fig:ppipim_rpi_analysis}
\end{figure}

Although the high quality data of two-pion production
are now available, the analysis in
terms of extraction of resonance parameters has not been fully developed.
The theoretical understanding of these complex processes,
 mainly based on the
tree-diagram isobar approach\cite{oset2}, is also very limited and
preliminary.
Recently, two distinctly different approaches have been applied to
analyze  the photoproduction and          
electroproduction data from JLab\cite{bellis04,mokeev01}
and CB-ELSA\cite{thoma02}. 
The first approach is to adjust the parameters of an isobar
 model\cite{mokeev01} to fit the fully extracted cross section and
polarization asymmetry data~\cite{mokeev01,mokeev03}.
The second one is to 
fit directly the unbinned 
data event-by-event\cite{thoma02}.  We describe them in the remaining part of this
subsection. 

The first approach  makes use of knowledge 
from hadronic production. The energy-dependence of non-resonant processes is parameterized, and   
resonance photocouplings and hadronic couplings are fixed, if known, e.g. from single pion 
processes. Resonances in specific partial waves can be introduced to search for 
undiscovered states. Model parameters are usually fitted to the one-dimensional projections of the 
multi-dimensional differential cross section. Such a model can lead to a qualitatively good 
description of the projected data as shown in Fig.~\ref{fig:diff_cs_ppippim}.
 The method has been used in the analysis of CLAS electroproduction data~\cite{ripani03}.
In this analysis a significant disprepancy was found near W = 1.7 GeV between the data and the 
resonance parametrizations implemented in the fit model. 
This discrepancy was attributed to either inaccurate hadronic couplings for the 
well known $P_{13}(1720)$ resonance determined in the analysis of hadronic experiments, 
or to an additional resonance with $J^P = {3 \over 2}^+$ with either $I = {1 \over 2}$ or 
$I = {3 \over 2}$.
The discrepancy is best visible in the total cross sections for electroproduction, shown in 
Fig. \ref{fig:total_cs_ppippim}.  The dotted line shows the model predictions using resonance 
parameters from single pion electroproduction and from the analysis of $\pi N \rightarrow N\pi \pi$ 
data~\cite{pdg04,kent,pittanl}. The solid line represents the fit when the hadronic coupling of the 
$P_{13}(1720)$ to $\Delta\pi$ and $N\rho$ are allowed to vary much beyond the ranges established 
in the analysis of hadronic data. Alternatively, a new state was introduced with hadronic couplings 
extracted from the data. Table~\ref{tab:clas_ppippim} summarizes results of the analysis using 
a single $P_{13}$ with modified hadronic couplings, and a new $P_{I3}$ state with undetermined isospin
while keeping the PDG $P_{13}(1720)$ hadronic couplings unchanged. In either case, the fit requires a resonance
with hadronic couplings that are significantly different from the ones of 
the $P_{13}(1720)$ state listed by PDG.

\begin{table}[htbp]
\tcaption{PDG parameters for the $P_{13}(1720)$ and parameters resulting from fits to $p\pi^+\pi^-$ 
electroproduction data. }
\centerline{\footnotesize\smalllineskip
\begin{tabular}{l l c c l }\\
\hline
{} &{Mass (MeV)} & $\Gamma$(MeV) & $\Gamma_{\pi\Delta}/\Gamma $(\%) & $\Gamma_{N\rho}/\Gamma $(\%) \\ 
\hline\\
PDG $P_{13}$ & 1725 $\pm$ 20 & 114 $\pm$ 19 & 63 $\pm$ 12 & 19 $\pm$ 9 \\
PDG  & 1650 - 1750 & 100 - 200 & - & 70 - 85 \\
New $P_{I3}$ & 1720 $\pm$ 20 & 88 $\pm$ 17 & 41 $\pm$ 13 & 17 $\pm$ 10 \\
\hline\\
\end{tabular}}
\label{tab:clas_ppippim}
\end{table}
The total photoproduction cross section in the left panel of Fig~\ref{fig:total_cs_ppippim} shows a 
W dependence that is very different from the electroproduction data
in the right panel. In particular, the photoproduction has a
much higher nonresonant contribution largely due
to increased non-resonant $\rho^{\circ}$ production at the photon point. 
Both data are, however, consistent with a strong resonance 
near $W = 1.72$ GeV in the $P_{13}$ partial wave~\cite{mokeev04}.        
The drawback of the approach described above is that when fitting one-dimensional projections of cross sections, correlations 
between the data sets are lost.

The second approach~\cite{bellis04} is based on a partial wave formalism starting from the
T matrix at a given photon energy E:
\begin{eqnarray} 
T_{fi}(E) &=& < p\pi^+\pi^-;\tau_f|T|\gamma p; E> \nonumber \\
 &=&\sum_{\alpha} < p\pi^+\pi^-;\tau_f|\alpha><\alpha|T|\gamma p; E> 
\nonumber \\
&=& \sum_{\alpha} \psi^{\alpha}(\tau_f)V^{\alpha}(E)~,
\end{eqnarray}  
where $\alpha$ denotes all intermediate states, and $\tau_f$ characterizes the final state 
kinematics. The decay amplitude 
$\psi^{\alpha}(\tau_f)=< p\pi^+\pi^-;\tau_f|\alpha>$ is 
calculated using an 
isobar model for specific decay channel, e.g. $\Delta^{++}\pi^-$, $\Delta^{-}\pi^+$, or $p\rho^{\circ}$. 
The production amplitude
 $V^{\alpha}=<\alpha|T|\gamma p; E> $ is then
fitted at fixed energy using
 an unbinned maximum likelihood procedure. This method makes use of 
all information contained in the data, and takes into account all correlations between the 
variables. 

In this analysis a total of 35 partial
waves were included in addition to t-channel processes with
 adjustable parameters. Figure \ref{fig:ppipim_rpi_analysis}
shows intensity distributions in different isobar channels, for the ${5 \over 2}(m= {1\over 2})$,
and  ${3 \over 2}(m = {1\over 2})$ partial waves. Clear signals of the $F_{15}(1680)$ and
the $P_{33}(1600)$ are seen, the latter being a not fully established 3-star resonance. 
In the final analysis the energy-dependence is fitted to a Breit-Wigner form to
determine masses and widths of resonant states. This method is closer to a model-independent approach, 
and can directly 'discover' new resonances in specific partial waves.

\subsubsection{Description of $\gamma p \rightarrow p \pi^{\circ} \pi^{\circ}$ in resonance analyses.} 
\label{sect:dalitz_fit}

\begin{figure}[tphb]
\vspace{60mm} 
\centering{\includegraphics{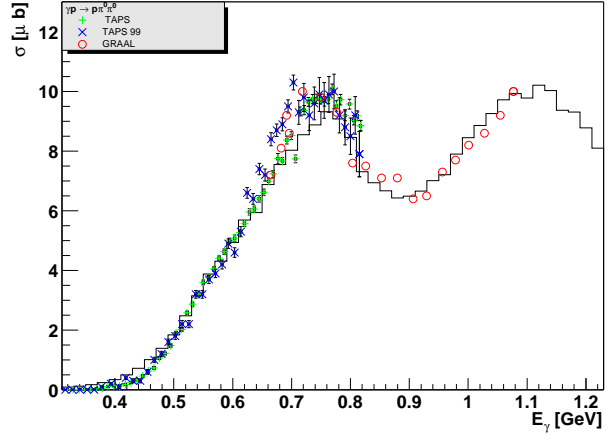}}
\caption{\small Total cross section of
$\gamma p \rightarrow p\pi^{\circ}\pi^{\circ}$ reaction.
The connected line is from the preliminary results of a partial wave analysis.
 The data are from
the measurements with TAPS and GRAAL.}
\label{fig:ppi0pi0_cs_tabs_graal}
\end{figure}

The CB-ELSA collaboration has analysed the $p\pi^{\circ}\pi^{\circ}$ final state using a
more model-dependent version of the partial-wave-analysis described above. Here s-channel 
Breit-Wigner distributions are fitted to the data on an event-by-event basis, therefore 
retaining the correlations in the data. However, the fit is constrained by the parametrized 
energy-dependence of the Breit-Wigner function. In Fig.~\ref{fig:ppi0pi0_cs_tabs_graal} the total cross section for 
$\gamma p \rightarrow \pi^{\circ}\pi^{\circ}$ is shown as extracted from the integral over all 
partial waves contributions in comparison to previous data from TABS~\cite{wolf00} and GRAAL~\cite{assafiri03}. 

\begin{figure}[thpb]
\vspace{8cm} 
\centering{\includegraphics{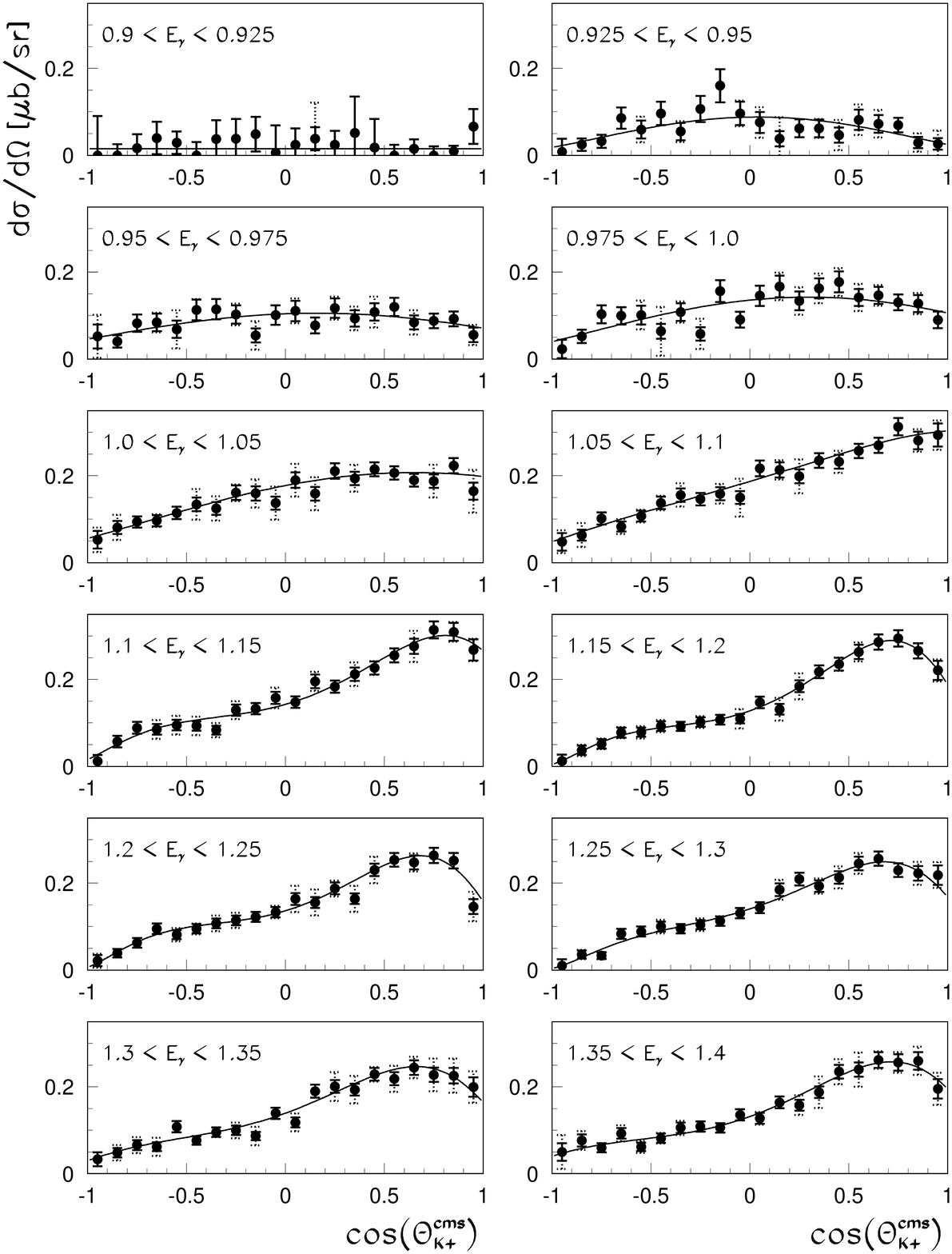}}
\centering{\includegraphics{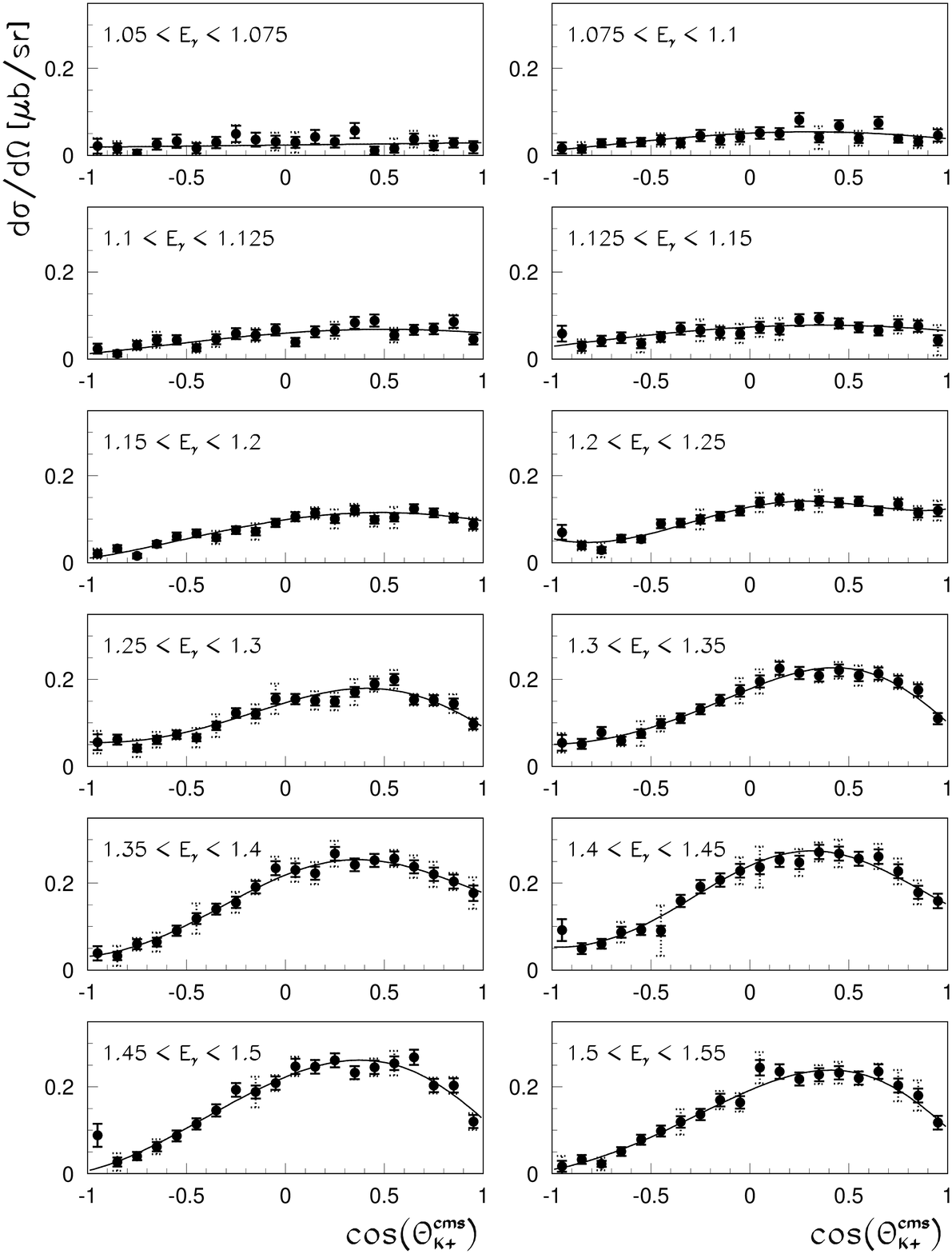}}
\caption{\small Left: Angular distributions of $K^+\Lambda$ photoproduction 
at fixed photon energy $E_\gamma$.
 Right: $K^+\Sigma^{\circ}$ angular distributions. Both data sets are from SAPHIR. The lines represent Legendre fits to the data.}
\label{fig:lambda_photo_angle}
\end{figure} 

\subsection{Kaon production.}

Production of kaons from nucleons has long been recognized 
as a potentially very sensitive tool in the search for excited baryon 
states \cite{caprob98}. 
Analyses of the $K\Lambda$ and $K\Sigma$ channels include the isospin selectivity; 
the $K\Lambda$ final state selects isospin $1 \over 2 $, similar to the $N\eta$ channel, while 
$K\Sigma$ couples to both $N^*$ and $\Delta^*$ resonances. 
An important tool in resonance studies is the measurement of polarization 
observables. The self-analyzing power of the weak 
decay $\Lambda \rightarrow p\pi^-$ can be utilized to measure the $\Lambda$ 
recoil polarization.  
To make full use of this unique feature large acceptance detectors are needed. 

For a long time, the lack of consistent data sets for
$K\Lambda$ and $K\Sigma$ production in a wide kinematics range  
 has hampered the
use of kaon production in the study 
of non-strange baryon resonances.
Moreover, the interpretation of these data, mainly for 
charged $K^+\Lambda$ and $K^+\Sigma^{\circ}$ channels, in terms 
of $N^*$ excitation is complicated by the fact that
they may be dominated by the nonresonant particle-exchange processes.
 Another drawback in comparison to $N\pi$ and $N\pi\pi$ 
is the relatively small cross section, and the lack of known strong resonances 
with a dominant coupling to kaon-hyperon channels. 
 This fact makes it more difficult 
to use strangeness production as a tool in the study of excited baryons, and specifically in 
the search for new resonances. 

Most of the available theoretical
 models\cite{adel,saghai1,will,saghai,saghai2,bennhold00,hbm98,mbh02,j02}
for kaon production are based on the tree-diagrams approach, as described in 
section 4.2. The validity of these tree-diagram models is questionable, as
discussed, for example, in a coupled-channel study\cite{chitab}
 of Kaon photoproduction. We therefore will mainly
focus on the status of the data, not on the results from these
theoretical models.

\subsubsection{Photoproduction of  $K^+\Lambda$ and $K^+\Sigma$} 

 High statistics data of kaon photoproduction
 covering the resonance region are now available 
from the SAPHIR~\cite{glander04} and the CLAS~\cite{mcnabb03} 
collaborations. These new data consist of high statistics
angular distributions as well as $\Lambda$ 
polarization asymmetries, as summarized in 
table \ref{tab:kaon_photo_electro}.

\begin{table}[htbp]
\tcaption{Summary of hyperon photo- and electroproduction data}
\centerline{\footnotesize\smalllineskip
\begin{tabular}{l c c c c l l}\\
\hline
{Reaction} &{Observable} &W range &$\cos\Theta_K^*$ range & $Q^2$  &Experiment \\
{} &\phantom0 &(GeV) &\phantom0 &($GeV^2$) & \phantom0 & \phantom0 \\ 
\hline\\
{$\gamma p \rightarrow \Lambda K$} &$d\sigma / d\Omega$  &$< 2.15$   &-0.95 - +0.95 &&SAPHIR \cite{tran98}\\
				 &$d\sigma / d\Omega$  &$< 2.6$   &-0.95 - +0.95 &&SAPHIR \cite{glander04}\\
				 &$d\sigma / d\Omega$  &$< 2.3$   &-0.85 - +0.85 &&CLAS \cite{mcnabb03}\\
{$\gamma p \rightarrow \Sigma K$}  &$d\sigma / d\Omega$  &$< 2.15$   &-0.95 - +0.95 &&SAPHIR \cite{tran98}\\
			           &$d\sigma / d\Omega$	 &$< 2.6$   &-0.95 - +0.95 &&SAPHIR \cite{glander04}\\
			           &$d\sigma / d\Omega$	 &$< 2.3$   &-0.85 - +0.85 &&CLAS \cite{mcnabb03}\\
{$\gamma p \rightarrow K^+ \vec{\Lambda},\vec{\Sigma} $} & P  &$< 2.3$   &-0.85 - +0.85 &&CLAS \cite{mcnabb03}\\
					 & P &$< 2.6$   &-0.95 - +0.95 &&SAPHIR \cite{glander04}\\
{$ep \rightarrow eK^+\Lambda,~\Sigma$} & $\sigma_{LT}$,$\sigma_{TT}$, $\sigma_T+\epsilon\sigma_L$ &$< 2.5$  &-1.0 - +1.0 &  
$< 3$  &CLAS\cite{mestayer99}\\
{$\vec{e}p \rightarrow eK^+\Lambda$} &$P_x^{\prime}$, $P_z^{\prime}$  &$< 2.15$  &-1.0 - +1.0&
0.3 - 1.5&CLAS\cite{carman03}\\
\hline\\
\end{tabular}}
\label{tab:kaon_photo_electro}
\end{table}

There are significant discrepancies between the CLAS data and 
the published SAPHIR~\cite{tran98} data, while the new data from SAPHIR \cite{glander04} are  
in much better agreement with the CLAS data.  
We therefore disregard the earlier published data from SAPHIR. Unfortunately, most of the 
model calculations have been fitted to the published results, and therefore can not be 
reliably compared to the new data. 

The angular distributions for $K^+\Lambda$ and $K^+\Sigma$ production
are shown in Fig.~\ref{fig:lambda_photo_angle}. We see that 
the  $K^+\Lambda$ data (left panel) show a strong 
forward peaking  for photon energies greater than 1 GeV,
 indicating  the large t-channel contributions. 
For the $K^+\Sigma^{\circ}$ channel (right panel)
 the angular distribution are more symmetric 
or ``resonance-like'' at low energies , but become
soemwhat more forward-peaked at energies above 1.3 GeV.

\begin{figure}[t]
\vspace{10cm} 
\centering{\includegraphics{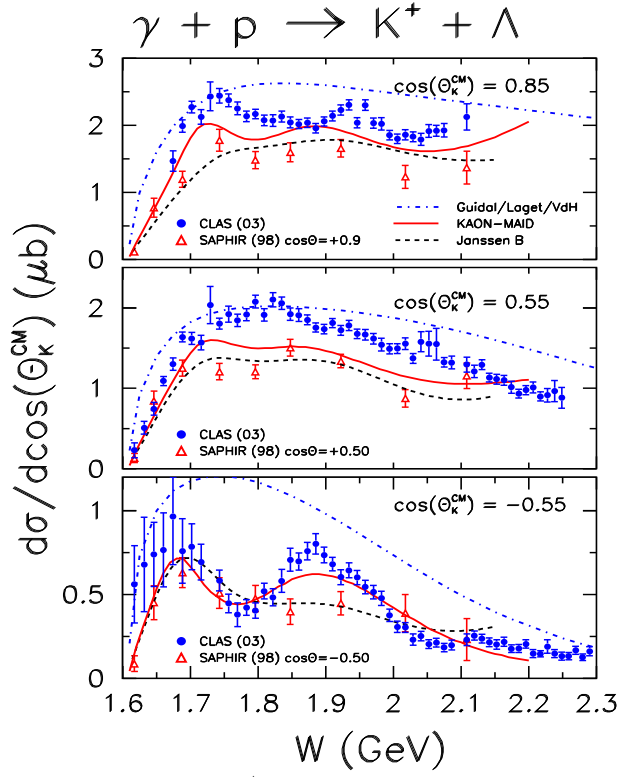}}
\caption{\small W-dependence of $K^+\Lambda$ production for several bins in $\cos\theta^*_K$.
Points with full circles are from CLAS. 
The triangles are older data from SAPHIR. The theoretical curves
are from Ref.201 (Guidal/Laget/vdH), Ref.194 (KAON-MAID), and
Ref.195 (Janssen).}
\label{fig:lambda_dcs2}
\end{figure}

The high statistics of these data allows, for the first time, to identify 
the structures in the 
differential cross section that hint the interference between the resonances 
and the nonresonant backgound. 
The presence of s-channel resonances is particularly evident in the W-dependence
of the differential cross section shown in Figure \ref{fig:lambda_dcs2}. At the most 
forward angles (upper panel), 
two resonance-like structures are visible at $W \approx 1.7$GeV, and at $W \approx 1.95~GeV$. 
The structure at 1.7~GeV could be accomodated by the known states $S_{11}(1650)$,
$P_{11}(1710)$, and $P_{13}(1720)$, if the $K\Lambda$ coupling of these 
states is allowed to vary. From  hadronic processes these couplings
are very poorly known \cite{pdg04}. At intermediate angles (middle panel)
 the data indicate 
a smoother falloff with W, while at backwards angles (lower panel)
another resonance-like 
structure near $W \approx 1.875$ GeV emerges, overlapping with the structure at 
the higher mass. These distributions reveal complex processes,  indicating contributions 
from more than a single resonance near $W = 1.9$~GeV.  

\begin{figure}[tbh]
\vspace{80mm} 
\centering{\includegraphics{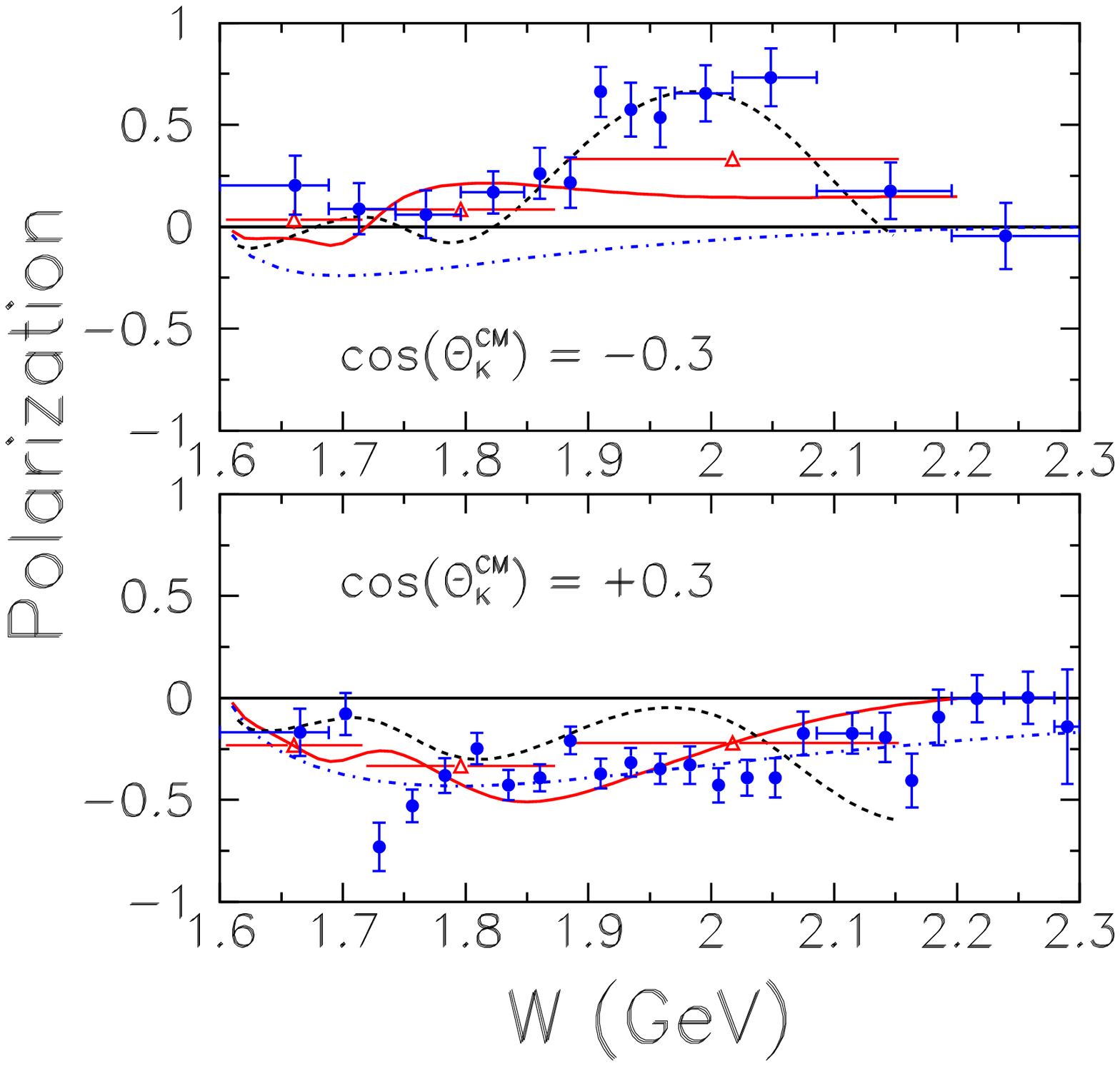}}
\caption{\small  $\Lambda$ polarization in 
$K^+\Lambda$ photoproduction measured with CLAS. The theoretical curves
are the same as those displayed in Fig.36.  }
\label{fig:clas_lambda_polar}
\end{figure}

Samples of the $\Lambda$ polarization measured with CLAS are shown in
 Figure \ref{fig:clas_lambda_polar}.
The data show 
a strong W dependence especially at backward angles (upper panel). 
Comparing the displayed data at two angles, we can conclude that
the  angle distributions of the $\Lambda$ polarization
 change sign from largely negative at forward angles to positive 
at backward angles.

The three theoretical results 
 displayed in
 Figs.\ref{fig:lambda_dcs2} and \ref{fig:clas_lambda_polar}
only  describe very  qualitatively the main features 
of the data in the region of the nucleon resonance region. 
\begin{figure}[tbhp]
\vspace{40mm} 
\centering{\includegraphics{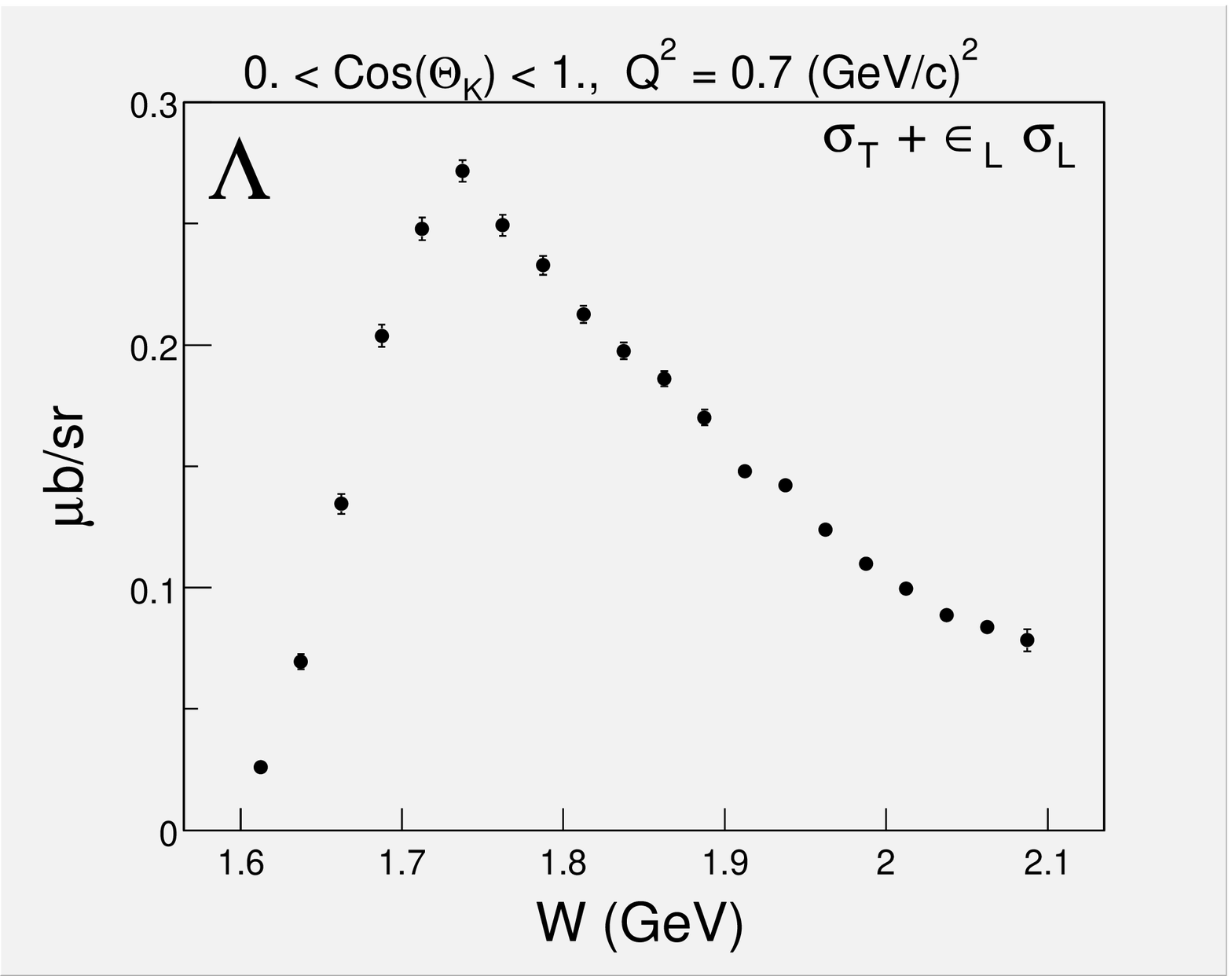}}
\centering{\includegraphics{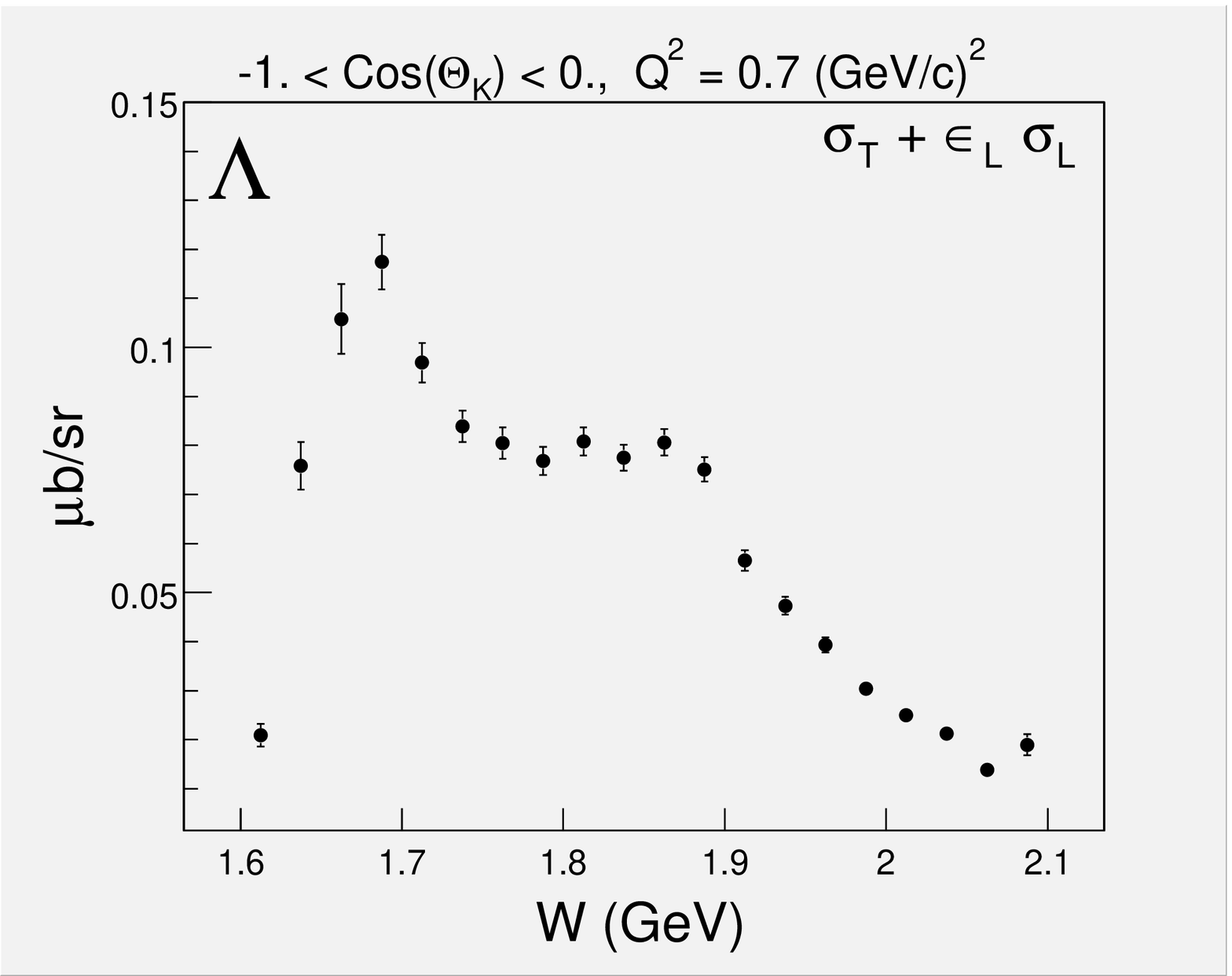}}
\caption{\small  Cross section  from CLAS for $K^+\Lambda$ electroproduction
 at $Q^2=0.7$GeV$^2$  integrated over to the forward hemisphere (left panel) and backward hemisphere 
(right panel) in the center-of-mass angle $\theta^*_{K}$.}
\label{fig:clas_k_lambda_electro1}
\end{figure}

\subsubsection{Electroproduction of $K^+\Lambda$ and $K^+\Sigma$.}

Kaon electroproduction is another tool in the study of non-strange 
nucleon resonances. 
 While the $K^+\Lambda$ and $K^+\Sigma^{\circ}$ photoproduction cross 
section exhibits a complex structure of 
resonant and nonresonant contributions that is difficult 
to disentangle, some of the resonance contributions 
in electroproduction may be enhanced 
at higher $Q^2$ due to their slower form factor falloff compared to other resonances, and 
compared to the background amplitudes. A significant amount of data has become available recently 
from CLAS~\cite{feuerbach,carman03}.
In these experiments the electron beam is polarized, 
and hence the virtual photon also 
has a net circular polarization. 

Figure \ref{fig:clas_k_lambda_electro1} shows samples of
the   $K^+\Lambda$ production cross sections integrated over either
the forward hemisphere (left panel) and backward hemisphere (right panel)
at fixed  $Q^2$. The results  
reveal resonant behavior near W=1.7 GeV
and 1.87 GeV at large angles while at the forward angles 
the resonant structures are masked by the large non-resonant contributions. 
The enhancements in the cross section appear in the same 
invariant mass $W$ range as in photoproduction, 
and are likely due to the same resonances contributions.  
\begin{figure}[hptb]
\vspace{50mm} 
\centering{\includegraphics{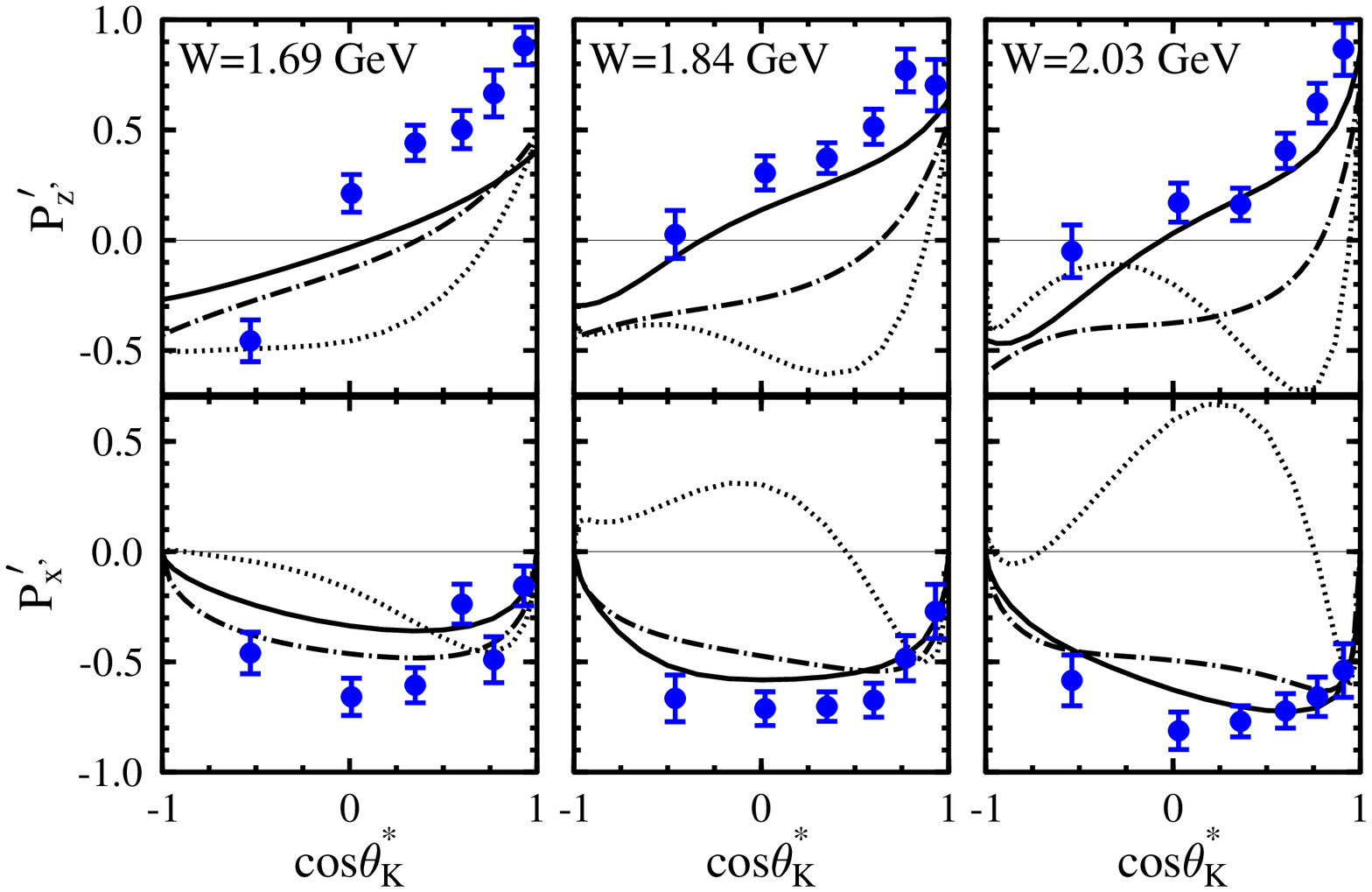}}
\centering{\includegraphics{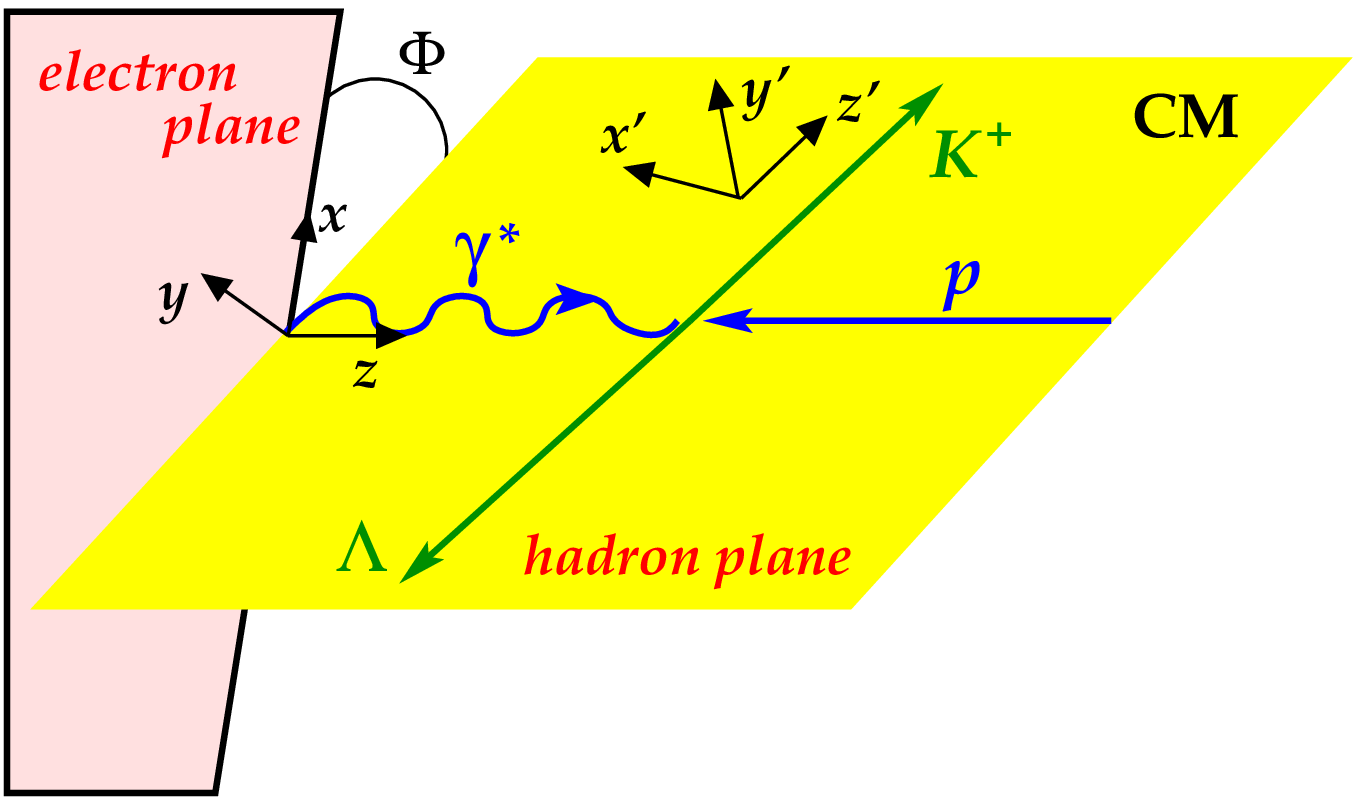}}
\caption{\small Left: Transferred $\Lambda$ polarization in electroproduction from CLAS. 
The data have been integrated over all $\phi$ angles. The curves are discussed in the text.
Right: Coordinate system defining the $\Lambda$ polarization projections. The curves are  
predictions of models in Ref. 81 (dotted),  Ref. 194 (solid), and Ref. 195 (dot-dashed).}
\label{fig:lambda_pol_axes}
\label{fig:clas_lambda_pol_electro1}
\end{figure}

The data of
$\Lambda$ recoil polarization have been obtained in measurements with polarized
electron beams.
The measured  total $\Lambda$ recoil polarization can  be  written as
\begin{eqnarray}
\vec{P}_{\Lambda} = \vec{P}^{\circ} \pm P_e\vec{P}^{\prime}
\end{eqnarray}
where $P_e$ is the electron beam polarization,
 $\vec{P}^{\circ}$ is the {\sl induced} polarization 
which is present without beam polarization, and $\vec{P}^{\prime}$ is
the {\sl transferred} polarization.
Figure \ref{fig:clas_lambda_pol_electro1} displays the data of the transferred
$\Lambda$ polarization integrated over all $Q^2$ for three bins in W. 
The considered  $P_{x^{\prime}}^{\prime}$ and $P_{z^{\prime}}^{\prime}$
are the projections of
 the polarization 
vector $\vec{P}^{\prime}$ onto the $x^{\prime}$ and $z^{\prime}$ 
axes which are also defined in Fig.~\ref{fig:lambda_pol_axes}.
The data show that the  $z^{\prime}$-polarization is large and
is rising with $\cos\theta^*_K$, indicating a t-channel mechanism. 
On the other hand, the $x^{\prime}$ polarization is
large and remains negative throughout the angular range.
None of the displayed theoretical results  from
tree-diagram models~\cite{will,mbh02,j02}
and a Regge model~\cite{guidal03} can give an adequate description of the data.  


To summarize, production of $K^+\Lambda$ and $K^+\Sigma^{\circ}$ from protons exhibit evidence of s-channel 
nucleon resonance contributions in the mass range where no $N^*$ or $\Delta^*$
resonances have been well established. However, resonances are masked by 
large t-channel processes. In order to extract reliable 
information on contributing resonances a better understanding of 
nonresonant processes is needed. 
Currently, the most important task is to continue experimentally to
establish a broad and solid base of consistent data in the strangeness sector, including extensive 
differential cross sections, beam and target polarization asymmetries, and polarization transfer
measurements.
A ``complete'' measurement of all observables which is needed to unambiguously extract all helicity 
amplitudes can be achieved~\cite{barker74,barker75}. This requires use of 
a polarized photon beam and of a polarized 
target and the measurement of the hyperon recoil polarization. 
Experimental effort in this direction
 will continue with a series of new measurements planned 
at JLab~\cite{jlab-e02-112}. 
 On the theoretical side, a dynamical coupled-channel approach, such
as that described in section 4.6,
must be developed to interpret the extracted $N^*$ parameters.

\subsubsection{Photoproduction and electroproduction of vector mesons.} 

\begin{figure}[hptb]
\vspace{65mm}
\centering{\includegraphics{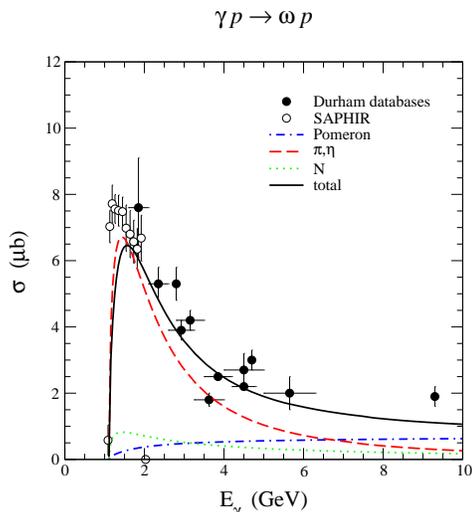}}
\caption{$\gamma p \rightarrow \omega p$ total cross sections. 
The theoretical curves are from the model of Oh, Titov, and Lee$^{86}$.}
\label{fig:omega_total}
\end{figure}

The early investigations of  photoproduction
 and electroproduction of vector mesons 
were mainly in the high energy region where the data can be explained largely
by the diffractive Pomeron-exchange mechanism.
For the study of nucleon resonances in the $p\omega$ channel, measurements have been
performed at ELSA, JLab, and GRAAL to obtain high quality data at energies 
from production thresholds to $W \sim 2.5$  GeV. In this low energy region, the meson-exchange
mechanism plays
 an important role and must be treated correctly for extracting $N^*$
resonance parameters from the data. This is illustrated in Fig.~\ref{fig:omega_total}.
We see that the diffractive Pomeron-exchange(dash-dotted curve)
 becomes negligible at energies near 
$\omega$ photoproduction threshold. The s- and u-channel nucleon terms, 
and $\pi$ and $\eta$ exchanges can account for the main part of the total
cross section. The results shown in Fig.~\ref{fig:omega_total} are from the tree diagram model 
described in section 4.2 and no $N^*$ excitations are included. 
                                                                                
\begin{figure}[hptb]
\vspace{80mm} 
\centering{\includegraphics{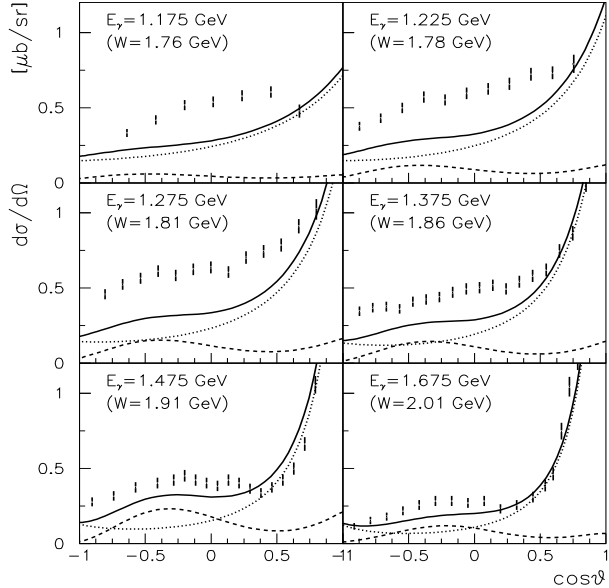}}
\caption{\small Differential cross section for $\gamma p \rightarrow p \omega$ in comparison
with model predictions of Oh, Titov, and Lee$^{86}$.
 The dotted curves  include diffractive production and the 
$\omega \rightarrow \pi^{\circ} \gamma$ vertex. The dashed curves
 are s-channel resonance contributions
using the quark model predictions of Capstick and Roberts.}
\label{fig:clas_omega_photo}
\end{figure}

We will here only describe the status of $\omega$ photoproduction in the 
resonance region( $W <$ 2.5 GeV). The $\rho$ production will not be discussed
since $\rho$'s width is very broad and the coupling of the
$\rho N$ channel to $N^*$ states can be meaningfully
defined only in the analysis involving two pion production channels discussed in section 5.4.
The $\phi$ photoproduction will also not be covered here since $\phi$ meson
has little, if any,
contributions from s-channel resonances, as the 
$s\bar{s}$ quark structure of the $\phi$ makes $N^* \rightarrow N \phi$ an OZI forbidden decay. 

Quark models that also couple to hadronic channels
 predict that $\omega$ photoproduction
off protons is a promising tool
 in the search for undiscovered $N^*$ states~\cite{caprob,capstick92}. As in the case of 
$N\eta$ and $K^+\Lambda$, the $p\omega$ final state, due to the isoscalar nature of the $\omega$, 
is only sensitive to isospin $1 \over 2$  $N^*$ resonances. 
Experimentally, $p\omega$ production has been measured in both magnetic detectors and in 
neutral particle detectors.
In magnetic detectors 
the $p\omega$ channel is usually identified through the $\omega \rightarrow \pi^+ \pi^- \pi^{\circ}$ decay. 
This channel has a 89\% branching ratio. 
Detectors with large acceptance for the detection of photons allows to use the 
$\omega \rightarrow \pi^{\circ} \gamma$ channel with an 8.5\% branching ratio.

The low energy $\omega$ photoproduction data have been obtained ELSA, JLab, and GRAAL.
In Fig. \ref{fig:clas_omega_photo} we show preliminary differential cross sections from CLAS in comparison
with predictions using the model of Oh, Titov, and Lee ~\cite{otl}. 
The model contains contributions from diffractive production, $\pi^{\circ}$ exchange, and 
s-channel $N^*$ contributions with the photocouplings from the constituent quark model 
of Capstick~\cite{capstick92}, and the $N^* \rightarrow p\omega$ couplings predicted in the model by
Capstick and Roberts~\cite{caprob}. At high W and forward angles the cross section is 
completely dominated by the t-channel processes, i.e. diffractive and pion-exchange contributions. 
Resonance contributions are evident at larger angles, and they seem to
play an important role in the entire mass range covered by the data. The quark model perhaps underestimates 
the resonance contributions.

\begin{figure}
\vspace{50mm}
\centering{\includegraphics{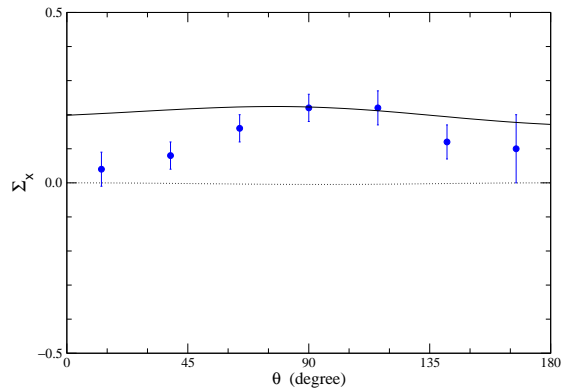}}
\caption{Photon Asymmetry of $\omega$ photoproduction at 1.125 GeV. 
The data are from GRAAL. The dotted curve is from a tree-diagram calculation.
The solid curve includes the one-loop coupled-channel effect calculated in Ref.87.}
\label{fig:omega_as}
\end{figure}

The theoretical models for investigating low energy $\omega$ production are still in a developing
stage. Most of the calculations, such as those displayed in
Fig.~\ref{fig:omega_total} and \ref{fig:clas_omega_photo},
 are based on tree-diagrams models,
It has been recognized that  coupled-channel effects must be accounted for before the model
 can be used reliably to extract resonance parameters from the data.
The importance of the coupled-channel effects 
on $\omega$ photoproduction has been demonstrated in a one-loop  calculation\cite{ohlee} 
based on the dynamical coupled-channel formulation Eqs.(52)-(59). 
As illustrated in Fig.~\ref{fig:omega_as}, the
 photon asymmetry at $E_\gamma =1.125$ GeV 
can be changed drastically if the coupling
with the $\pi N$ channel is included in the calculation. 
The K-matrix coupled-channel model of the Giessen
group\cite{giessen}, based on Eq.(71), 
has also been used to investigate the data from GRAAL. 
More efforts are needed to improve these theoretical approaches.

\subsection{Comments on the search for ``missing'' baryon resonanes}

 The search for all baryon states predicted by the $SU(6)\otimes O(3)$ constituent quark model
is without question of the highest importance for the field. This model has the largest number 
of excitation degrees of freedom of any quark model based on constituent quarks and must form 
the basis for this search.  
Hints of possible new states, even claims of discovery have been presented in the analyses of 
single channel 
processes~\cite{bennhold00,saghai01,ripani03,mokeev04,saghai03,ijr03}
as well as in a coupled-channel analysis~\cite{giessen}.

The analysis of the Giessen group is  currently the
most extensive in searching for new states. They employ
 a coupled-channel K matrix model, as described in section 4.4.2,
and include
all available pion and real photon induced reaction channels, $\gamma N$, 
$\pi N$, $N\pi\pi$, $N\eta$, $K\Lambda$, $K\Sigma$, $N\omega$.
This analysis finds evidence for several new states. 
While there are clear 
indications of new resonances near 1900 MeV in some data sets, 
various analyses do not allow to draw 
definite conclusions on the partial waves that are needed to explain the data.
The Giessen model may be the most promising approach in the search for new states,
however, the large amount of data fitted simultaneously with many fit parameters involved makes 
it difficult to assess the systematic uncertainties in the fit. The dynamical model approaches should
be complemented by more experiment-oriented techniques as the one described in 
section 5.4. Here the binning of data and the evaluation of one-dimensional 
projections of the multi-dimensional parameter space is replaced by an event-by-event analysis 
in specific partial waves that retains the correlations in the data for all variables. Theoretical efforts
are needed to provide realistic background amplitudes that could strengthen the reliability of 
these techniques.

\section{Concluding remarks and outlook}

In the past few years, we have witnessed very significant
progress in the study of $N^*$ physics.
We now have fairly extensive data for $\pi$, $\eta$, $K$, $\omega$ and
$\pi\pi$ production channels. Much more data will soon be available.
The theoretical
models for interpreting these new data and/or  extracting the $N^*$ parameters 
have also been developed accordingly. 

>From the analyses of the single pion data in the
$\Delta$ region, 
quantitative information about the $\gamma N \rightarrow \Delta$ transition
form factors have been obtained. 
With the development of dynamical reaction models, the 
role of  pion cloud effects in determining  the $\Delta$ excitation
has been identified as the source of
 the long-standing 
discrepancy between the data and  the constituent quark model predictions.
Moreover, the $Q^2$-dependence of the $\gamma N \rightarrow \Delta$ 
from factors has also been determined up to about
$Q^2 \sim 6$ (GeV/c)$^2$ and found to be in good agreement with the predictions
from a dynamical model.
The extracted $M1$, $E2$ and $C2$ $\gamma N \rightarrow \Delta$ transition
form factors should be considered along with the proton and neutron
form factors as benchmark data for testing various hadron models as well as
Lattice QCD calculations. 

The combined analyses of the $\pi$ and
$\eta$ production data had led to a rather quantitative, perhaps nearly 
model independent,
determination of several $N^*$ parameters in the second resonance
region. However, a correct interpretation of the extracted $N^*$ parameters 
in terms of the current hadron model predictions requires a rigorous
investigation of the dynamical coupled-channel effects which are not 
included in the employed amplitude analyses based on either the
K-matrix isobar model or the dispersion relations approach.

The analyses of the $K$, $\pi\pi$ and $\omega$ channels are still in
the developing stage. So far, most of the analyses are based on the
tree-diagram models with the isobar parameterization for
the $N^*$ excitations.
The final state interactions, as required by the unitarity condition,
are either neglected completely or
calculated perturbatively using effective Lagrangians.
The coupled-channel K-matrix effective Lagrangian model, pioneered by
the Giessen Group, looks very promising
for extracting the $N^*$ parameters from a combined analysis of
all channels. But much works are needed to reduce 
the uncertainties in their non-resonant parameters 
and to account for the $\pi\pi N$ unitarity condition.  
For a rigorous interpretation of the extracted resonance parameters in
terms of predictions from hadron models or Lattice QCD calculations, 
the analyses based on the
dynamical coupled-channel model, as given in section 4, are indispensable.
Progress is being made in this direction.

%
In the search for new baryon states, progress is being made in developing 
partial wave analysis procedures that make full use of correlations in the  
multi-dimensional phase space presented in the complex final states. This effort 
must be supported by the development of the theory to obtain improved descriptions
of background contributions to specific partial waves.
 
To end, we mention that the progress we have made in the past decade resulted
from rather close collaborations between experimentalists and theorists. 
With much more complex data to be analyzed and interpreted,
 such collaborations must be continued and extended in order to
bring the study of $N^*$ physics to a complete success.

\vspace{1cm}\noindent 
{\bf Acknowledgments}
This work was supported in part by the U.S. Department of Energy,
Office of  Nuclear Physics, under Contract No. W-31-109-ENG-38, and
in part under Contract No. DE-AC05-84ER40150.


\end{document}